# Electronic properties and phase transitions in low-dimensional semiconductors

**A. M. Panich**
Department of Physics, Ben-Gurion University of the Negev, P.O. Box 653, Beer Sheva 84105, Israel
e-mail: pan@bgu.ac.il

**Abstract**
We present the first review of the current state of the literature on electronic properties and phase transitions in TlX and TlMX$_2$ (M = Ga, In; X = Se, S, Te) compounds. These chalcogenides belong to a family of the low-dimensional semiconductors possessing chain or layered structure. They are of significant interest because of their highly anisotropic properties, semi- and photoconductivity, non-linear effects in their *I-V* characteristics (including a region of negative differential resistance), switching and memory effects, second harmonic optical generation, relaxor behavior and potential applications for optoelectronic devices. We review the crystal structure of TlX and TlMX$_2$ compounds, their transport properties under ambient conditions, experimental and theoretical studies of the electronic structure, transport properties and semiconductor-metal phase transitions under high pressure, and sequences of temperature-induced structural phase transitions with intermediate incommensurate states. Electronic nature of the ferroelectric phase transitions in the above-mentioned compounds, as well as relaxor behavior, nanodomains and possible occurrence of quantum dots in doped and irradiated crystals is discussed.

**Contents** page

## 1. Introduction

Over the past decades there has been considerable interest in the physics of low-dimensional materials that exhibit highly anisotropic properties. In these crystals, the atomic arrangement is such that the electrons are constrained to move preferentially in only one or two directions, and thus the systems are described as having reduced dimensionality, which has some unusual consequences that are responsible for the intense development in this field.

The aim of this article is to provide the readers with the first review of the electronic properties and phase transitions in TlX and TlMX$_2$ (M = Ga, In; X = Se, S, Te) compounds. These binary and ternary chalcogenides belong to $A^3B^6$ and $A^3B^3C^6_2$ families of low-dimensional semiconductors possessing chain or layered structure. They are of significant interest because of their highly anisotropic properties, semiconductivity, photoconductivity, and potential applications for optoelectronic devices. They exhibit non-linear effects in their *I-V* characteristics (including a region of negative differential resistance), switching and memory effects. Electrical conductivity of several chain-type crystals exhibits time oscillations and intermittency. Second harmonic optical generation has been reported in TlInS$_2$. Layered TlMX$_2$ compounds were the first low-dimensional semiconductors in which a series of phase transitions with modulated structures was discovered. Furthermore, being doped with some impurity atoms or subjected to gamma- irradiation, TlInS$_2$ compound exhibits relaxor behavior and formation of the nano-sized polar domains. Thallium sulphide and thallium selenide nanorods, which show the quantum confinement effects, have recently been synthesized and studied.

Owing to the significant scientific interest, a lot of experimental techniques, such as x-ray and neutron diffraction, specific heat and dielectric measurements, nuclear magnetic resonance and electron paramagnetic resonance, dielectric sub-millimeter spectroscopy, IR spectroscopy, Raman and Mandelshtam-Brillouin scattering, inelastic neutron scattering, Mossbauer spectroscopy, etc. have been used to study these compounds. The great advantage in these studies is the opportunity of growing the sizable single crystals, which expanded the experimental potentialities of the investigators. Growing of mixed crystals (solid solutions) allows the design of materials with tailored properties for eventual applications in electronics and optoelectronics.

The review covers the recent literature and author's data on electronic properties, pressure- and temperature-induced phase transitions and incommensurate states in the TlX and TlMX$_2$ compounds.

## 2. Crystal structure at ambient conditions

The family of the TlX and TlMX$_2$ (M = Ga, In; X = Se, S, Te) compounds belongs to a group of low-dimensional semiconductors possessing chain or layered structure [1-13]. Room temperature x-ray diffraction (XRD) measurements showed that TlGaSe$_2$, TlGaS$_2$ and TlInS$_2$ are layered compounds [1-6].

For example, the structure of TlGaSe$_2$ belongs to monoclinic symmetry, the space group is C2/c-$C_{2h}^6$, $a$=10.772 Å, $b$=10.771 Å and $c$=15.636 Å, $\beta$=100.6°, Z=16 [2,5,6]. It crystallizes as a structure with two anion layers stacked along [001] ($c^*$ direction) in the unit cell. The structural motive of the layers (Figure

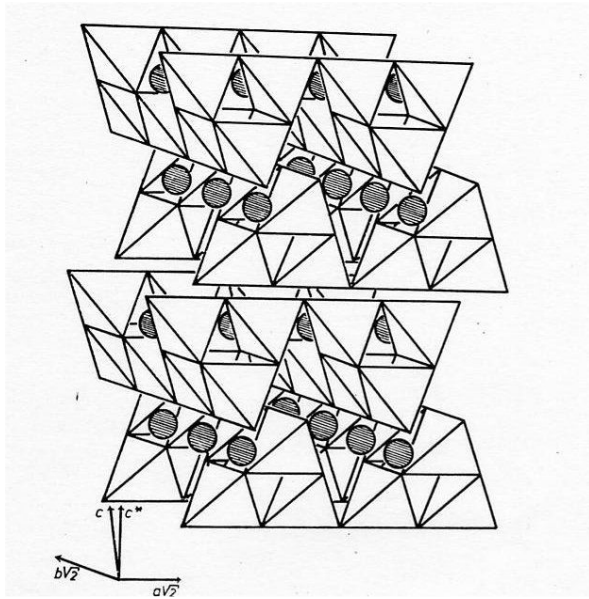

Figure1. Sketch of the layered structure of monoclinic TlGaSe$_2$. The layers are composed of corner-linked Ga$_4$Se$_{10}$ tetrahedra. Tl ions are shown by black circles (From [6]).

1) comprises large corner-linked Ga$_4$Se$_{10}$ tetrahedra consisting of four corner-linked GaSe$_4$ tetrahedra [2,6]. The average Ga-Se distance, 2.39 Å, is close to the sum of the covalent radii of Ga (1.26 Å) and Se (1.17 Å), respectively, and the average Se-Ga-Se angle is 109.5° [2], supporting occurrence of covalent $sp^3$ Ga-Se bonds. Two adjacent layers are turned relative to each other by 90° and kept together by Tl$^{1+}$ ions, which are located in trigonal prismatic voids between the layers, on straight lines along the [110] and [1$\bar{1}$0] directions that are parallel to the edges of the Ga$_4$Se$_{10}$ groups. Each Tl atom is surrounded by six Se atoms, forming trigonal-prismatic TlSe$_6$ polyhedra. The average Tl-Se bond length is 3.45 Å, a little shorter than the sum of the ionic radii of Tl$^{1+}$ (1.50 Å for coordination number CN=6 [14]) and Se$^{2-}$ (1.98 Å). The average Tl-Tl spacing in chains is of 3.81 Å.

Table 1. Structures of layered compounds at ambient temperature.

| Compound | Structure | Space group | $a$, Å | $b$, Å | $c$, Å | $\beta$ | Z | Ref. |
|---|---|---|---|---|---|---|---|---|
| TlGaS$_2$ | monoclinic | $C_{2h}^6 - C2/c$ | 10.2990 | 10.2840 | 15.175 | 99.603° | 16 | 4, 6 |
| TlGaSe$_2$ | monoclinic | $C_{2h}^6 - C2/c$ | 10.772 | 10.771 | 15.636 | 100.6° | 16 | 2, 6 |
| TlInS$_2$ | monoclinic | $C_{2h}^6 - C2/c$ | 10.90 | 10.94 | 15.18 | 100.21° | 16 | 3, 6 |
| TlS | monoclinic | $C2$ | 11.018 | 11.039 | 60.16= 4×15.039 | 100.69° | 128 | 11, 13 |
| TlS | tetragonal | P4$_1$2$_1$2 | 7.803 | 7.803 | 29.55 | | 32 | 12 |

TlGaS$_2$ and TlInS$_2$ are isostructural to TlGaSe$_2$ [1,3-6]. The structural parameters of the compounds are given in Table 1.

In contrast, TlSe, TlGaTe$_2$, TlInTe$_2$ and TlInSe$_2$ show chain structure [7-10], often called in literature as B37 TlSe type. The crystal structure of TlSe [7,10] belongs to the tetragonal symmetry, the space group is $D_{4h}^{18}$-I4/mcm, and the lattice parameters are $a=b=8.02$ Å and $c=6.79$ Å, Z=4. TlSe is a mixed valence compound. Its formula should be more accurately written as Tl$^{1+}$Tl$^{3+}$(Se$^{2-}$)$_2$. The trivalent and univalent thallium ions occupy two crystallographically inequivalent sites. The Tl$^{3+}$ cations form covalent ($sp^3$) Tl-Se bonds and are located at the centers of Tl$^{3+}$Se$_4^{2-}$ tetrahedra, which are linked by common horizontal edges and form linear chains along the $c$-axis. The Tl$^{3+}$-Se distance, 2.67 Å, is close to the sum of the covalent radii of Tl (1.49 Å) and Se (1.17 Å), respectively; the Se-Tl-Se angle is 115°. Each univalent Tl$^{1+}$ cation is surrounded by eight chalcogen atoms, which form slightly deformed Thomson cubes, which are skewed by a small angle. Columns of Thomson cubes with common square faces are parallel to the $c$ axis and alternate with the columns of the aforementioned Tl$^{3+}$Se$_4^{2-}$ tetrahedra. Tl$^{1+}$-Se distances are 3.43 Å, a little shorter than the sum of the ionic radii of Tl$^{1+}$ (1.59 Å for CN=8 [14]) and Se$^{2-}$ (1.98 Å). The Tl$^{1+}$ - Tl$^{1+}$ and Tl$^{3+}$ - Tl$^{3+}$ distances in the chains are 3.49 Å [7], while the distances between these atoms in the (001) plane are 5.67 Å. Each Tl$^{1+}$ ion has four Tl$^{3+}$ neighbors at 4.01 Å in the $a,b$-plane, and, in reverse, each Tl$^{3+}$ ion has four Tl$^{1+}$ neighbors at the same distance. Projection of the TlSe structure on the (001) plane is shown in Figure 2.

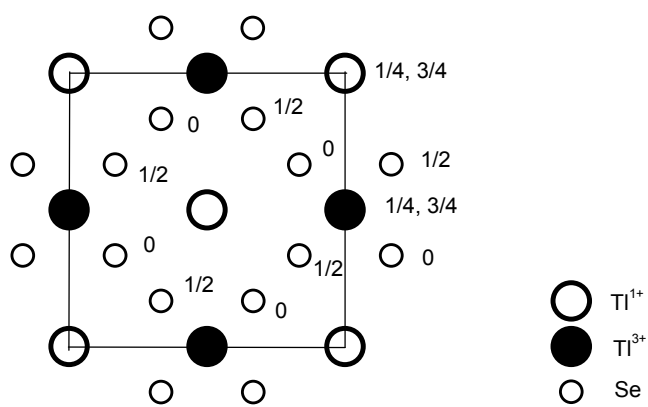

Figure 2. Projection of the TlSe chain-type structure on the $a,b$-plane. The numbers 1/4, 1/2 and 3/4 are the $z/c$ numbers.

TlGaTe$_2$, TlInTe$_2$ and TlInSe$_2$ are isostructural to TlSe. The structural parameters of the chain TlMX$_2$ compounds are given in Table 2. For comparison, the data on InTe [15,16], which is isostructural to TlSe and is described by the formula In$^{1+}$In$^{3+}$(Te$^{2-}$)$_2$, are also included.

Table 2. Structures of chain-type compounds at ambient temperature.

| Compound | Structure | Space group | $a$, Å | $c$, Å | Z | Ref. |
|---|---|---|---|---|---|---|
| TlSe | tetragonal | $D_{4h}^{18} - I4/mcm$ | 8.02<br>8.020(2) | 7.00<br>6.791(2) | 8 | 7<br>10 |
| TlS | tetragonal | $D_{4h}^{18} - I4/mcm$ | 7.77<br>7.785 | 6.79<br>6.802 | 8<br>8 | 8<br>12 |
| TlGaTe$_2$ | tetragonal | $D_{4h}^{18} - I4/mcm$ | 8.429 | 6.865 | 4 | 9 |
| TlInSe$_2$ | tetragonal | $D_{4h}^{18} - I4/mcm$ | 8.075<br>8.02 | 6.847<br>6.826 | 4 | 9<br>6 |
| TlInTe$_2$ | tetragonal | $D_{4h}^{18} - I4/mcm$ | 8.494 | 7.181 | 4 | 9 |
| TlTe | tetragonal | $D_{4h}^{18} - I4/mcm$ | 12.961<br>12.953 | 6.18<br>6.173 | 16<br>16 | 17<br>18 |
| InTe | tetragonal | $D_{4h}^{18} - I4/mcm$ | 8.444 | 7.136 | 8 | 15,16 |

For thallium monosulfide, TlS, both chain (tetragonal) and layered (monoclinic) modifications have been obtained [8, 11-13]. The former one is a mixed valence compound that is isostructural to TlSe and belongs to the tetragonal symmetry [8,12]. The latter modification of TlS [11-13] is isostructural to TlGaSe$_2$, with Tl$^{3+}$ ions instead of Ga$^{3+}$ ions, and belongs to the monoclinic symmetry. Furthermore, a layered tetragonal modification of TlS was also reported [12]. The structural parameters of the chain and layered TlS modifications are given in Tables 1, 2.

The structure of TlTe (Figure 3) also belongs to the tetragonal symmetry and space group $D_{4h}^{18} - I4/mcm$ [17,18]. However, it is considerably different from the structure of TlSe. Instead of crystallizing as mixed valence compound Tl$^{1+}$Tl$^{3+}$(Te$^{2-}$)$_2$, it forms a poly-anionic structure fragments, $Tl^+(Te_n)_{1/n}^-$, revealing univalent Tl$^+$ cations and a polytelluric counterpart with linear equidistant Te chains in the [001] direction at distances 3.0863 Å. One half of these chains is unbranched; the other one consists of linear [Te$_3$]$_n$ units stacked cross-shaped one upon the other.

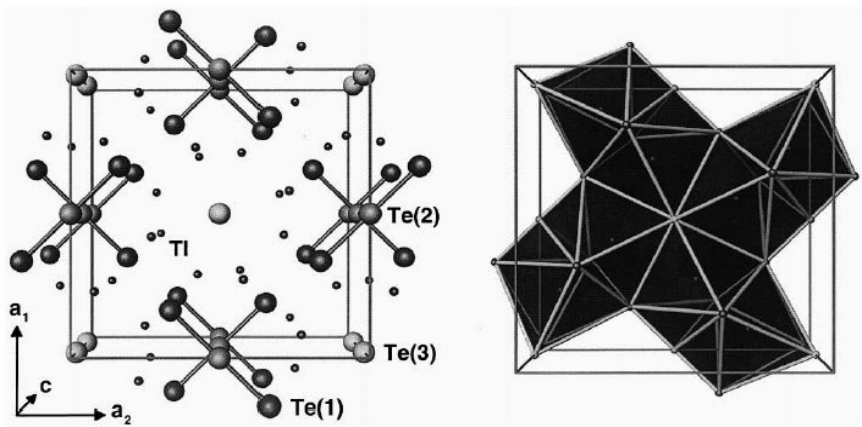

Figure 3. Crystal structure of room temperature phase of TlTe along [001] direction (From [18]).

We note that TlGaSe$_2$, TlGaS$_2$, TlInS$_2$, TlInSe$_2$, TlGaTe$_2$ and TlInTe$_2$ crystals allow a substitution

among (i) Ga and In cations and among (ii) Se, S and Te anions, respectively, and therefore form a continuous series of mixed crystals (solid solutions) of $TlIn_xGa_{1-x}S_2$, $TlIn_xGa_{1-x}Se_2$, $TlInS_{2(1-x)}Se_{2x}$, etc. in the whole range of concentrations ($0 \leq x \leq 1$). These compositions show a variation of the lattice parameters depending on *x*. At that, when the first and the last member of series belong to different symmetry and space group (say, monoclinic $TlGaSe_2$ and tetragonal $TlInSe_2$), a structural phase transformation from monoclinic to tetragonal phase is observed at some value of *x*. The composition variations of the lattice parameters in $TlGa(S_{1-x}Se_x)_2$ layered mixed crystals have been reported by Gasanly et al [19]. For more detailed information about the structure of such mixed compounds, we refer the reader to references [19-26]. Furhtermore, composition variations of the lattice parameters in $Tl_{2x}In_{2(1-x)}Se_2$ layered mixed crystals have been reported by Hatzisymeon et al [27].

### 3. Transport properties under ambient conditions

Numerous electrical conductivity, photoconductivity and optical measurements [28-81] showed that all TlX and $TlMX_2$ compounds examined in this review, except for TlTe, are semiconductors at ambient conditions. At that, layered crystals exhibit significant (and also temperature dependent) anisotropy of the electric conductivity [28-33]. Mustafaeva et al. [28,30] and Aliev et al. [29] reported that the difference in the $\sigma_\perp$ and $\sigma_\parallel$ values in thallium indium sulfide is of the order of magnitude [29,30], in thallium gallium sulfide – almost three orders of magnitude [29], while in thallium gallium selenide the ratio of $\sigma_\perp/\sigma_\parallel$ varies from $10^8$ to $10^6$ in the temperature range 90 to 250 K [28] (here $\sigma_\perp$ is the in-plane conductivity and $\sigma_\parallel$ is the conductivity along the *c* axis, i.e. perpendicular to the *a,b* plane). At that, Mustafaeva et al. [28,30] suggested occurrence of the hopping conductivity between the states localized near the Fermi level both along and across the layers in the layered crystals $TlGaSe_2$, $TlGaS_2$ and $TlInS_2$. Hanias et al [31], who also measured temperature dependence of the conductivity along and perpendicular to the *c* axis (Figure 4), reported $\sigma_\perp/\sigma_\parallel$ ratios as $\sim 10^4$, $\sim 10^5$ and $\sim 10^1$ for $TlInS_2$, $TlGaSe_2$ and $TlGaS_2$, respectively. Monoclinic, layer-type semiconductor TlS crystal also show anisotropic conduction [32,33]; the conductivity within the layer is about two orders of magnitude higher than that normal to the layer. These findings correspond to the two-dimensional-like (2D) behavior. A disagreement in the $\sigma_\perp/\sigma_\parallel$ values in the same compounds obtained by different authors is surprising and is presumably owing to the presence of different kinds of uncontrolled structural defects and impurities. Besides this, proper orientation of crystals for measurements along the crystallographic axes, and knowledge of the real amount of chalcogen atoms are among the other factors that strongly influence the results of measurements. Possible existence of polytypes may also be a reason for some controversies concerning the optical and other properties of these crystals.

The values of the energy band gaps [31-39,41,58] of the layered compounds are given in Table 3. The band gap variation in the mixed crystals with layered $TlGaSe_2$-type structure has been reported in Ref. [22,23,42-44]. E.g., in the mixed $(TlGaSe_2)_{1-x}(TlInS_2)_x$ single crystals, the energy band gap varies linearly with *x* at $0 \leq x \leq 0.4$; it deviates from linearity at $x = 0.6$ [43].

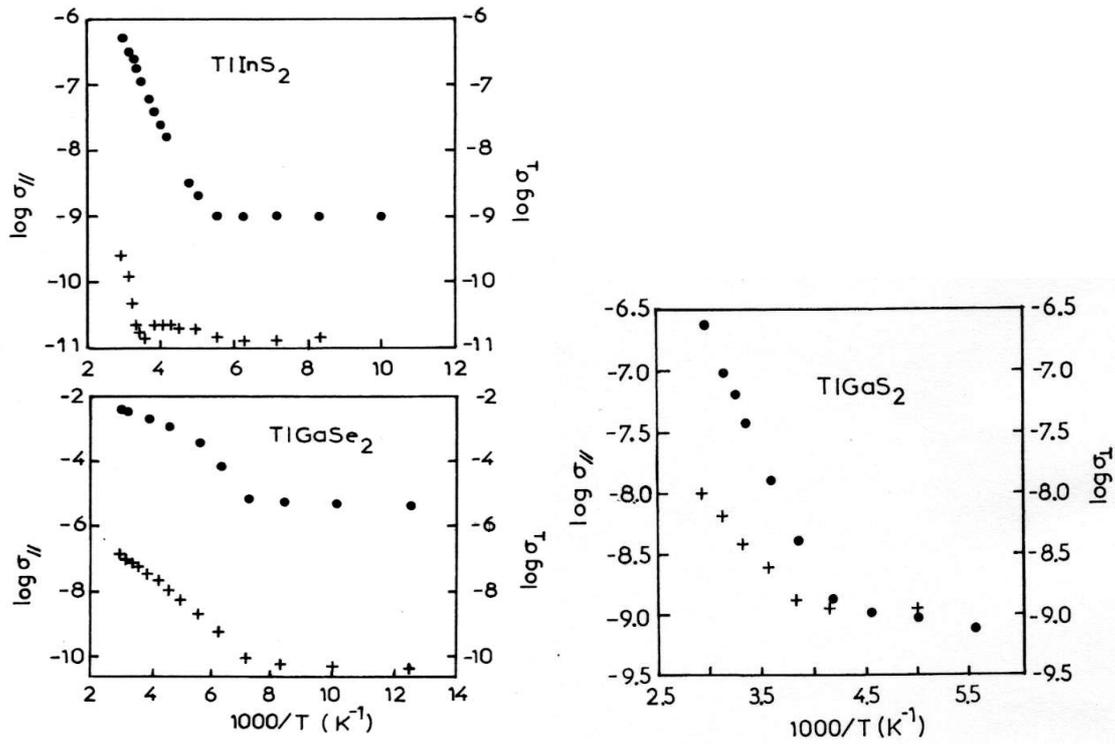

Figure 4. Temperature dependence of conductivity of $TlInS_2$, $TlGaSe_2$, and $TlGaS_2$ crystals along the $c$ axis ($\sigma_\parallel$, crosses) and in the $a,b$ plane ($\sigma_\perp$, circles) (From [31]).

Table 3. Band gaps of TlX and $TlMX_2$ compounds.

| Compound | Thermal band gap, eV | Optical band gap, eV | Ref. |
|---|---|---|---|
| $TlGaSe_2$ | 2.1 – 2.2 | 1.83- 2.23 | 31, 34-36, 41 |
| $TlGaS_2$ | 2.55 - 2.64 | 2.38 – 2.54 | 31, 36, 37, 41 |
| $TlInS_2$ | 2.45 – 2.56 | 2.28 –2.55 | 31,36, 38, 39, 58 |
| TlS layered monoclinic | 0.9 | 1.1 | 32, 33 |
| $TlInSe_2$ | 1.12 | 1.07 – 1.44 | 39, 56,58 |
| TlSe | 0.56 - 0.71 | 0.72 – 1.03 | 46-49 |
| TlS chain tetragonal | 0.94 | 1.16 – 1.57 | 32, 40, 46, 55 |
| $TlGaTe_2$ | 0.84 | | 59 |
| $TlInTe_2$ | 0.7 - 0.8 | 0.97 – 1.1 | 39, 56-58 |
| InTe | 0.34 ? | 1.16 | 66-68 |

Let us now turn to the transport properties of the chain-type compounds. Electrical transport study and optical measurements [45-49] have shown that TlSe is a semiconductor with the energy gap measured by different authors as 0.6 to 1.0 eV at 300 K. Qualitatively, such behavior may be explained by the existence of structural constraints upon electron transfer between the chemically distinct $Tl^{1+}$ and $Tl^{3+}$ ions that occupy two different crystallographic positions. Guseinov et al. [50] and Hussein et al. [51] observed anisotropy of the electrical conductivity in [110] and [001] directions. The Hall mobilities derived from these measurements at room temperature were of the order $\mu_\perp$ = 42.66 cm$^2$ V$^{-1}$ s$^{-1}$ and $\mu_\parallel$ = 112.2 cm$^2$ V$^{-1}$

s$^{-1}$. Abdullaev et al. [52], who studied resistance and magnetoresistance of TlSe single crystals at 1.3 to 300 K, also observed a difference in resistivity in two directions, i.e. parallel ($\rho_{\parallel}$) and perpendicular ($\rho_{\perp}$) to the *c*-axis, in particular at low temperatures, from 1.3 to 5 K. However, this difference varied from sample to sample, showing both $\rho_{\parallel} > \rho_{\perp}$ and $\rho_{\parallel} < \rho_{\perp}$ cases. A "metallic-like" behavior, observed in some samples, was attributed to the impurity conductivity. According to the measurements of Allakhverdiev et al. [53], TlSe shows $\sigma_{\parallel} < \sigma_{\perp}$ under ambient conditions, while Rabinal et al. [54] reported that $\rho_{\parallel}/\rho_{\perp} = 1.92$, also under ambient conditions. In any case, one is led to a conclusion that TlSe nevertheless exhibits a three-dimensional-like (3D) electronic nature rather than one-dimensional (1D) nature, in spite of its chain-like structure. The explanation of such a behavior will be done in the next sections.

Semiconductor properties of the chain-type thallium monosulfide have been established by Kashida et al. [32,33] and Nagat [55]. Rabinal et al [56] showed that the chain-like TlInSe$_2$ and TlInTe$_2$ crystals exhibit $\rho_{\parallel}/\rho_{\perp} = 0.004$ and 6.0, respectively, at ambient conditions. While the former can be attributed to 1D behavior, the latter definitely cannot be.

The energy band gap values [32,39,40,46-49,55-59] for the chain-type crystals are given in Table 3. Measurements of the band gap variation in the mixed crystals with TlSe-type structure have been reported by Allakhverdiev et al. [60].

Chain TlInSe$_2$, TlGaTe$_2$, and TlInTe$_2$ crystals exhibit current-voltage (I-V) characteristics that consist of two parts: a linear (Ohmic) regime at low current densities and a non-linear (S-type) regime at higher current densities [39, 61-64]. In the latter regime, a well-formed region of negative differential resistance appears. Additionally, voltage oscillations were observed in TlInTe$_2$ [39, 62-64], revealing two components in the signal, a quasi-periodic component and a chaotic component. Such oscillations were explained suggesting that the conductivity signal is formed by two concurring effects: jumping between different levels of conductivity and fluctuations on these levels [64]. Abdullaev and Aliev [65] reported on switching and memory effect in thallium indium selenide, TlInSe$_2$.

InTe crystal is a semiconductor [66-68] with the optical band gap is 1.16 eV [68]. (The thermal band gap value in InTe determined by Hussein [66] from Hall coefficient studies, 0.34 eV, seems to be incorrect). Analogously to the compounds mentioned above, InTe exhibits noticeable anisotropy of the conductivity [67,68]. The $\sigma_{\parallel}/\sigma_{\perp}$ ratio was shown to be temperature dependent. The carrier mobility µ is also anisotropic and temperature dependent; at that, the mobility perpendicular to c-axis, $\mu_{\perp}$, which increases with temperature exponentially above 140 K with an activation energy of 0.03 eV, was attributed to the hopping mechanism due to the barriers between the chains.

In contrast to semiconductors mentioned above, the room temperature phase of thallium telluride exhibits semimetallic behavior [69,70].

The reviewed semiconductor compounds reveal pronounced photoconductive properties. The photoconductivity occurs due to the excitation of carriers from the valence band and impurity levels to the conduction band. For more detailed information about photoconductive characteristics of TlX and TlMX$_2$ crystals and their solid solutions, the reader is referred to Ref. [22,23,32,33,42,43,58,71-81].

# 4. Electronic structure: experimental studies

Electronic structure of the semiconductor compounds under review was studied experimentally and theoretically in a number of papers. Experimental data have mainly been obtained by means of photoemission spectroscopy and nuclear magnetic resonance (NMR) techniques.

## 4.1. Photoemission measurements

X-ray photoelectron spectroscopy (XPS) is a quantitative spectroscopic technique in which a sample is irradiated with a beam of monochromatic x-rays, and the energies and number of the resulting photoelectrons, escaping from the surface (typically from the depth of ~1 to 10 nm), are measured under ultra-high vacuum conditions. The collected photoelectrons result in a spectrum of electron intensity as a function of the measured kinetic energy. The kinetic energies of the emitted photoelectrons are converted into binding energy values, which are characteristic of the chemical bonding and molecular orbital structure of the material. Besides X-ray, the typical beam origins are helium gas source of ultraviolet light and synchrotron radiation. The experimental photoemission spectra usually exhibit only several pronounced peaks whose resolution is left much to be desired, and sometimes the information acquired is rather scanty in comparison with the detailed band structure calculations. Noticeable improvement may be reached by measuring the angle-resolved spectra. Very useful approach in the spectra interpretation is comparison of the data on (i) isostructural compounds such as TlSe, InSe, TlInSe$_2$ and on (ii) series of compounds with variation of ionic/covalent state of the same element [82, 83].

The first XPS study of the thallium chalcogenides Tl$_2$S, TlS, Tl$_2$Se, TlSe, Tl$_5$Te$_3$, TlTe and Tl$_2$Te$_3$ has been carried out by Porte and Tranquard [82], who reported the measurements from both the core and the valence levels. In the thallium telluride series, two intensive doublets coming from Te $4d$ and Tl $5d$ states are accompanied by two weak valence bands. The relative variations of the core level binding energies and the evolution in the valence band structure are consistent with an increase of ionic contribution from Tl(III) to Tl(I) compounds. In thallium sulfides, the authors observed an intensive doublet coming from Tl $5d_{3/2}$ and Tl $5d_{5/2}$ core states, accompanied by two weak valence bands in the range from 0 to 10 eV. Valence band structures of thallium sulfides and selenides were analyzed with regard to the crystal structures. Particular attention was devoted to the structure derived in major part from the Tl $6s$ level. An explanation of the variations observed for this structure in various compounds was advanced, taking into account the peculiarity of Tl $6s$ level participation in the chemical bond.

Kilday et al [83] measured photoelectron energy distribution curves (EDCs) of TlInSe$_2$ and compared them with those of isostructural compounds TlSe and InSe (Figure 5). The main peaks of the TlSe and TlInSe$_2$ spectra, at 1.4 eV below the top of the TlInSe$_2$ valence band E$_v$ that was assigned to zero energy, correspond primarily to Se $4p$ states. The second feature at larger binding energies, from −5 to -7 eV, is primarily due to Tl $6s$ states. Its position is different for TlSe and for TlInSe$_2$ owing to two inequivalent Tl sites in TlSe. One can see in Figure 5 also a feature at -3.5 eV below E$_v$ for TlInSe$_2$, which is not present for TlSe. On the other hand, a similar feature is present in the InSe XPS spectrum (peak C in Figure 5). That feature is due to hybridized $p_x$, $p_y$ states related to In-Se bonds. Thus, one can tentatively interpret the –3.5 eV feature of as due to hybridized In and Se orbitals. Analogously, the positions of the Tl 5*d* and In 4*d* core-level peaks in TlInSe$_2$ measured at photon energy 29 eV, were found to be –12.15, -14.4 eV for Tl 5*d* and -16.7, -17.5 eV for In 4*d* [83]. We note that the Tl 5*d* positions for TlSe are –12.9, -15.1 eV [82]. The above difference between TlInSe$_2$ and TlSe is in qualitative agreement with the replacement of trivalent Tl with trivalent In. The 4*d* level positions in TlInSe$_2$ are close to those in InSe, -16.9 and –17.7 eV. This is the further evidence that In atoms in TlInSe$_2$ are involved in the formation of covalent bonds with Se, like in InSe.

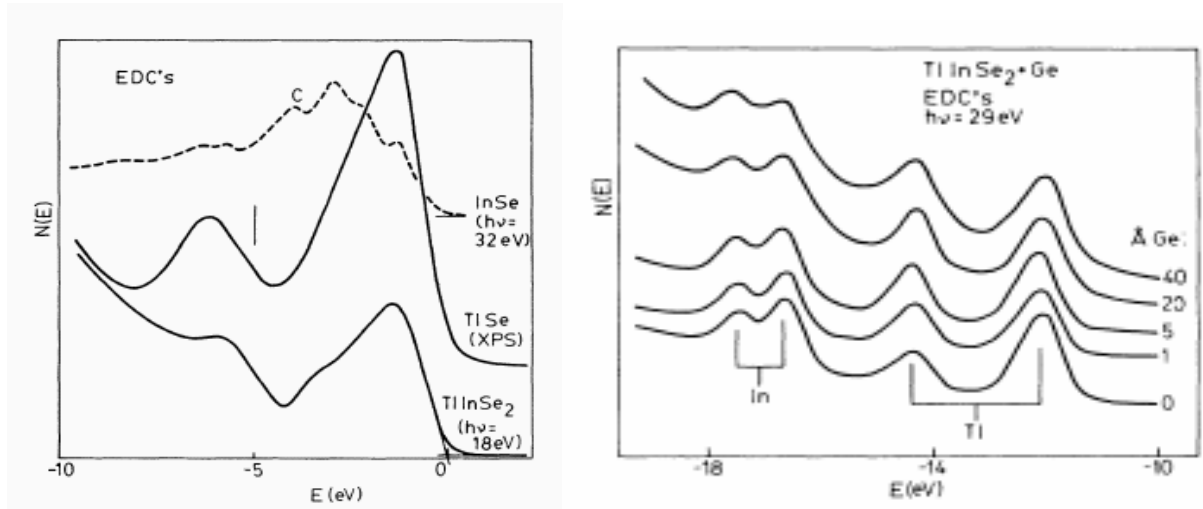

Figure 5. Left panel: Photoelectron EDCs of TlInSe$_2$, taken with photon energy 18 eV, and x-ray photoemission EDCs of TlSe and InSe, taken with photon energy 32 eV. The letter C labels one of the features of the spectrum. The horizontal scale is referred to the top of the clean TlInSe$_2$ valence band. Right panel: EDC's taken on clean and Ge-covered TlInSe$_2$. The nominal thickness of the Ge overlayer is shown at the right side of each curve (From [83]).

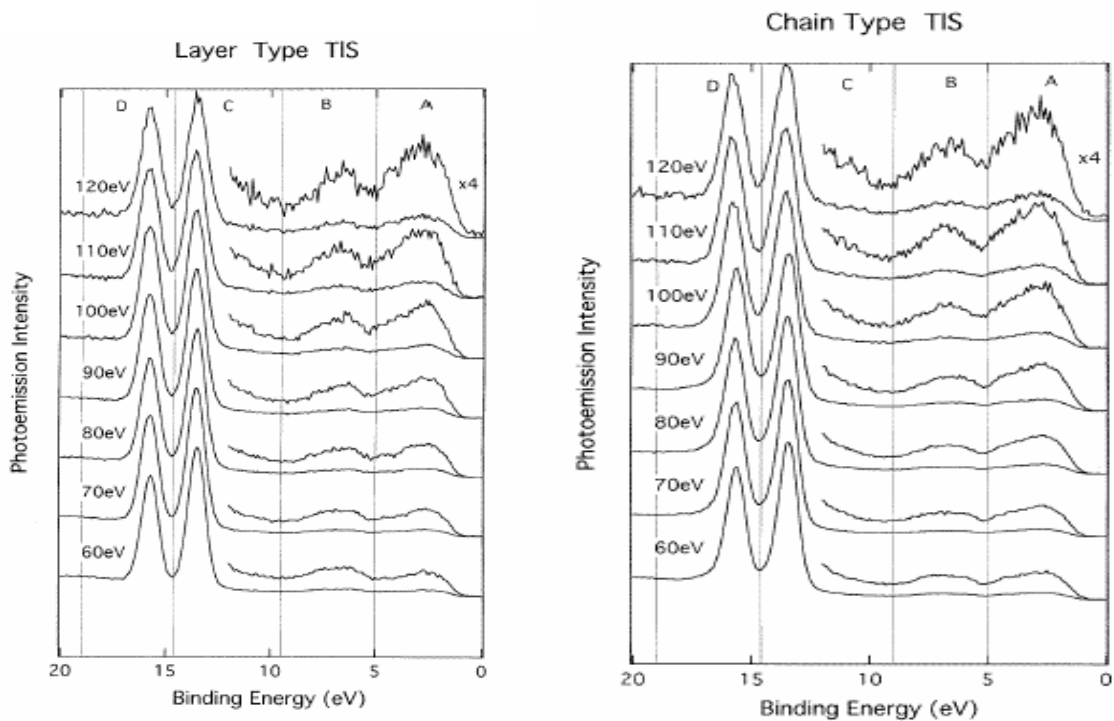

Figure 6. Photoelectron EDCs of layer-type and chain-type TlS taken with different incident photon energies. The spectra are normalized so that the Tl 5*d* levels have the same height. (From [32]).

Table 4. Binding energies of the Tl 5d core levels. (From [32]).

| Sample | Tl valence | Binding energy (eV), Tl $5d_{5/2}$ state | Binding energy (eV), Tl $5d_{3/2}$ state |
|---|---|---|---|
| Tl metal | 0 | 14.81 | 12.54 |
| Tl$_2$S | 1 | 15.55 | 13.36 |
| TlS chain | 1, 3 | 15.72 | 13.50 |
| TlS layered | 1, 3 | 15.81 | 13.59 |

Kashida et al. [32] reported on the measurements of photoemission spectra of chain and layered modifications of TlS using synchrotron photon source (Figure 6). The EDCs were taken with different incident photon energies, from 40 to 120 eV. For comparison, the EDCs of Tl metal were also measured. The binding energies were determined from the Fermi edge $E_F$ of the Tl metal sample. The most prominent peaks observed around -13 and -15 eV come from Tl $5d_{3/2}$ and $5d_{5/2}$ core states. In the chain and layer types of TlS, the positions of these 5d core levels coincide within the experimental uncertainty. As the average valence of Tl atoms increases, the levels shift to higher binding energies (Table 4). The shift measures the charge transfer from the cations to the anions. However, the splitting of the 5d doublet corresponding to the two inequivalent cation sites in TlS, for Tl$^{1+}$ and Tl$^{3+}$, is not observed, as already reported in reference [82]. A qualitative explanation for this finding is the rise in the Madelung energy; that is, around the Tl$^{3+}$ ion the anions are in closer distances than those around the Tl$^{1+}$ ion, which compensates the charge transfer effect. The more asymmetric line shapes of the 5d levels in the chain-type TlS than those in the layer-type TlS were attributed to a higher density of states near the Fermi level in the chain-type TlS. This is consistent with the fact that the electrical conductivity of the chain-type TlS is higher than that of the layer-type TlS.

Above the core 5d levels, broad valence band peaks are observed (Figure 6), which are composed of two sub-bands. The upper valence band edges are found around 0.95–1.05 eV below $E_F$. The edges are seen at almost the same positions in both the chain and layer types of TlS. The valence band A is composed mainly of the S 3p states, while the valence band B is derived from the Tl 6p states as well as from the S 3p and Tl 6s states. As the photon energy increases, the intensity of these peaks increases relative to that of the Tl 5d peaks. This change is due to the rise in the photo-ionization cross sections of the Tl 6s and S 3p states, relative to that of the Tl 5d states.

Let us now discuss the angle-integrated photoemission (AIPES) spectra of TlGaTe$_2$ [84] taken with He radiation (Figure 7), which show five features labeled as A – E in the He I spectrum. Two pronounced features A and C are not seen in the He II spectrum, and hence these structures are mainly attributed to Te 5p states, because the Te 5p cross section is dramatically reduced in going from He I to He II. On the other hand, structures D and E at 5 – 8 eV are seen in the He II spectrum, and hence these are assigned to Tl 6s and Ga 4s states. This assignment is confirmed by the muffin-tin projected partial density of states (DOS) given at the bottom of Figure 7. The He II spectrum, for which the cross sections of the Te 5p, Tl 6s and Ga 4s states are not so different as those for the He I spectrum, is in rather good agreement with the broadened total DOS. Very important information was obtained from the angle-resolved photoemission (ARPES) spectra (Figure 8). In the following, **k** denotes the electron momentum in the solid and $\mathbf{k}_\parallel$ and $\mathbf{k}_\perp$ denote the components parallel and perpendicular to the (110) surface, respectively. In spite of the chain-like structure running along the *c* axis, one can clearly see dispersive features for both $\mathbf{k}_\parallel \parallel c$ and $\mathbf{k}_\parallel \perp c$ arrangements, i.e., the band dispersions in TlGaTe$_2$ depend on the momentum not only parallel but also perpendicular to the chain direction. This weighty finding, indicating 3D rather than 1D character of the electronic structure, will be discussed below along the NMR data and calculated electronic band structure. Recently, angle-resolved photoemission study of quasi-one-dimensional TlInSe$_2$ [85] also showed noticeable band dispersion in the direction normal to the chains.

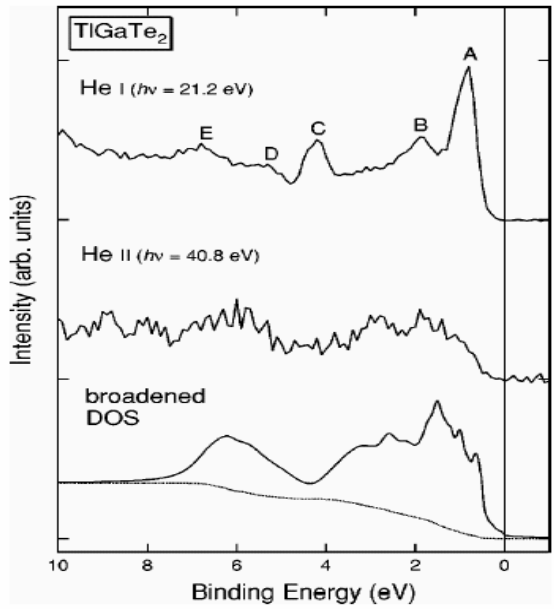

Figure 7. Comparison of the AIPES spectrum of TlGaTe$_2$ with the calculated DOS (bottom). The dotted line shows the integral background (From [84]).

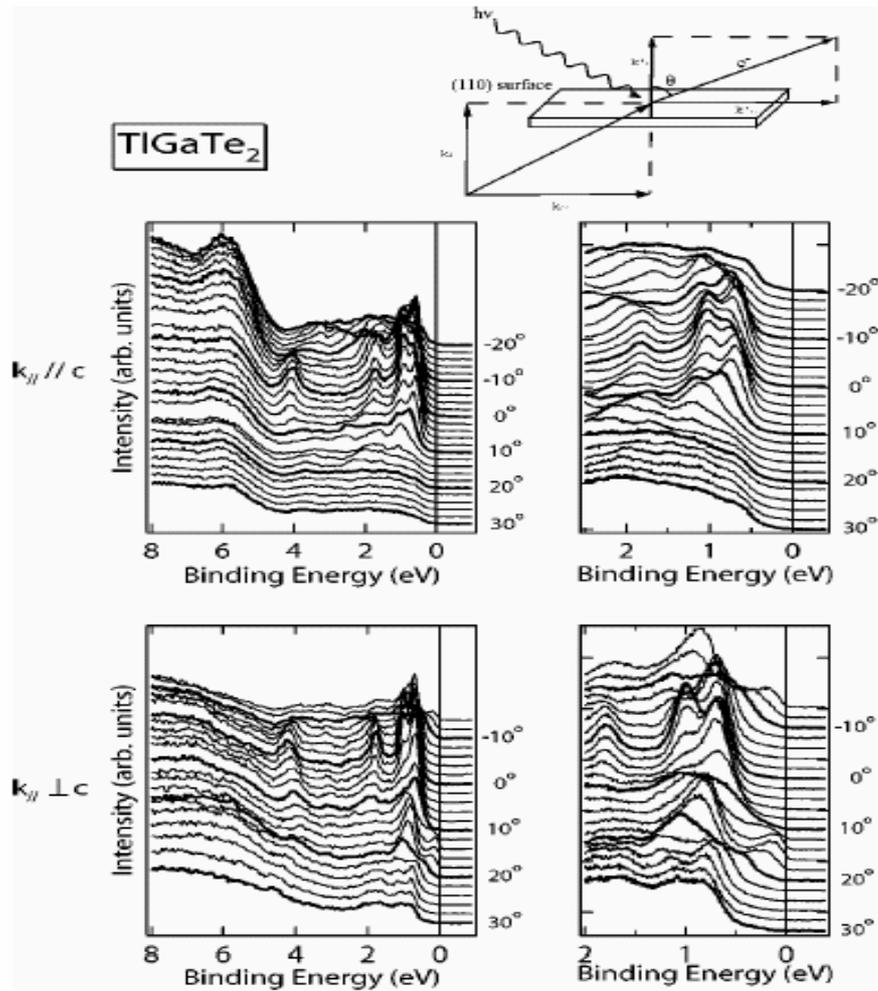

Figure 8. ARPES spectra of TlGaTe$_2$ in wide region (left panels) and narrow region (right panels). Top: for $k_\parallel \parallel c$. Bottom: for $k_\parallel \perp c$. Inset: measurement geometry showing the definition of $k_\parallel$ and for $k_\perp$ (From [84]).

## *4.2. Nuclear Magnetic Resonance measurements*
### *4.2.1. NMR in thallium-contained compounds*

Nuclear Magnetic Resonance (NMR) is the resonance absorption of electromagnetic wave by a nuclear spin system subjected to the external magnetic field $B_0$. NMR is an element-selective, inherently quantitative tool for studying the electronic structure, local crystal structure, dynamics and phase transitions in solids at the atomic level. For the present series of compounds, $^{203}$Tl and $^{205}$Tl are the most attractive nuclear probes for NMR measurements. These nuclei are particularly sensitive to effects of chemical bonding because of the strong indirect exchange coupling between the nuclear spins, $J_{12}\mathbf{I}_1\mathbf{I}_2$, which is realized across the overlapping electron clouds. In the solid thallium compounds, this coupling dominates over the dipole-dipole one and determines the line shape in single crystals and also in powder samples measured in low magnetic field, when chemical shielding anisotropy is negligible. In the early 1980s, the author discovered [86] that the indirect exchange between nuclei could arise from the electron shell of a bridging atom or atomic group, by analogy with the Kramers mechanism of electron-spin exchange via a nonmagnetic bridge ion [87]. Just such effect is realized in the compounds under review. For these systems, the scalar exchange term of the spin Hamiltonian is given as

$$\hat{H} = J_{11} \sum_{i,j} I_i^I I_j^I + J_{33} \sum_{i,j} I_i^{III} I_j^{III} + J_{13} \sum_{i,j} I_i^I I_j^{III} \qquad (1)$$

Here spins $I^I$ and $I^{III}$ belong to the Tl$^{1+}$ and M$^{3+}$ (sites I and III) respectively, $J_{11}$ and $J_{33}$ are the Tl$^{1+}$ - Tl$^{1+}$ and M$^{3+}$ - M$^{3+}$ exchange coupling constants among the spins of univalent and trivalent ions, respectively, and $J_{13}$ is the Tl$^{1+}$ - M$^{3+}$ exchange interaction. (Note that here M = Tl(III), Ga and In). Due to low natural abundance of the $^{33}$S ($f = 0.76\%$), $^{77}$Se ($f = 7.56\%$), $^{123}$Te ($f = 0.87\%$) and $^{125}$Te ($f = 6.99\%$) isotopes having nuclear spins, one can neglect the spin-spin coupling between thallium and chalcogen nuclei. Van Vleck has shown [88] that in the crystal that contains two different types of the exchange-coupled spins $I$ and $I'$, the contribution to the second moment of the NMR line $S_2$ comes from the exchange interaction with the unlike nuclei only and is proportional to the abundance of the unlike isotope. Therefore the ratio of the second moments of two different isotopes is inversely proportional to the ratio of their abundances. For thallium, the natural abundances are $f$=29.5% for $^{203}$Tl and (1-$f$)=70.5% for $^{205}$Tl with (1-$f$)/$f$=2.39, which makes the aforementioned effect readily observable [86, 89-93].

*4.2.2. Indirect nuclear exchange in chain-type compounds*

The most impressive manifestation of the exchange coupling is observed in the single crystal of the chain semiconductor TlSe [94]. The low-field thallium spectrum at $B_0 \perp c$ is given at the left panel of Figure 9. In this orientation the chemical shifts of Tl$^{1+}$ and Tl$^{3+}$ ions coincide, and all thallium atoms are equivalent. Both $^{203}$Tl and $^{205}$Tl isotopes show single Lorentzian-like resonances with the second moments $S_2$ = 360 and 150 kHz$^2$ for $^{203}$Tl and $^{205}$Tl, respectively. The values of $S_2$ are more than two orders of magnitude larger than the contributions of the dipole-dipole interactions of nuclear spins, estimated from the structure of TlSe as ~1 kHz$^2$. The ratio of the second moments of two thallium isotopes, $S_2(^{203}$Tl$)/S_2(^{205}$Tl$)$ = 2.4, is inversely proportional to the ratio of their natural abundances that is characteristic for the exchange coupling among Tl nuclei. The effective exchange constant $J_0 = (J_{11}^2/2+J_{33}^2/2+J_{13}^2)^{1/2}$, calculated from the $S_2$ values, is 45.1 kHz.

The interchain exchange is readily seen in the high field NMR measurements (Figure 9, right panel), when the difference in chemical shifts of each isotope, belonging to the Tl$^{1+}$ and Tl$^{3+}$ ions, exceeds the exchange coupling between them. At $B_0 \parallel c$, the $^{205}$Tl spectrum shows two separate lines attributed to the Tl$^{1+}$ and Tl$^{3+}$ ions. When the applied magnetic field $B_0$ is tilted from the $c$ axis, two lines move to each other due to the angular dependence of the Tl$^{1+}$ and Tl$^{3+}$ chemical shifts, broaden and finally collapse. Such a behavior is characteristic for the exchange interaction between the nuclei at inequivalent sites. It means that besides the common intrachain Tl-Tl exchange interaction, the exchange coupling between nuclei of structurally inequivalent $Tl^{1+}$ and $Tl^{3+}$ ions, which reside in neighboring chains, is realized in TlSe. Analysis of the spectra in terms of the theory of exchange processes in NMR yields the value of $J_{13}$ = 39.4 kHz. Comparing this value with $J_0$ = 45.1 kHz and assuming that $J_{11}$ and $J_{33}$ are equal, one can calculate the intrachain exchange constants $J_{11} = J_{33}$ =21.9 kHz. Thus intra- and interchain wave function overlaps are comparable, and that is why TlSe is not a highly anisotropic compound as it was noticed in the section 3.

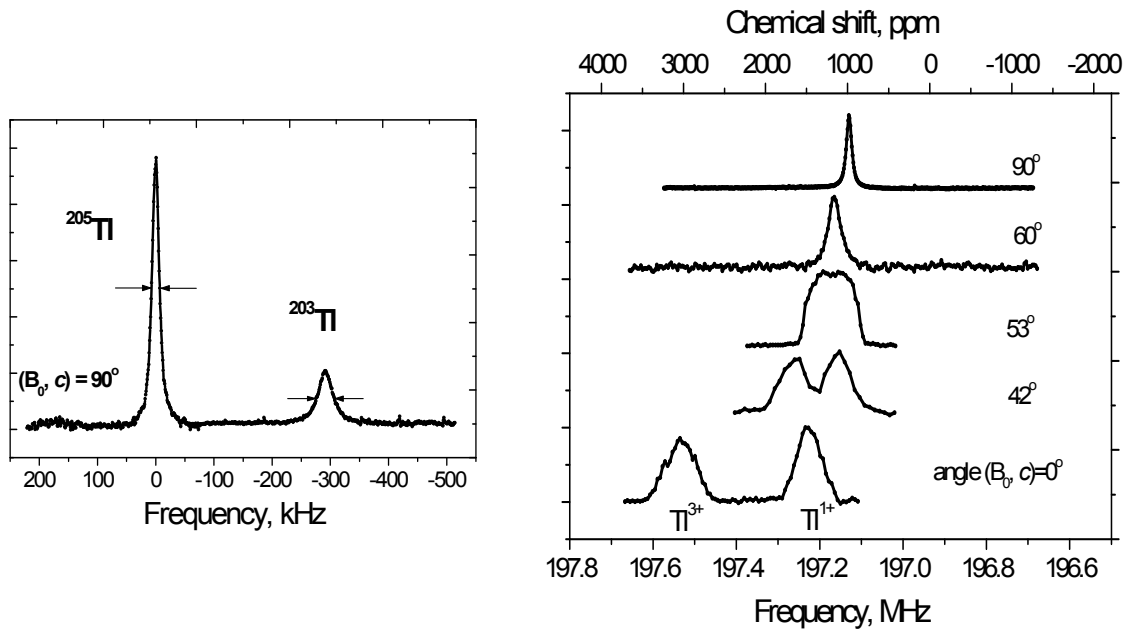

Figure 9. Left panel: Room temperature $^{203}$Tl and $^{205}$Tl NMR spectra of the TlSe single crystal at the resonance frequency 30.3 MHz ($B_0$ = 1.2 T). Applied magnetic field $B_0$ is perpendicular to the $c$ axis.
Right panel: Angular dependence of the room temperature $^{205}$Tl NMR spectra of the single crystal of TlSe in high magnetic field ($B_0$= 8.0196 T). (From [94]).

Analogous effect of the indirect nuclear exchange interaction has been observed in the chain semiconductors TlS [95] and $TlGaTe_2$ [96]. The scalar terms of the exchange coupling constants $J_{ij}$ were evaluated from the $S_2$ values of the $^{205}$Tl and $^{203}$Tl resonances; at that, separation of intrachain (Tl-Tl) and interchain (Tl-Ga) contributions to $S_2$ was shown to be possible [96] in $TlGaTe_2$ with neighboring Tl(I) and Ga(III) chains.

*4.2.3. Indirect nuclear exchange in layered compounds*

Low-field $^{203}$Tl and $^{205}$Tl measurements of the powdered samples of layered TlX and $TlMX_2$ compounds [97-99] show broad singlet lines (Figure 10, left panel). Their second moments are much larger than those resulted from the contributions of dipole-dipole interactions of nuclear spins and chemical shielding anisotropy and are indeed characteristic for the indirect exchange coupling among nuclei [86,89-99]. At that, the experimental line widths and second moments of $^{203}$Tl and $^{205}$Tl isotopes are close to each other.

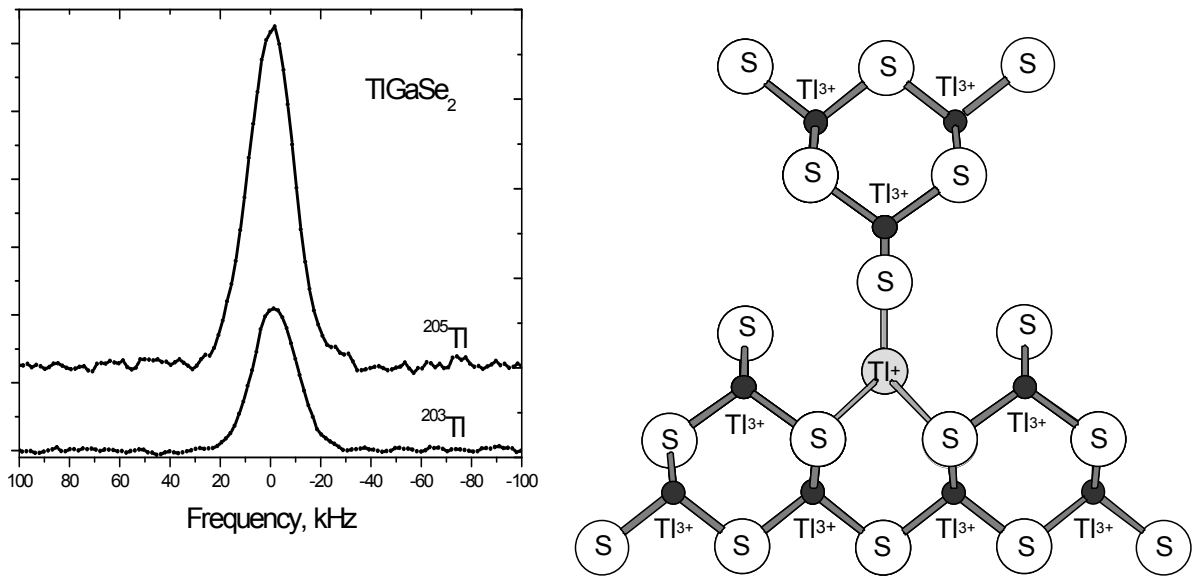

Figure 10. Left panel: room temperature $^{203}$Tl and $^{205}$Tl NMR spectra of the powder $TlGaSe_2$ in magnetic field $B_0$ = 1.228 T. (From [99]). Right panel: arrangement of atoms around the $Tl^{1+}$ ion showing one of the $Tl^{1+}$ - S - $Tl^{3+}$ bonds in the layer-type TlS structure (view along the $Tl^{1+}$ channel). (From [98]).

For example, powder layered TlS [98] shows the ratio of the second moments of $^{203}$Tl and $^{205}$Tl isotopes at the resonance frequency 21.4 MHz of 1.23 instead of 2.39 (after subtracting the chemical shielding anisotropy contribution). It means that the indirect exchange between structurally inequivalent $Tl^{1+}$ and $Tl^{3+}$ ions plays a significant role and possibly dominates over the $Tl^{3+}$ - $Tl^{3+}$ and $Tl^{1+}$ - $Tl^{1+}$ exchange. In such a case, all nuclei are unlike ones, and line broadening is realized not only for $^{203}Tl$-$^{205}Tl$ but also for $^{205}Tl^{1+}$-$^{205}Tl^{3+}$ and $^{203}Tl^{1+}$-$^{203}Tl^{3+}$ exchange interactions. Therefore the $S_2(Tl^{203})/S_2(Tl^{205})$ ratio differs from 2.39. In the case that the resonances of uni- and trivalent Tl ions are not well resolved (e.g., in powder samples), two approaches for evaluation the exchange couplings $J_{ij}$ are used. The interlayer $Tl^{1+}$ - $Tl^{3+}/M^{3+}$ interaction is extracted (i) from the second moment values of the $^{205}Tl$ and $^{203}Tl$ resonances and (ii) from the field dependence of the line width at low resonance frequencies, when the case of the "fast" exchange ($J>>\delta$) is realized (here $\delta$ is the NMR frequency separation between two sites). In the latter case, the additional line broadening caused by the exchange interaction among the spins of the Tl(I) and B(III) atoms is proportional to $\delta^2$ [100]:

$$\Delta\nu = \Delta\nu_0 + \delta^2\nu^2/(4\ J_{13}) \qquad (2)$$

Such a behavior is readily observed in the experiment [94,98]. For the layered TlS, using both these approaches and assuming that $J_{11}$ and $J_{33}$ are equal, we found $J_{11} = J_{33} = 11$ kHz, and $J_{13} = 12$ kHz [98]. These results show that the interlayer exchange coupling is comparable with the intralayer one.

Similar results have been obtained in the layered semiconductors $TlInS_2$, $TlGaS_2$ and $TlGaSe_2$ [97] and have indicated an overlap of the wave functions of univalent $Tl^{1+}$ and trivalent $Ga^{3+}$ and $In^{3+}$ ions through the intervening S or Se atoms. Such interaction implies a formation of weak $M^{3+}$ - X - $Tl^+$ chemical bonds (here X is the chalcogen atom) by means of directed $sp$- and $p$ - orbitals. One of such bonds in the TlS structure [98] is shown in Figure 10 (right panel).

### 4.2.4. Wave function overlap and electronic structure

As it was mentioned above, the indirect nuclear exchange coupling is realized due to overlap of the electron clouds of atoms. In the aforementioned compounds, the $Tl^{1+}$-$Tl^{1+}$, $Tl^{3+}$-$Tl^{3+}$ and $Tl^{1+}$- $M^{3+}$ distances exceed the sum of the ionic radii of the corresponding $Tl^{1+}$, $Tl^{3+}$ and $M^{3+}$ ions and are therefore rather long to guarantee a significant Tl - Tl and Tl - M overlap. Since the chalcogen atoms are the first neighbors of Tl, one can conclude that the interchain and interlayer exchange couplings of nuclei are mostly caused by the overlap of the $Tl^{1+}$ and $M^{3+}$ electron wave functions of the $Tl^{1+}$ - X - $M^{3+}$ type across the intervening chalcogen atom. (The $Tl^{3+}$ - X - $Tl^{3+}$ coupling within the chains in TlSe, TlS and $TlGaTe_2$ is evidently realized by means of the $Tl^{3+}$ - X covalent bonds). The obtained wave function overlap should be an important mechanism in the formation of the uppermost valence bands, lower conduction bands and entire electronic structure of the aforementioned compounds. The band structure of the semiconductors under review is consistent with a long-range indirect nuclear exchange coupling via an intervening chalcogen atoms, analogous to the Kramers mechanism [87] of electron spin exchange via a nonmagnetic bridge ion. Common wave functions of thallium and chalcogen guarantee, via electron-nuclear hyperfine interaction, an effective correlation of Tl nuclear spins.

As shown by Bloembergen and Rowland [89], the indirect exchange coupling of nuclei is put into effect via intermediate excited electronic states. Thus, to describe the indirect nuclear exchange interaction via a bridge atom, we should discuss the excited electronic states of $Tl^{1+}$ and $M^{3+}$ mixed with the states of the bridge chalcogen ion. For $Tl^{1+}$ in TlSe, such interaction may be realized by means of mixing of Tl $6s^2$ electron states with unoccupied $4p$-states of Se. The Tl(I) orbitals are likely $sp$-hybridized Tl wave functions, which increases the orbital overlap, since Tl $p$ orbitals span a large range. Such a mixing of the empty $Tl^{1+}$ $6p$ orbital into the filled Tl $6s$ level was predicted by Orgel [101]. Though formally the $Tl^{3+}$ ion has a configuration $5d^{10}$, covalent $Tl^{3+}$ - $Se^{2-}$ bonds with $sp^3$ hybridization are realized, since the $Tl^{3+}$-Se distance in tetrahedra is close to the sum of the covalent radii of Tl and Se.

On the other hand, exchange interaction between nuclear spins of atoms $A$ and $B$ is of the order of

$$(8\pi/3)\gamma_{nA}\gamma_e h^2|\Psi_A(0)|^2 \times (8\pi/3)\gamma_{nB}\gamma_e h^2|\Psi_B(0)|^2 /\Delta E \qquad (3)$$

Here $\Delta E$ is a suitable average of the energy difference between the conduction and valence bands. The interaction is realized by means of the $s$-parts of wave functions having a nonzero value $|\Psi(0)|^2$ at the nucleus site. Thus an assistance of Tl $6s$ states is necessary, and one is led to consider a role of the outer $6s^2$ electron pair in interatomic interactions. Such interaction involves excited states with the electronic configuration $6s^2$ due to their mixing with the empty $6p$ (and perhaps $6d$) states of Tl. The wave functions of these thallium $6s6p$ and $6s6d$ states overlap with $p$-orbitals of chalcogen. This model is in accordance with the recent calculations [102] demonstrating that lone pairs cannot be completely localized and exhibit a presence in the bonding regions to varying degrees, from 10-15% to 44% of the total covalent bond order. In the case of the chain-like TlSe and TlS, we suggest that the outer $6s^2$ lone pair electrons of $Tl^{1+}$ are delocalized and actually shared between the uni- and trivalent thallium ions in order to guarantee the exchange coupling of $Tl^{3+}$ ion. Presence of some portion of the $s$-electron at the $Tl^{3+}$ atom causes the electron-nuclear hyperfine interaction and explains the indirect exchange coupling of its spin. Analogous effect is suggested for the $Tl^{1+}$ and $Ga^{3+}/In^{3+}$ ions in the chain and layered TlX and $TlMX_2$ compounds. It seems that such overlap and $6s^2$ pair activity are common properties of thallium chalcogenides.

The obtained interchain wave function overlap should reduce the anisotropy of the physical properties of TlSe in comparison to the layered semiconductors $A^{III}B^{VI}$. That is why TlSe crystals posses a three-dimensional electronic nature in spite of its chain-like structure. This can be seen from the experimental values of conductivity that are not much different along and normal to the $c$ axis [46,52,54]. The same is right for chain-like $TlInSe_2$ compound [56]. This finding will be discussed below along the band structure calculations. We note that the considerable interlayer and interchain overlap observed in the aforementioned compounds affects their physical properties, e.g., it reduces the anisotropy of the elastic coefficient in comparison to the layered semiconductors $A^{III}B^{VI}$ [103]. As shown above, the two-dimensional square lattice of TlSe in the (001) plane, with alternating univalent and trivalent ions, is not metallic at ambient temperature. However, significant overlap of electron wave functions in this plane allows us to expect an electron hopping between $Tl^{1+}$ and $Tl^{3+}$ ions at higher temperature, possibly accompanied by a phase transition into a metallic state.

The similar nuclear exchange effects were also observed in the layered $Tl_2Te_3$ [93], $Tl_2Se$ [104] and chain $TlTaS_3$ [105] semiconductor compounds. In Table 5, all up-to-now known data on the indirect exchange coupling in the thallium-contained semiconductors are collected for the convenience of the readers. We stress that undistinguishing between the scalar and pseudo-dipolar interactions does not affect our conclusions, since both these interactions are realized due to overlap of the electron clouds of atoms and imply an occurrence of a weak Tl - X- M chemical bond.

Table 5. Parameters of indirect nuclear exchange in solid thallium semiconductors and Tl metal.

| Compound | T, K | $J_{11}$, kHz | $J_{33}$, kHz | $J_{13}$, kHz | Ref. |
|---|---|---|---|---|---|
| TlSe | 295 | 21.9 | 21.9 | 39.4 | [94] |
| TlS (chain) | 295 | 27 | 27 | 5 | [95] |
| TlS (layered) | 200 | 11 | 11 | 12 | [98] |
| $TlGaTe_2$ | 295 | 7.7 | | 1.6 | [96] |
| $TlGaSe_2$ | 295 | 7.3 | | | [97] |
| $Tl_2Te_3$ | 295 | 22.9 | | | [93] |
| $Tl_2Se$ nanorods | 295 | 21 | | | [104] |
| $TlTaS_3$ | 295 | 8.6 | | 1.4 | [105] |
| Tl (metal) | 77 | 17.5 | | | [89] |
| Tl (metal) | 4.2 | 37.5 | | | [90] |

We note that both univalent and trivalent Tl atoms in the reviewed compounds show essential chemical shielding anisotropy (Figures 9 and 11) despite their formal spherically symmetric $5d^{10}6s^2$ and $5d^{10}$ electron configurations. Thus one is led to consider *sp* (or *dsp*) hybridization of the Tl wave functions, which interact with the *p*-orbitals of the chalcogen atom and yield strong deviation of the Tl electron cloud and electronic charge distribution from the spherical form. Thus the chemical shielding data support the aforementioned conclusions based on the analysis of the indirect exchange coupling. (Taking into account the coordination polyhedron around $Tl^{1+}$, one can speculate that *d*-orbitals, perhaps in the form of *dsp* or *d⁴sp*, are also included, yielding weak interaction with the neighbors). We note that chemical shielding results from the electron-nuclear interaction and is inversely proportional to the band gap [100]; the main contribution comes therefore from the states near the top of the valence band and bottom of the conduction band. This is also right for the indirect spin-spin coupling according to Eq. 3. The above findings on spin-spin coupling, chemical shielding and wave function overlap yield new physical insight into the electronic structure and properties of the chain and layered semiconductor compounds. The aforementioned experimental data are in good qualitative agreement with the band structure calculation of the above-mentioned compounds to be discussed in the next section.

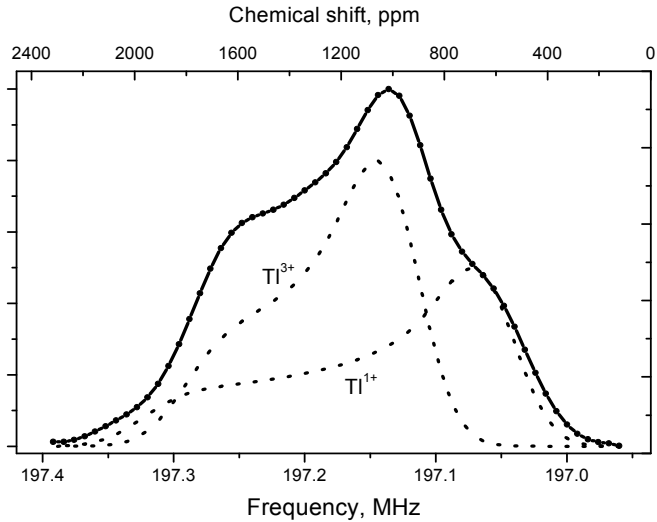

Figure 11. Room temperature $^{205}$Tl NMR spectrum of the powder sample of chain-type TlS in the magnetic field $B_0 = 8.0196$ T (solid line) and calculated spectra of the Tl$^+$ and Tl$^{3+}$ components (dashed lines). The line shape is mainly caused by the chemical shielding anisotropy. (From [95]).

## 5. Electronic structure: band structure calculations

Let us start from the band structure calculations of the tetragonal compounds having chain-type structure. The first calculations of the band structure in TlSe-type compounds - TlSe, TlGaTe$_2$, TlInTe$_2$, TlInSe$_2$, have been carried out by Gashimzade et al. [106-108], using (i) the concept of band representation and continuity chord and (ii) pseudopotential method; at that, some unknown parameters of the atomic pseudopotential were determined by fit of the experimental energy gaps near the fundamental optical edge to the theoretically calculated values. The 5$d$ states of Tl atoms were not taken into account. The top of valence band and the bottom of conduction band were shown to be localized at different points on the surface of the Brillouin zone of the body-centered tetragonal lattice, namely at the symmetry point T(0, π/a, 0) and on symmetry line D(π/2α, π/2α, κ), respectively. The direct band transitions were found to be forbidden according to the symmetry rules of selection. Therefore the compounds are indirect gap semiconductors with the band gap ~ 1 eV; these findings are in agreement with the existing experimental data [50]. Lowest four valence bands originate from $s$-states of chalcogen atoms. Two next valence bands were attributed to $s$-states of the trivalent cation located in the tetrahedral environment of chalcogen atoms. Then, a large group of ten valence bands follows, presumably taking their origin from the hybridized $p_z$-states of chalcogen atoms and the $p_x$-, $p_y$-, $p_z$-states of trivalent cations. At last, the highest two valence bands were attributed to the $s$-state of univalent Tl atoms. The band's assignment is given in Table 6. (Here and below, to ease the work of the reader in comparison the results of different authors, we collected the band structure data in a series of Tables). Orudzhev calculated [109] charge distribution in TlSe using pseudopotential wave functions determined from the band structure calculation. This computation (Table 7) confirmed that four lowest valence bands are composed from the $s$-states of Se and showed that the uppermost valence bands and the first two conduction bands near the Tl(I) and Tl(III) atoms originate from *s*-states of the Tl(I) and Tl(III) atoms, respectively. The charge distribution showed a noticeable maximum on the line joining the Tl(III) and Se atoms, indicating covalent bonds between these atoms. However, neither interchain Tl(I)-Se-Tl(III) overlap nor wave function delocalization, observed in the NMR experiments (section 4), were discussed in the aforementioned paper [109].

Table 6. Results of band structure calculations of TlSe, TlGaTe$_2$, TlInTe$_2$ and TlInSe$_2$ by Gashimzade et al. [106-108].

| Energy, eV | Number of bands | Bands | Electronic states |
|---|---|---|---|
| -3 to 0 | 2 | valence bands | *s*-state of univalent Tl atoms. |
| -5 to -2.5 | 10 | valence bands | hybridized $p_z$-states of chalcogen atoms and the $p_x$-, $p_y$-, $p_z$-states of trivalent cations. |
| -6 to -5 | 2 | valence bands | *s*-states of the trivalent cation |
| -15 to -14 | 4 | valence bands | *s*-states of chalcogen atoms (Se, S, Te). |

Table 7. Results of band structure calculations of TlSe by Orudzhev et al. [109].

| Energy, eV | Number of bands | Bands | Electronic states |
|---|---|---|---|
| | 2 | first conduction bands | *s*-states of Tl(I) and Tl(III) |
| | - | uppermost valence bands | *s*-states of Tl(I) and Tl(III) |
| | 4 | valence bands | *s*-states of Se |

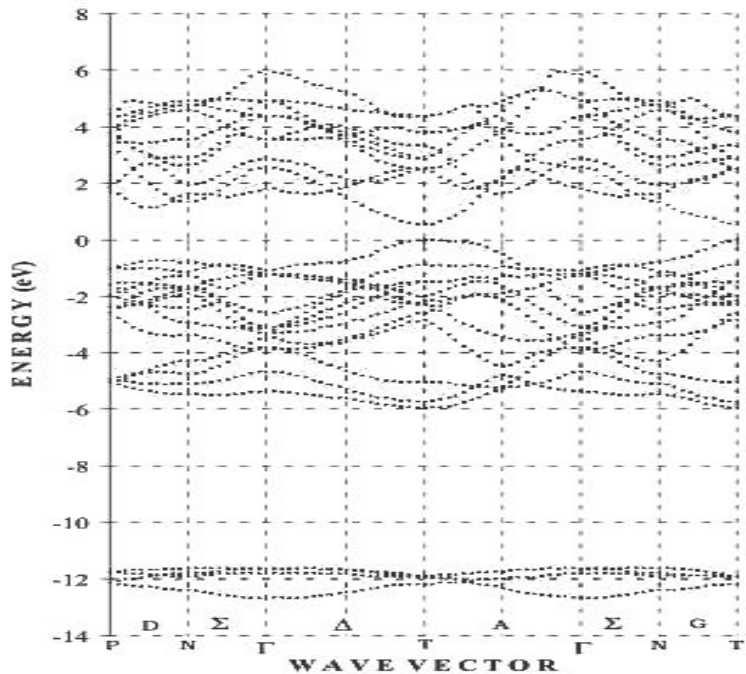

Figure 12. Band Structure of TlInSe$_2$. The top of the valence band is taken to be zero. (From [110]).

Band structure calculation of the ternary chain-type TlInSe$_2$ by Orudzhev et al. [110] (Figure 12), using a pseudo-potential method with allowance for non-locality of ionic pseudo-potentials, showed that the top of the valence band and the bottom of the conduction band in this compound are located in the symmetry point T (0.5, -0.5, 0.5) on the surface of the Brillouin zone. Thus $TlInSe_2$ is a direct gap semiconductor with band gap of 0.60 eV, in contradiction with the previous findings [106-108] showing it to be the indirect gap semiconductor. The experimental data presented in section 3 also favor indirect scenario with twice larger thermal and optical band gaps. The calculated valence bands of $TlInSe_2$ may be sorted into three groups (Table 8).

Table 8. Results of band structure calculations of $TlInSe_2$ by Orudzhev et al. [110].

| Energy, eV | Number of bands | Bands | Electronic states |
|---|---|---|---|
| - 4 to 0 | 10 | valence bands | mainly $p$-states of Se and In(III) |
| - 6 to - 4 | 4 | valence bands | $s$-states of Tl(I) and In(III) |
| -11.8 to -13 | 4 | valence bands | $s$-states of Se |

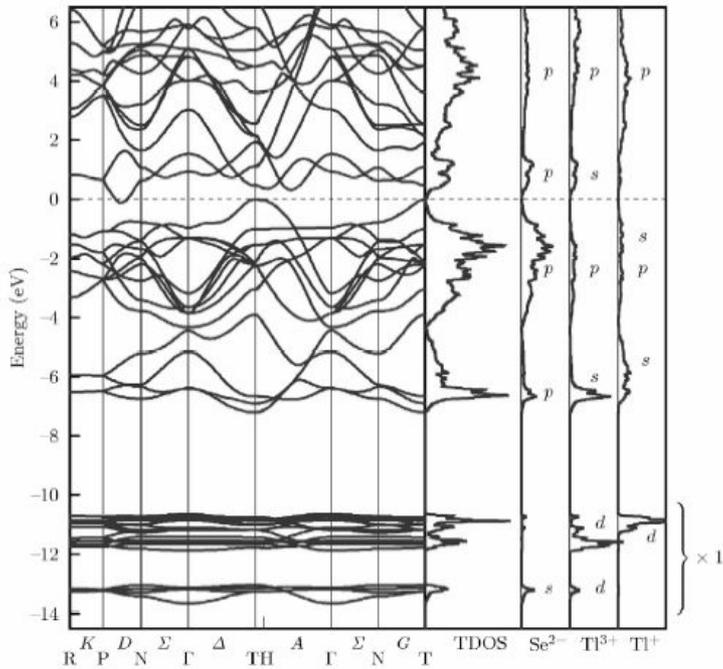

Figure 13. Energy bands for TlSe along the high symmetry lines of the Brillouin zone, total DOS, and local DOSs for $Se^{2-}$, $Tl^{3+}$, and $Tl^{+}$, in panels from left to right, respectively [111]. DOSs for the lower valence states due to Tl $5d$ and Se $4s$ electrons are shown in a scale reduced by a factor of 10 in order to fit into the same frame with the rest of the densities above −9 eV. The top of the valence band is taken to be zero. (From [111]).

Recently, detailed calculations of the electronic structure of TlSe and $TlGaTe_2$ have been reported by Ellialtoglu et al [111], Kashida [112] and Okazaki et al [84]. The first of them (Figure 13) has been made by means of *ab-initio* pseudopotential method using density functional theory within the local-density approximation. The calculated valence and conduction bands of TlSe may be sorted into six groups (Table 9). The bottom of the conduction band is located almost at the midway $D_1 = (\pi/a, \pi/a, \pi/2c)$ along the line $D$ joining the points $P$ and $N$, and corresponds to the irreducible representation $D_1$.

Two additional minima are situated along the symmetry line *A* that connects the points G and *H*. The band at point *T* is a little higher in energy. The energy gap is underestimated relative to the experimental value (~ 0.7 eV) due to the well-known artifact of the local density approximation (LDA) calculations. As a result the bottom of the conduction band crosses the top of the valence band, and the indirect gap appears to be negative, leading to a semimetal band structure with a hole pocket at *T* and electron pocket at *D* points, respectively. Ellialtoglu et al. [111] have also built several charge density plots of TlSe, in which the $Tl^+$ ions, having lost their 6*p* electrons, show *s*-like character due to the outermost $6s^2$ electrons participating in the valence bands. $Tl^{3+}$ ions, on the other hand, donate their 6*p* and 6*s* electrons to bond formation and show some charge density extending towards $Se^{2-}$ ions and a negligible amount of *d* influence. Most of the charge is accumulated on the $Se^{2-}$ ion rather than on the $Tl^{3+}$-$Se^{2-}$ bond that is therefore more ionic than covalent. The monovalent cation seems to be not bound to the chalcogens. These findings are in contrast with the pronounced charge accumulation at the $Tl^{3+}$-$Se^{2-}$ covalent bonds found in the empirical calculation [109] and NMR data [94]. Absence of the overlap of thallium and selenium wave functions contradicts to the experimentally observed [94] overlap and nearly 3D behavior of TlSe [52,54].

Table 9. Results of band structure calculations of TlSe by Ellialtoglu et al. [111].

| Energy, eV | Number of bands | Bands | Electronic states |
| --- | --- | --- | --- |
| Above 1.5 | 12 | conduction bands | *p* states of Se, Tl(I) and Tl(III) ions |
| 0 to 2 | 2 | conduction bands | antibonding mixture of Se 4*p* and Tl(III) 6*s* states; intermix slightly (around *H*) with the 12 upper conduction bands |
| 0 | 1 | uppermost valence band (tops at *T*) | Mainly non-bonding Se 4*p* states and 6*s* states of Tl(I) |
| -4 to 0 | 10 | valence bands | Se 4*p* states and 6*p* states of Tl(I) and Tl(III) |
| -7 to -4 | 4 | valence bands | mostly 6*s* states of Tl(I) and some Se 4*p* states mixed with the 6*s* states of Tl(III) |
| -12 to -10.6 | 20 | valence bands | 5*d* states of Tl |
| -13.7 to -13 | 4 | valence bands | 4*s*-states of Se |

Another band structure calculation of TlSe has recently been made by Kashida [112], using the full-potential linear-muffin-tin-orbital (LMTO) program LMTART and LDA and taking into account the spin-orbit interaction. As the base functions, *s, p* and *d* orbitals were taken for each atom (5*d*, 6*s* and 6*p* for Tl, and 4*s*, 4*p* and 4*d* for Se). The space was divided into muffin-tin spheres and the interstitial region. Within the muffin-tin spheres, the wave function and potential are expanded using spherical harmonics, and for the interstitial region, the wave function and potential are Fourier transformed. The calculated band structure and DOS's of TlSe [112] are shown in Figure 14. The calculated valence and conduction bands of TlSe are sorted into five groups (Table 10). The bottom of the conduction bands is located along the P($\pi/a,\pi/a,\pi/c$)–N($\pi/a,\pi/a,0$) line around W(($\pi/a,\pi/a,\pi/2c$)), while the top of the valence bands is located at T($2\pi/a,0,0$). Though the measurements show that TlSe is a semiconductor, the calculated electronic band structure (Figure 14), however, suggests that TlSe is a semimetal that has an electron pocket on the PN line and whose valence band touches Fermi surface at T point. This discrepancy results from the well-known underestimation of the band gap characteristic of LDA calculations.

Table 10. Results of band structure calculations of TlSe by Kashida et al. [112].

| Energy, eV | Number of bands | Bands | Electronic states |
|---|---|---|---|
| 0 to 3 | | conduction bands | mixture Se 4$s$, 4$p$ and Tl$^{3+}$, Tl$^{1+}$ 6$s$, 6$p$ anti-bonding states |
| - 5 to 0 | | valence bands | mainly Se 4$p$ states mixed with Tl 6$s$, 6$p$, states |
| –6 to –5 | | valence bands | Tl$^{3+}$ 6$s$ states, Se 4$s$, 4$p$ bonding states, Tl$^{1+}$ 6$s$ nonbonding states |
| Around -10 | | valence bands | Se 4$s$, Tl$^{3+}$ and Tl$^{1+}$ 5$d$ states |
| -13 to -11 | | valence bands | Se 4$s$, Tl$^{3+}$ and Tl$^{1+}$ 5$d$ states |

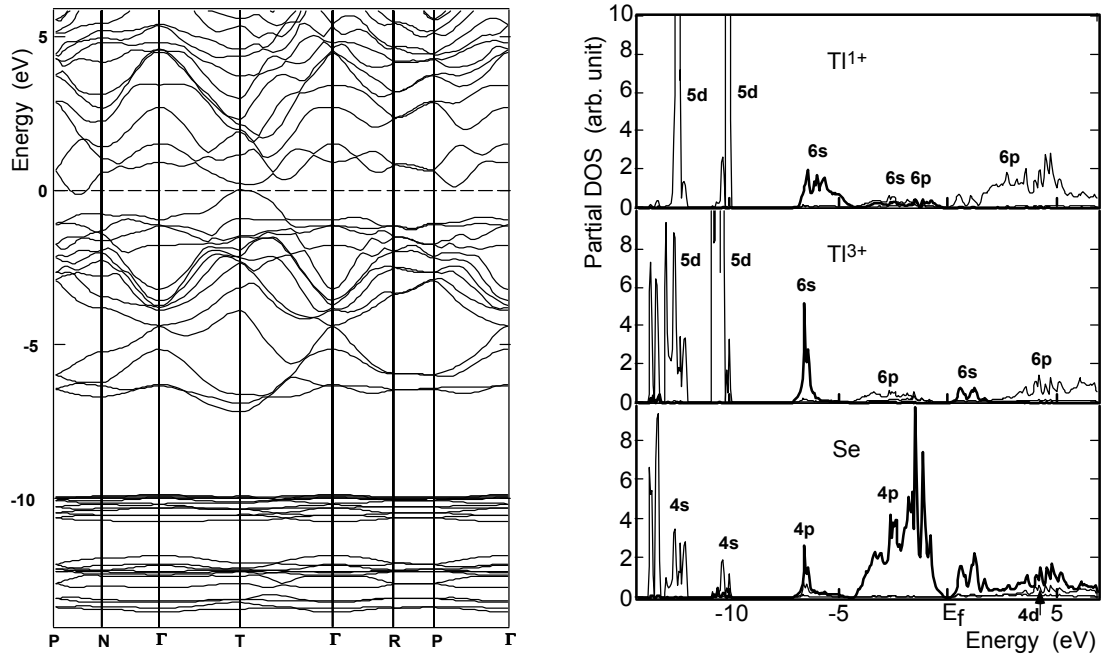

Figure 14. Electronic band structure (left panel) and angular momentum resolved densities of states of the chain semiconductor TlSe (right panel). Maximum of the valence band at T is set to zero. (From [112]).

The calculated band structure shows that the bands near the Fermi energy have a pronounced dispersion not only along the chains but also normal to the chain axis; the latter reflects the interchain interaction. Occurrence of such dispersion is in good agreement with the experimental NMR data [94] on the wave function hybridization and intra- and interchain overlap. Indeed, as it was noticed above, the main contribution to indirect exchange and chemical shielding comes from the states on the top of the valence band and at the bottom of the conduction band. The calculated angular momentum resolved wave functions at P-N line (W) and T point [112] show that near the top of the valence band, the density of states results mainly from the contributions of $Tl^{1+}$ *s*, Se *p* and some amount of $Tl^{3+}$ *s* and *p* states. Near the bottom of the conduction band, the density of states results mainly from contributions of $Tl^{1+}$ *p*, Se *s*, *p*, and $Tl^{3+}$ *s* and *p* states; some amount of $Tl^{1+}$ *s*-states also presents. This result reflects the occurrence of the $Tl^{3+}$ 6*s*4*p* – Se 4*s*4*p* - $Tl^{3+}$ 6*s*6*p* and $Tl^{3+}$ 6*s*6*p* – Se 4*s*4*p* - $Tl^{1+}$ 6*s*6*p* mixed states, yielding the intra- and interchain overlap of the $Tl^{3+}$ - Se - $Tl^{3+}$ and $Tl^{3+}$ - Se - $Tl^{1+}$ types. These electronic states give rise to the effective coupling of Tl nuclear spins observed in the NMR experiment and implies a formation of weak $Tl^{3+}$ - S - $Tl^{1+}$ chemical bonds by means of directed *p*, *sp* (and perhaps *spd*) orbitals. This calculation also demonstrates the presence of some portion of the *s*-electron at the $Tl^{3+}$ atom that causes the electron-nuclear hyperfine interaction and explains the indirect exchange coupling of its spin. The calculation also agrees with the experimental fact that TlSe possess 3D electronic nature (rather than 1D) in spite of its chain-like structure.

Band structure and DOS calculations in the chain-type $TlGaTe_2$ [84] by Okazaki et al. has been carried out in the local-density approximation using a full-potential, scalar-relativistic implementation of the linear augmented plane-wave (LAPW) method. At that, the band structure was also studied experimentally by means of photoemission spectroscopy (section 4), focusing on the anisotropy of the electronic structure. The calculated DOS and band dispersion in $TlGaTe_2$ [84] are shown in Figure 15 and Table 11. Although the experimental results show that $TlGaTe_2$ is a semiconductor, the present calculations suggest that $TlGaTe_2$ is a semimetal, which has a hole pocket at the *M* point and an electron pocket on the *W* line. This is due to the inherent deficiency of LDA. Really, because of the short Ga-Te bond length (~2.70 Å) within the $GaTe_4$ tetrahedra, the strong Ga 4*s*–Te 5*p* interaction raises two of the 12 Te 5*p* bands above $E_F$, thereby opening a band gap at the $TlGaTe_2$ Fermi level. At the *M* point, for example, these antibonding Ga 4*s*–Te 5*p* bands occur at energies 0.9 and 2.3 eV, respectively. These unoccupied antibonding bands are overlapped by the Tl 6p and Ga 4p type states. Note that the lowest groups of the valence bands, between –15 to –8 eV, were not shown (maybe not calculated) by the authors. Analysis of the wave function for the valence-band maximum at the *M* point reveals that ~50% of the weight consists of Te $5p_{x,y}$ and ~20% Tl 6*s*. The interaction between the Tl 6*s* and Te 5*p* orbitals is analogous to that of the Ga-Te interaction, but is reduced by the fact that the Tl-Te bond length (~3.55 Å) is significantly larger than the corresponding Ga-Te value. As a result, this Tl-Te antibonding band falls at a lower energy and forms the valence-band maximum. The electronic properties near $E_F$ of *p*-type $TlGaTe_2$ are determined by a combination of the interchain Tl-Te interactions as well as intrachain Te-Te hopping. Although the Tl-Te bond length is rather large, the interchain interactions are enhanced by the fact that each Tl has eight Te nearest neighbors. As a result, the expected one-dimensional features in the $TlGaTe_2$ band structure are masked by interchain interactions. From the strong band dispersion perpendicular to the *c* axis around the *M* point, it is concluded that TlGaTe$_2$ has a three-dimensional electronic structure in spite of the chain structure at least for the transport properties associated with the *p*-type carriers. This conclusion is in good agreement with the NMR data on TlGaTe$_2$ [96]. The calculated electronic structure of TlGaTe$_2$ also accords the photoemission spectroscopy data mentioned above.

Table 11. Results of band structure calculations of TlGaTe$_2$ by Okazaki et al. [84].

| Energy, eV | Number of bands | Bands | Electronic states |
|---|---|---|---|
| 0 to 4 | - | conduction bands | two Te 5*p* bands overlapping with Tl 6*p* and Ga 4*p* states |
| -4 to 0 | 10 | valence bands | Te 5*p* states |
| -7 to -4 | 4 | valence bands | mainly Tl 6*s* and Ga 4*s* |

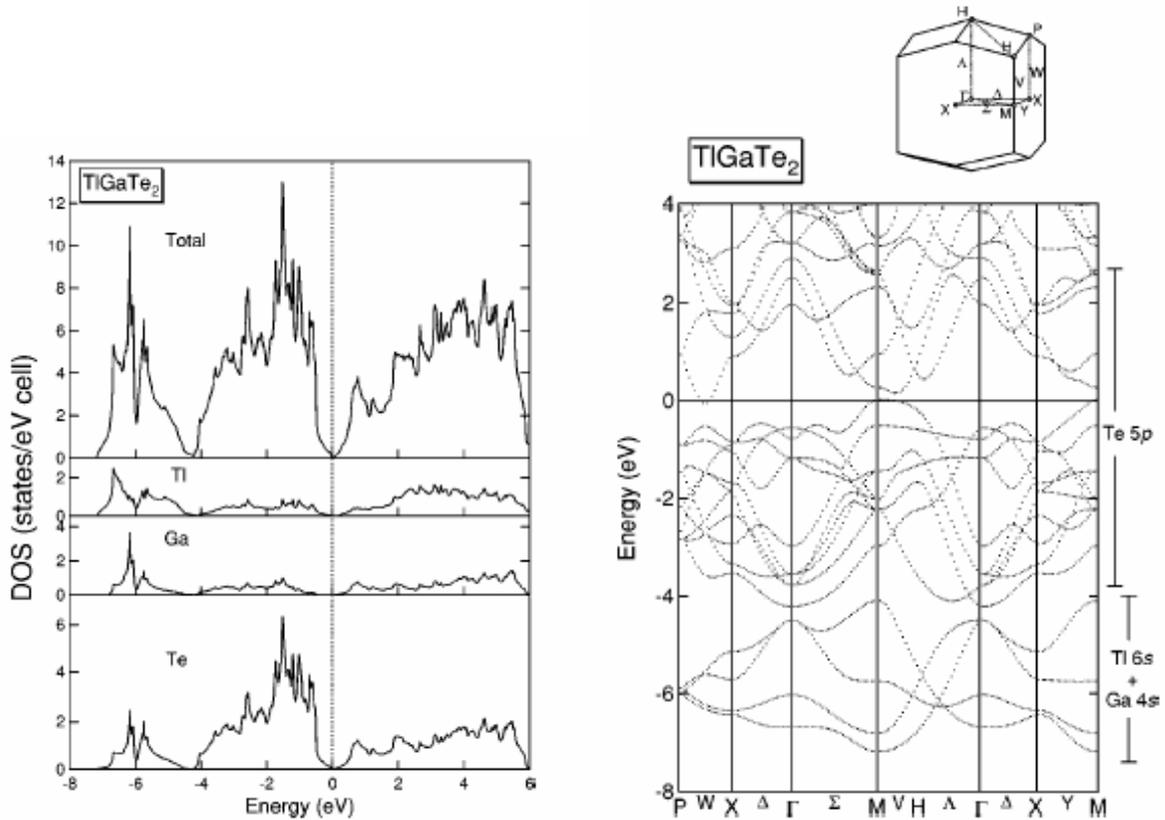

Figure 15. Total and muffin-tin projected DOS (left panel) and calculated band structure (right panel) of TlGaTe$_2$. The top of the valence band is taken to be zero. (From [84]).

Electronic structures of the tetragonal (chain-type) and monoclinic (layer-type) thallium monosulfde TlS have been studied [98,113] using linear-muffin-tin-orbital (LMTO) calculations. Spin–orbit interaction was taken into account in calculation of the chain compound, while the layered compound was calculated neglecting this interaction. The calculated Brillouin zones, band structures and DOS's of these two crystals are shown in Figures 16-18, respectively; the corresponding band assignments are given in Tables 12 and 13. In the chain TlS, the bottom of the conduction band is located along the P–N line around W ($\pi/a$, $\pi/a$, $\pi/c$), while the top of the valence bands is located at T($2\pi/a$, 0, 0). Thus this compound is an indirect gap semiconductor with the indirect gap Eg = 0.07 eV from T ($2\pi/a$, 0, 0) to W ($\pi/a$, $\pi/a$, $\pi/c$), while the direct gap at T is 0.76 eV. Experimental gap estimated from the electrical conductivity is 0.94 eV [32]. The discrepancy may be attributed to drawback of the LDA, which tends to underestimate the energy gap. An inspection of the LMTO wave functions shows that the valence band top at T is mostly composed of S $3p_{x,y}$ and Tl$^{1+}$ $6s$ with weights 0.81 and 0.17, while the conduction band bottom at W is mainly composed of S $3p_z$, Tl$^{3+}$ $6s$ and Tl$^{1+}$ $6p_{x,y}$ with weights 0.46, 0.36 and 0.26. This fact suggests that the top of the valence bands has some character of the Tl$^{1+}$ $6s$ – S $3p$ antibonding state, and that the bottom of the conduction bands has some character of the Tl$^{1+}$ $6p$ - S $3p$ - Tl$^{3+}$ $6s$ antibonding states. The present band structure is very similar to that of tetragonal TlGaTe$_2$ calculated by the LAPW method [84] where the valence band top is located at T and the conduction band bottom is located along the P–N line. Analogously to ref. [84], it can be argued that the band dispersion is fairly strong not only along the chain but also perpendicular to the chains; the latter is ascribed to the interchain interaction. These findings correlate with the NMR data [95] on the wave function overlap.

Table 12. Results of band structure calculations of the chain-type TlS by Shimosaka et al. [113].

| Energy, eV | Bands | Electronic states |
| --- | --- | --- |
| 0 to 3 | conduction bands | S $3p$, Tl(III) $6s$ and Tl(I) $6p$ states |
| - 4 to 0 | valence bands | mainly S $3p$ states mixed somewhat with the Tl(I) and Tl(III) 6 $p$ states |
| -7 to – 4.5 | valence bands | Tl(I) $6s$ and the Tl(III) $6s$ – S $3s$, $3p$ states |
| around - 11 | valence bands | S $3s$ states mixed with Tl(III) $5d$ states |
| -14 to -12 | valence bands | Tl $5d$ |

Table 13. Results of band structure calculations of the layer-type TlS by Shimosaka et al. [113].

| Energy, eV | Bands | Electronic states |
| --- | --- | --- |
| 0 to 3 | conduction bands | S $3p$ and Tl(I) $6p$ states, with some contribution of Tl(III) $6s$, $6p$ states |
| -3.5 to 0 | valence bands | mainly S $3p$ and some Tl(III), Tl(I) $6s$, $6p$ - S $3p$ states |
| - 8 to -5 | valence bands | Tl(III) and Tl(III) 6s and S 3s, $3p$ states |
| -12 to –10.5 | valence bands | mainly S $3s$ states mixed with Tl(III) $5d$ states |
| - 14 to -13 | valence bands | Tl(III) $5d$ states mixed with S $3s$ states, and Tl(III) $5d$ nonbonding states |
| Around -14.7 | valence bands | Tl(I) $5d$ |

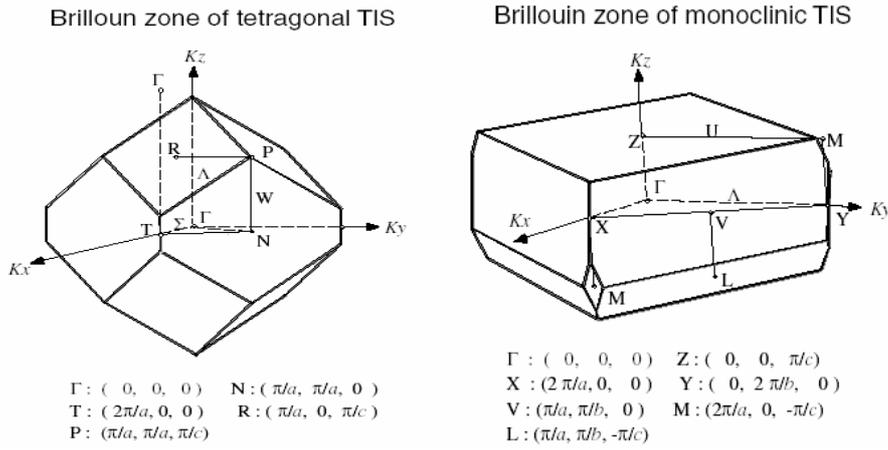

Figure 16. Brillouin zones of tetragonal chain-type (left panel) and monoclinic layer-type (right panel) TlS. The symmetry points and lines used to calculate the dispersion relation are labeled. (From [113]).

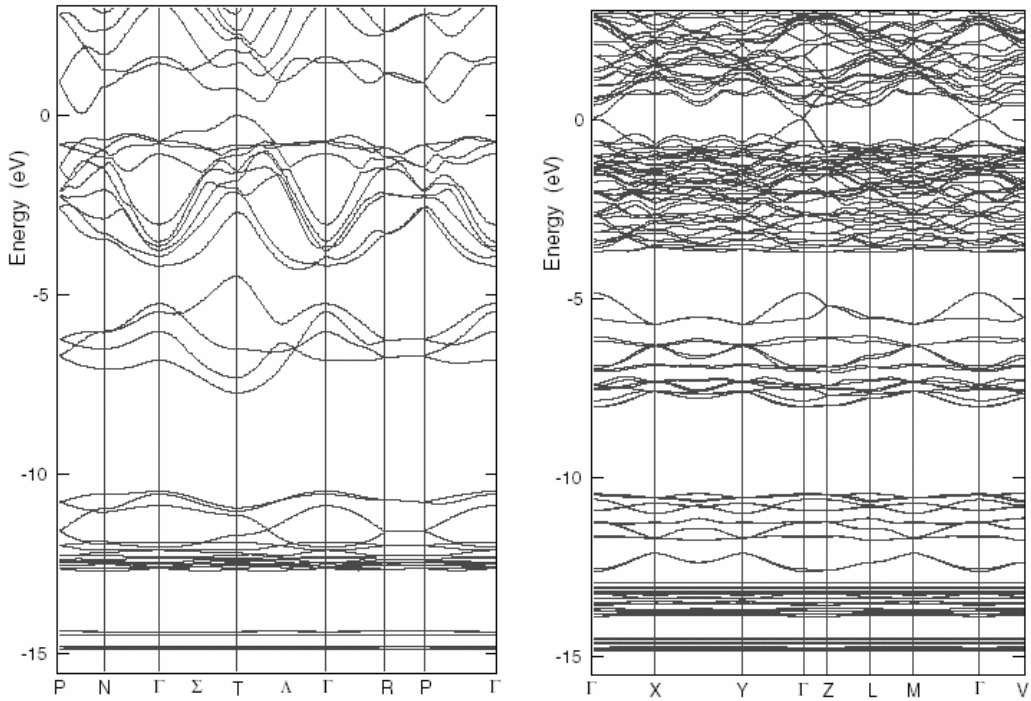

Figure 17. Band structures of tetragonal TlS (left panel) and of monoclinic TlS (right panel). The maximum of the valence band is set to zero. (From [113]).

In the layered TlS (Figure 17, Table 13), the top of the valence band and the bottom of the conduction bands are located at the Γ point. Thus monoclinic TlS is a direct (at Γ point) gap semiconductor with $E_g$= 0.06 eV. This value is much smaller than the experimental value of 0.9 eV [32], which was determined from the temperature dependence of the electrical conductivity. The obtained result is an expected manifestation of the well-known LDA underestimate of the band gap. We note, however, that LDA gives relatively correct eigenstates and wave functions. The calculated DOSs seem to qualitatively reproduce the experimental photoemission data [32].

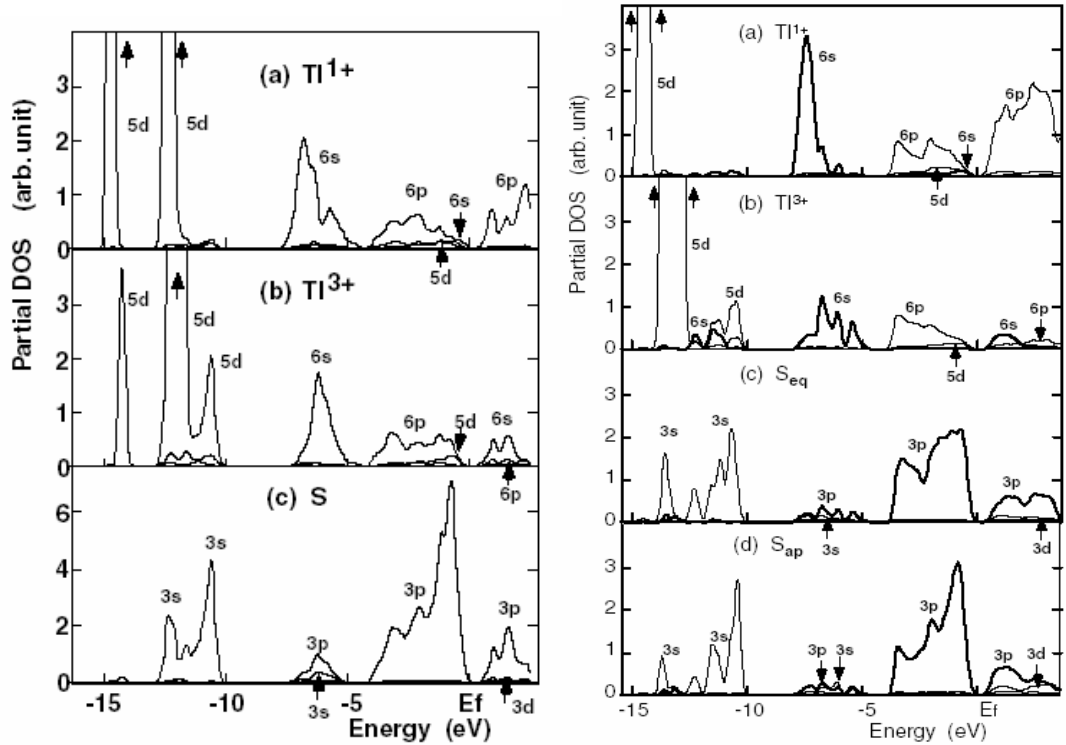

Figure 18. Calculated partial DOS for tetragonal TlS (left panel) and for monoclinic TlS (right panel). Thick lines represent in (a) and (b) the Tl 6s states, and in (c) and (d) the S 3p states, respectively. At the right panel, note in (a) small $Tl^{1+}$ 6s DOS peak just below $E_F$, and in (d) relatively large 3p peak just below $E_F$. (From [113]).

Monoclinic TlS crystals show anisotropic conduction, the conductivity within the layer is about two orders of magnitude higher than that normal to the layer direction [33]. The above band calculation yields relatively large dispersion along the Γ–Z line, which does not explain the above anisotropy though does not disclaim it as well. An inspection of the LMTO wave functions reveals that the valence band top at the Γ point is mainly composed of S $3s$, $3p$, $Tl^{1+}$ $6s$, $6p$, and $Tl^{3+}$ $6s$, $6p$, $5d_{x2-y2}$ states, while the conduction band bottom - of S $3s$, $3p$, $Tl^{1+}$ $6s$, $6p$ and $Tl^{3+}$ $6s$, $6p$ states, in which the aforementioned wave function are likely mixed into each other. The corresponding atomic orbital coefficients are given in Table 14. One can find that on the top of the valence band the contribution of the $Tl^{1+}$ $6s$ states exceeds that of the $Tl^{3+}$ $6s$ ones, while at the bottom of the conduction band the $Tl^{3+}$ 6s wave function dominates here over the $Tl^{1+}$ 6s one. The aforementioned wave function's structure correlates well with the NMR data [98], i.e. with the experimentally observed indirect exchange coupling among thallium nuclei due to the overlap of the $Tl^{1+}$ and $Tl^{3+}$ electron wave functions across the intervening chalcogen atom discussed in the previous section.

Table 14. Wave functions and corresponding atomic orbital coefficients [98] (a) on the top of the valence band and (b) at the bottom of the conduction band for the layer-type TlS. Indexes *x*, *y*, and *z* correspond to the *a*, *b*. and *c*-axes.

(a) valence band top:

$Tl^{1+}$

| 6s | $6p_x$ | $6p_y$ | $6p_z$ | $5d_{yz}$ | $5d_{zx}$ | $5d_{z2}$ | $5d_{xy}$ | $5d_{x2-y2}$ |
|---|---|---|---|---|---|---|---|---|
| 0.153 | 0.002 | 0.002 | 0.031 | 0.000 | 0.000 | 0.000 | 0.002 | 0.015 |

$Tl^{3+}$

| 6s | $6p_x$ | $6p_y$ | $6p_z$ | $5d_{yz}$ | $5d_{zx}$ | $5d_{z2}$ | $5d_{xy}$ | $5d_{x2-y2}$ |
|---|---|---|---|---|---|---|---|---|
| 0.029 | 0.001 | 0.001 | 0.040 | 0.001 | 0.002 | 0.0142 | 0.001 | 0.060 |

$S^{2-}$

| 3s | $3p_x$ | $3p_y$ | $3p_z$ |
|---|---|---|---|
| 0.022 | 0.0168 | 0.0168 | 0.173 |

(b) conduction band bottom:

$Tl^{1+}$

| 6s | $6p_x$ | $6p_y$ | $6p_z$ | $5d_{yz}$ | $5d_{zx}$ | $5d_{z2}$ | $5d_{xy}$ | $5d_{x2-y2}$ |
|---|---|---|---|---|---|---|---|---|
| 0.018 | 0.018 | 0.018 | 0.189 | 0.000 | 0.003 | 0.009 | 0.003 | 0.011 |

$Tl^{3+}$

| 6s | $6p_x$ | $6p_y$ | $6p_z$ | $5d_{yz}$ | $5d_{zx}$ | $5d_{z2}$ | $5d_{xy}$ | $5d_{x2-y2}$ |
|---|---|---|---|---|---|---|---|---|
| 0.151 | 0.006 | 0.006 | 0.018 | 0.002 | 0.004 | 0.018 | 0.004 | 0.007 |

$S^{2-}$

| 3s | $3p_x$ | $3p_y$ | $3p_z$ |
|---|---|---|---|
| 0.093 | 0.020 | 0.020 | 0.097 |

Let us now discuss the band structure of the other layered compounds. Owing to the complexity of crystal structure, the published data show noticeable discrepancies in the calculated band structure. The first attempt to calculate the band structure of TlGaSe$_2$ was done by Abdullaeva et al. [114] using empirical pseudopotential method. The authors considered the monoclinic modification of TlGaSe$_2$ as a deformed tetragonal structure. It was found that the bottom of the conduction band is located at the point T (0, π/*a*, 0) of the Brillouin zone, while the maxima of the valence band are located at three points, N (π/2*a*, π/2*a*, 0), T (0, π/*a*, 0) and A (0, 0, π/2*c*), approximately at equal energies. Next pseudopotential calculation of the band structure of TlGaSe$_2$ by Abdullaeva et al. [115] showed that the top of the valence band is located at the **Γ** point. The bottom of the conduction band is located at the **Γ**-Y line. The direct band gap was about 2.1 eV. No assignments of the electronic states have been done in those two papers. Recently, the electronic structure of the ternary thallium chalcogenides TlGaSe$_2$ and TlGaS$_2$ has been studied in more detail by Kashida et al. [116], who used LMTO method without taking into account the spin–orbit interaction. Results of these band structure calculations are shown in Figures 19-21 and Table 15. The calculated band dispersion shows that both compounds are indirect gap semiconductors. For both TlGaSe$_2$ and TlGaS$_2$, the top of the valence bands is situated at Γ, where the wave function is mostly composed of mixed Tl 6*s* - Se 4*p* (for TlGaSe$_2$) or Tl 6*s* - S 3*p* (for TlGaS$_2$) states. For TlGaSe$_2$, the bottom of the conduction band is located along the Z(0, 0, –0.5)–L(0.5, 0.5, –0.5) line, while for TlGaS$_2$ the bottom of the conduction band is situated along the Γ−Y(0, 1, 0) line. The calculated orbital decomposition of the states near the Fermi level [116] shows that the wave function at the bottom of the conduction bands at Γ is composed of the mixed Tl 6$p$, Ga 4$s$ and Se 4$s$ (for TlGaSe$_2$) or S 3$s$ (for TlGaS$_2$) states. The bottom of the conduction bands has the character of Tl 6$p$–Se 4$p$ or S 3$p$ antibonding state. These states have relatively stronger character of Tl 6$p$ states, than the corresponding states at Γ.

Table 15. Results of band structure calculations of TlGaSe$_2$ and TlGaS$_2$ by Kashida et al. [116]. $E_F$ = 9.68 and 9.49 eV, respectively.

| Energy, eV | Bands | Electronic states |
|---|---|---|
| 11 to 15 | conduction bands | Tl 6$p$ states, mixed somewhat with Se 4$p$ and Ga 4$s$ states (for TlGaSe$_2$) or with S 3$p$ and Ga 4$s$ states (for TlGaS$_2$) |
|  | Bottom of the conduction band at Γ | Tl 6$p$, Ga 4$s$ and Se 4$s$ (for TlGaSe$_2$) or S 3$s$ (for TlGaS$_2$) |
|  | Top of the valence band at Γ | Se 4$p$ (for TlGaSe$_2$) or S 3$p$ (for TlGaS$_2$) states mixed with the Tl 6$s$ states |
| 5 to 10 | valence bands | Se 4$p$ (for TlGaSe$_2$) or S 3$p$ (for TlGaS$_2$) nonbonding states mixed somewhat with Tl 6$p$ and Ga 4$p$ states |
| 3 to 5 | valence bands | Ga 4$s$ and Tl 6$s$ – Se 4$p$ bonding states for TlGaSe$_2$, Ga 4$s$ and Tl 6$s$ – S 3$s$ bonding states for TlGaS$_2$ |
| around – 1.5 | valence bands | Ga 3$d$ and Tl 5$d$ states |
| – 4 to – 2.5 | valence bands | Se 4$s$ for TlGaSe$_2$ and S 3$s$ states for TlGaS$_2$ |
| around – 5 | valence bands | Ga 3$d$ and Tl 5$d$ states |

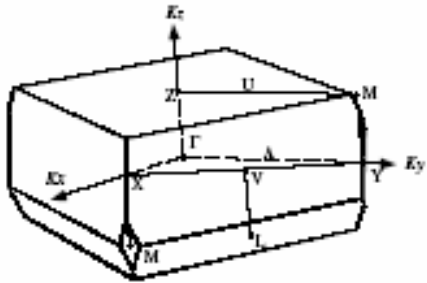

Figure 19. Brillouin zone of TlGaSe$_2$ and TlGaS$_2$. (From [116]).

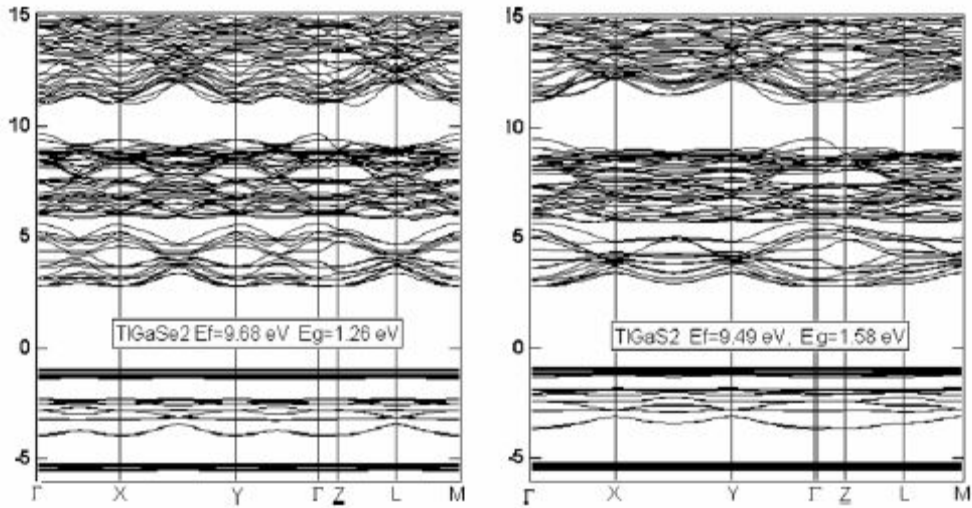

Figure 20. Electronic band structures of TlGaSe$_2$ and TlGaS$_2$. $E_F$ values are given in the figure. (From [116]).

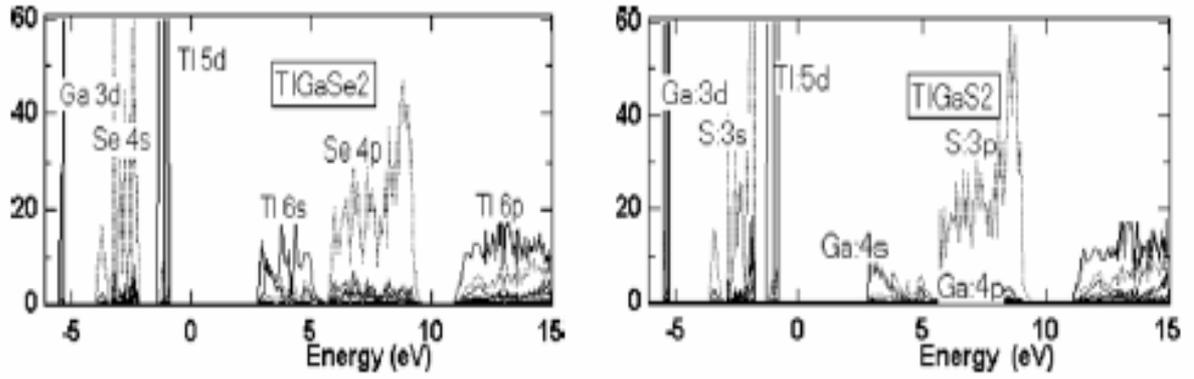

Figure 21. Partial densities of states of TlGaSe$_2$ and TlGaS$_2$ (From [116]).

TlGaSe$_2$ and TlGaS$_2$ are known as highly anisotropic semiconductors, where the conductivity within the layer is of several orders of magnitude higher than that normal to the layer. The present band calculation shows that, due to the relatively strong hybridization between the Tl 6$s$ and Se 4$p$ or S 3$p$ states, the dispersion of the valence band top state along the Γ–Z line is rather steep and does not explain the observed anisotropy, but explains interlayer overlap observed in NMR [97,99]. Figure 20 shows that TlGaSe$_2$ is an indirect gap semiconductor with the indirect gap from Γ to Z–L of 1.24 eV, while the direct gap at the Γ-point is 1.25 eV. Next, it shows that TlGaS$_2$ is also an indirect gap semiconductor, the direct gap at the Γ-point is 1.70 eV and the indirect gap from Γ to Γ–Y is 1.58 eV. These results are qualitatively in accord with the previous empirical pseudo potential band calculation [114,115] and those of optical absorption studies [31], where TlGaSe$_2$ and TlGaS$_2$ are assigned as indirect gap semiconductors. The experimentally measured by Hanias et al. [31] energy gaps are $E_g^d$ = 2.11 eV, $E_g^i$ = 1.83 eV for TlGaSe$_2$ and $E_g^d$ = 2.53 eV, $E_g^i$ = 2.38 eV for TlGaS$_2$ where $E_g^d$ and $E_g^i$ denote the direct and indirect gap, respectively. These values are larger than those calculated by Kashida et al. [116]. Preliminary calculations [116] showed that in TlInS$_2$ both the top of the valance band and the bottom of the conduction band are located at Γ, the Fermi level is around 9.55 eV and the energy gap is 1.58 eV. This fact suggests that, in contrast to TlGaSe$_2$ and TlGaS$_2$, TlInS$_2$ is a direct gap semiconductor.

It is worth to mention the paper by Yee and Albright [117] who investigated bonding and structure of TlGaSe$_2$ by tight binding calculations with an extended Hückel Hamiltonian. This calculation draws attention to the *sp*-hybridization of the Se and Tl wave functions and the role of lone electron pairs on the Se and Tl atoms. The bonding between Tl and Se was found to be reasonably covalent; the region around the Fermi level consists primarily of Tl 6$s$ states antibonding to Se lone pairs. The authors suggested that the large dispersion of the atomic orbitals signals some Tl-Tl communication. They found that the empty Tl $p$ orbitals mix with the Se lone pairs and create a net bonding situation, and that the Tl $p$ orbitals play a decisive role in the Tl-Tl interactions. Based on the paper by Janiak and Hoffmann [118], who has shown that $Tl^{1+}$ - $Tl^{1+}$ interactions in molecular and solid-state systems can be turned into a net bonding situation by Tl $p$ mixing into filled Tl $s$ orbitals, Yee and Albright [117] suggested the same situation in $TlGaSe_2$. The shortest Tl-Tl contact between two channels is 4.38 A, and certainly there is no direct communication. However, this might occur though bond coupling in the Ga-Se framework. Furthermore, the authors found that a soft, double well potential exists for the Tl atoms to slide away from a trigonal prismatic to a (3+3) environment, and discussed the electronic factors which create this distortion and leads to ferroelectric phase transition. This mechanism will be considered in more detail in the section 7.

Wagner and Stöwe [119] reported on self-consistent *ab initio* LMTO-ASA calculations of the electronic band structure and the crystal orbital Hamiltonian population function in the semimetallic TlTe. The calculations support a view of TlTe as a univalent Tl compound with two polyanionic partial structures, i.e. linear branched and unbranched chains. The branched Te2 chains show weaker Tl – Te interactions compared to Te3 chains. It was shown that in the energy range of 1 eV below $E_F$, Tl mixing with Te-centered bands, which are not involved in strong homoatomic σ bonding interactions, is quite strong. The role of Te - Tl orbital interactions in formation of the electronic structure and in the electronic nature of phase transition was discussed.

Summarizing, we conclude that though some results of different calculations are, in general, similar, the other ones exhibit noticeable differences. The results depend on the method of calculation, number of the wave functions, taking into account the spin–orbit interaction, etc. The more complicate the crystal structure, the more approximate the calculation. At least several results of calculations correlate well with the experimental NMR data on the spin-spin coupling, wave function overlap and chemical shielding effects and thus reflect real features of the electronic structure of the reviewed compounds. However, most of the band structure calculations did not place high emphasis on the stereo-chemical activity of $s^2$ lone pair. At that, the calculations often show qualitative agreement with the XPS curves.

## 6. Transport properties, semiconductor-metal phase transitions and band structure under high pressure.

### 6.1. Tetragonal (chain-type) crystals

In this subsection, we review the influence of high pressure on electronic properties of the chain-type $TlMX_2$ crystals. The effect of uniaxial pressure applied along and perpendicular to the tetragonal $c$ axis, as well as isotropic hydrostatic pressure on the electronic band structure of TlSe, was first reported by Gashimzade and Orudzhev [120]. The authors calculated the variation of the energies of the main extremums of the conduction and valence bands depending on the lattice constants $a$ and $c$. The calculations predict decrease of the band gap under pressure and, finally, semiconductor-metal phase transition in TlSe between 2.2 and 2.7 GPa under uniaxial pressure and ~5 GPa under isotropic hydrostatic pressure. Then Valyukonis et al [121] and Allakhverdiev et al [47, 122-124] measured the pressure dependence of the direct and indirect energy gaps $E_g^d$ and $E_g^i$ in TlSe, TlInSe$_2$, TlInS$_2$, and TlInSe$_{2(1-x)}$S$_{2x}$ (0 ≤ x ≤ 0.25) by analyzing the shifts of the fundamental absorption edges under applied pressure up to 5.5 GPa. (Here indices $d$ and $i$ denote direct and indirect gaps, respectively). The absorption coefficient was calculated from the transmission spectra using the value of the refractive index. Both direct and indirect gaps were found to linearly decrease with increasing pressure (Figure 22, left panel), i.e., showing negative pressure coefficients dE$_g$/dP. These coefficients for TlSe were determined [121] as $dE_g^d/dP$ = -0.15 eV/GPa for $\bar{E} \perp c$, $dE_g^d/dP$ = -0.17 eV/GPa for $\bar{E} \parallel c$, $dE_g^i/dP$ = -0.09 eV/GPa for $\bar{E} \perp c$, $dE_g^i/dP$ = -0.11 eV/GPa for $\bar{E} \parallel c$, respectively. (Here $\bar{E}$ is the electric field vector of the electromagnetic wave). These data are in satisfactory agreement with the calculated ones [123]. Some differences in $dE_g/dP$ measured by different authors (e.g., $dE_g^d/dP$ = -0.125 eV/GPa, and $dE_g^i/dP$ = 0.2 eV/GPa in TlSe [47,122]) maybe caused by different quality of the investigated crystals. The pressure dependences of the band gap in the TlInSe$_2$ crystal were determined as $dE_g^d/dP$ = -0.11 eV/GPa, and $dE_g^i/dP$ = 0.15 eV/GPa [122,125]. Since TlInSe$_2$ and TlInS$_2$ form a continuous series of mixed crystals in the whole range of concentrations (0 ≤ x ≤ 1), it was interesting to study the properties of these crystals. In the mixed crystals TlInSe$_{2(1-x)}$S$_{2x}$, $dE_g^i/dP$ was shown to slightly decrease with increasing $x$, from –0.145 to –0.13 eV/GPa for x = 0.05 to 0.25, respectively [122,123]. No phase transitions were observed up to pressures ~ 0.8 GPa in all aforementioned crystals [47,122,123]. However, a structural phase transition, accompanied by a reversal of the sign of $dE_g^i/dP$, was reported in TlInSe$_{0.2}$S$_{1.8}$ under pressure P ~ 0.6 GPa [123]. The transition was shown to be of the first order and reversible with pressure. Valyukonis et al [121] ascribed the aforementioned pressure dependence of the band gap to the changes in the interchain interactions with pressure, which, in turn, may influence the positions of energy minima and maxima at the bottom of the conduction band and on the top of the valence band. Next, an abrupt change of $dE_g^i/dP$ at $x$~0.3 [125] indicates a structural transformation, most probably from the tetragonal TlInSe$_2$-like crystal to the monoclinic TlInS$_2$-like crystal. Furthermore, TlInSe$_{1.4}$S$_{0.6}$ shows a linear reduction of the band gap with increased pressure up to ~0.72 GPa; at this pressure the band gap changes abruptly. The authors [125] assigned this change to the structural transformation under pressure. It was found that increasing amount of Se in the TlInSe$_2$-TlInS$_2$ system shifts pressure-induced phase transition to higher pressure. We note that Allakhverdiev et al [47,122,123] measured also the temperature dependence of the direct and indirect energy gaps.

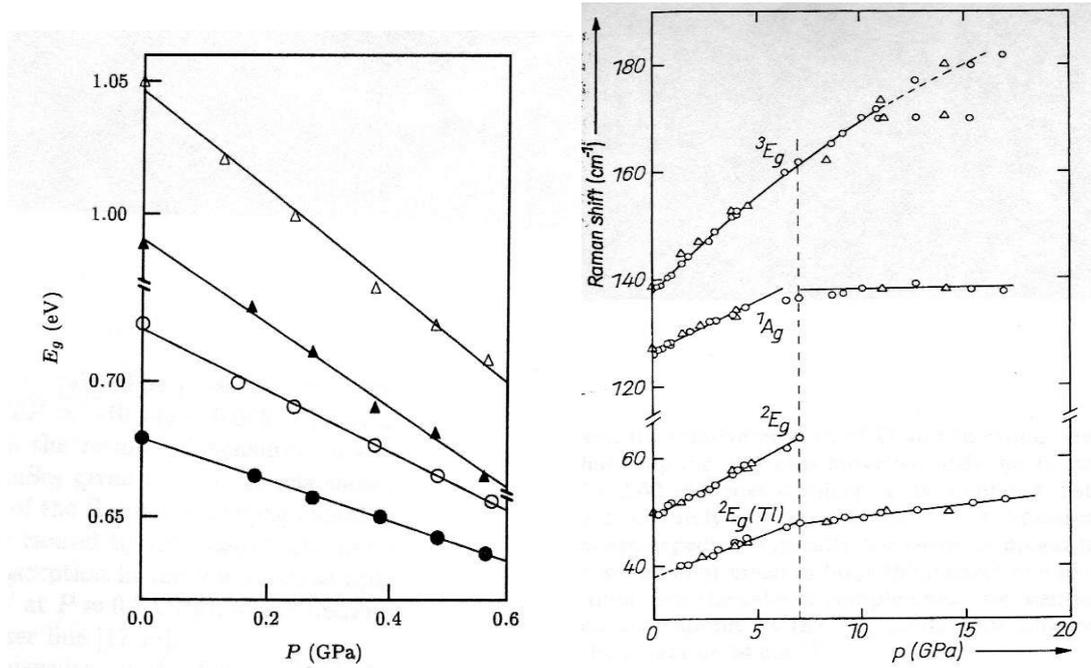

Figure 22. Left panel: Pressure dependence of the direct (triangles) and indirect (circles) band gaps in TlSe at room temperature measured for $\bar{E} \parallel c$ (open symbols) and for $\bar{E} \perp c$ (bold symbols). (From [121]). Right panel: Raman shifts of phonon peaks in TlInTe$_2$ at room temperature as a function of pressure. The vertical dashed line delineates the pressure-induced phase transformation at about 7 GPa. (From [126]).

Ves [126] has measured the Raman spestra of the chain thallium indium telluride (TlInTe$_2$) crystal up to high pressure of 17 GPa at room temperature, using a diamond anvil cell. Three Raman peaks were observed in the low-pressure tetragonal phase. The pressure coefficients and the corresponding mode Grueneisen parameters $\gamma$ of their frequencies were obtained. The modes were tentatively assigned by comparison to Raman spectra of related TlSe-type compounds. The frequency dependence of $\gamma$ indicated that a hierarchy of bonding forces is present. The appearance and disappearance of a new Raman peak at about 1.75 GPa was presumably attributed to a gradual, pressure induced mutual replacement of In atoms by Tl atoms but not to indication of the occurrence of a structural phase transition. However, noticeable change in the slope of the pressure dependence of the Raman shift near 7.0 GPa (Figure 22, right panel) was definitely assigned to a phase transition; furthermore, at even higher pressures indications for a second phase transition were found. We notice that, to our knowledge, the aforementioned explanation of the pressure evolution of the Raman modes, based on a pressure-induced mutual interchange of In and Tl atoms [126], was never supported by the x-ray measurements.

Investigation of the crystal structures of thallium sulfide and thallium selenide under very high pressure, up to 37 GPa, has been carried out by Demishev et al. [127] using XRD technique and diamond anvil cell. Three first-order phase transitions were found in TlS: TlS I (TlSe-type) → TlS II (α-NaFeO2-type) → TlS III (distorted α-NaFe2O-type) → TlS IV (CsCl-type) at 5, 10, and 25 GPa, respectively. The transition sequence is reversible. The space group of the first phase is $D_{4h}^{18}$-I4/mcm, and the lattice parameters were determined as $a$ = 7.77 Å, $c$ = 6.79 Å. For the second phase, the space group is $D_{3d}^5$-R3m, and the lattice parameters are $a$ = 3.945 Å, $c$ = 21.788 Å, Z = 6. This phase II of TlS is metastable under normal conditions and reveals semiconductor properties. At that, the phase II ↔ phase III transition exhibits a hysteresis: the pressure of the direct transition II → III is 10 GPa, while that of III → II transition is 5.5 GPa. Phase IV of TlS appears near P = 25 GPa, being mixed with the phases I and III in the pressure range from 25 to 30 GPa. At P = 35.5 GPa, its x-ray pattern corresponds to the pure phase IV structure of CsCl type with $a$ = 3.202 Å.

The compression of thallium selenide up to P = 21 GPa [127] results in the first-order structural transition from the tetragonal TlSe I phase to the cubic TlSe II phase. The latter is of the CsCl-type. The transformation is realized by a shift of the selenium atoms from the position with $x = 0.18$, $y = 0.68$, $z = 0$ to that with $x = 0.25$, $y = 0.75$, $z = 0$. The authors suggest destroying of the covalent chains under phase transitions [127]. The pressure-induced transition from TlSe I into TlSe II phase is accompanied by the reduction of the $a/c$ ratio from 1.16 to 1 and reduction of the relative volume of the unit cell $V/V_0$ on 40%. TlSe II phase shows $a = c$ ~3.78 Å. The Se - $Tl^{3+}$ and Se - $Tl^+$ distances become equal to 2.9 Å, $Tl^+$ - $Tl^+$ distance shortens as well, and hence the bond character is changed. Note that Se – Tl distance of 2.9 Å is longer than the sum of the covalent radii of Tl (1.49 Å) and Se (1.17 Å), 2.66 Å, but much shorter than the sum of the ionic radii of $Tl^{1+}$ (1.59 Å for CN=8 [14] in the CsCl lattice) and $Se^{2-}$ (1.98 Å), i.e. 3.57 Å, and is close to the sum of the ionic radii of $Tl^{3+}$ (0.98 Å for CN=8 [14]) and $Se^{2-}$, i.e. 2.96 Å, respectively. Such a bond is intermediate between the ionic and covalent ones.

Since $Tl^+$ and $Tl^{3+}$ ions in the phases I and II of TlS and phase I of TlSe occupy different crystallographic positions, the charge transfer between them is structurally forbidden, and these phase are suggested to show semiconductor properties. One might assume metallic properties for the TlS IV and TlSe II phases since in the CsCl-type lattice $Tl^+$ and $Tl^{3+}$ ions are structurally equivalent, and free charge transfer between $Tl^+$ and $Tl^{3+}$ is structurally allowed. This hypothesis has to be verified by the electric conductivity measurements of TlS and TlSe under high pressure. We note that Demishev et al. [127] also determined the equations of state for TlS and TlSe.

Pressure dependence of the electrical conductivity (as a rule, at room temperature) was reported by several authors. Kerimova et al [125] obtained a gradual increase in the conductivity along the $c$ axis in the undoped $TlInSe_2$ crystal, from $1.4 \times 10^{-6}$ Ohm$^{-1}$cm$^{-1}$ at ambient pressure to $7.5 \times 10^{-5}$ Ω$^{-1}$cm$^{-1}$ at P=1.4 GPa. The authors realized that pressure behavior of the conductivity may be well fit by the equation

$$\ln \sigma(P) = \ln \sigma(0) + AP , \qquad (4)$$

where A = d ln σ(P)/dP = $0.34 \times 10^{-2}$ Ω$^{-1}$cm$^{-1}$GPa$^{-1}$. Assuming exponential variation of the electrical conductivity with pressure

$$\sigma(P) = \sigma(0) \exp(-GP/2kT) , \qquad (5)$$

where $G = dE_g^i/dP$, $k$ is the Boltzmann constant and T is the temperature, the authors found that $G = 2kTA$ [125]. To satisfy Eq.5, parameter G should be negative, since A>0, and thus the band gap should decrease with increasing pressure, in accordance with that observed in the experiment.

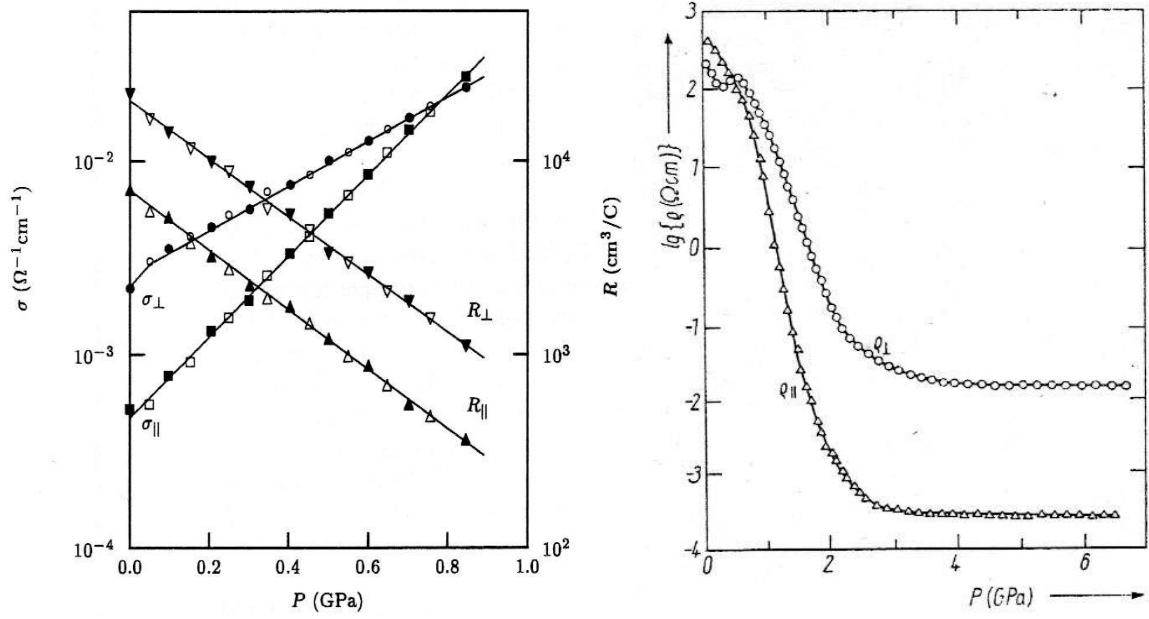

Figure 23. Left panel: Pressure dependence of the parallel ($\sigma_\parallel$) and perpendicular ($\sigma_\perp$) components of the electrical conductivity of TlSe (subscripts $\parallel$ and $\perp$ denote current direction relative to the $c$ axis) and of Hall coefficient $R_\parallel$ and $R_\perp$ (subscripts $\parallel$ and $\perp$ denote magnetic field direction relative to the $c$ axis) (From [53]). Right panel: Pressure dependence of the electrical resistivities of TlSe parallel (triangles) and perpendicular (cicles) to the $c$ axis at ambient temperature (From [54]).

Anisotropy of the crystal structure of the TlMX$_2$ compounds causes anisotropy of their electronic properties. Pressure dependence (up to ~0.8 GPa) of the anisotropy of the electrical conductivity $\sigma$ and Hall coefficient $R$ in TlSe single crystal was measured by Allakhverdiev et al. [53,128]. This study showed gradual (nearly linear) increase in conductivity and decrease in Hall coefficient with increased pressure (Figure 23, left panel). While the slopes of the $R_\parallel(P)$ and $R_\perp(P)$ curves are practically the same, the slopes of $\sigma_\parallel(P)$ and $\sigma_\perp(P)$ dependences are different, so that the anisotropy of the conductivity decreased with increasing P, and the $\sigma_\parallel(P)$ and $\sigma_\perp(P)$ curves cross each other at P ~ 0.7 GPa. Analogous decrease in ($\sigma_\parallel$ - $\sigma_\perp$) with increased temperature has also been observed [128].

`Rabinal et al [54] measured electrical resistivity $\rho$ of TlSe up to 8 GPa and reported that TlSe undergoes a pressure-induced semiconductor-metal transition. Figure 23 (right panel) shows the variation of electrical resistivities $\rho_\parallel$ and $\rho_\perp$ parallel and perpendicular to the $c$ axis, respectively, as a function of pressure at room temperature. At ambient conditions, $\rho_\parallel$ and $\rho_\perp$ are 400 and 208 $\Omega\times$cm, respectively. Both $\rho_\parallel$ and $\rho_\perp$ components decrease continuously with pressure and reach metallic values at about 2.7 GPa. Under ambient conditions, the $\rho_\parallel / \rho_\perp$ ratio is 1.92. Under high pressure, $\rho_\parallel/\rho_\perp$ becomes unity at about 0.6 GPa, decreases continuously, and becomes as low as 0.016 at 5.0 GPa (since $\rho_\parallel$ drops more rapidly than $\rho_\perp$). We note a correlation of this finding with the data of ref. [53] that shows $\sigma_\parallel < \sigma_\perp$ at ambient conditions. Furthermore, according to ref. [53], $\sigma_\parallel$ increases faster than $\sigma_\perp$ with increasing pressure up to 0.8 GPa, and the components become equal around 0.76 GPa. Measurements of Rabinal et al. [54] show a decrease in the band gap with increased pressure. The authors affirm that TlSe reveals positive temperature coefficient of the resistivity above 2.7 GPa, indicating the metallization of the samples [54]. These findings are in good agreement with the calculations of Gashimzade et al. [120] who predicted semiconductor-metal transition in TlSe at pressures of 2.2 – 5 GPa, but, however, conflict with the findings of Demishev et al. [127] who obtained pressure-induced phase transition in TlSe at P = 21 GPa, suggesting it to be the semiconductor-metal transition. Furthermore, Rabinal et al. [54] found the anisotropy of the resistivity in TlSe at ambient conditions ($\rho_\parallel / \rho_\perp > 1$) to be opposite to that expected from the chain structure ($\rho_\parallel / \rho_\perp << 1$); thus 1D behavior is realized only under high pressure. Let us remind that Demishev et al. [127] found that the high-pressure phase at P > 21 GPa is of the cubic CsCl type and therefore should not exhibit anisotropy at all. Thus one is led to a conclusion that Demishev et al. [127] and Rabinal et al. [54] reported on quite different properties of TlSe that hardly become reconciled with each other. The metallization of the thallium selenide observed by Rabinal et al. [54] definitely occurs in the Demishev's TlSe-I tetragonal phase, in which the chemically distinct $Tl^+$ and $Tl^{3+}$ occupy two different crystallographic positions, preventing free transfer of electrons from the $Tl^{1+}$ to the $Tl^{3+}$. Pressure-induced metallic-like properties of this phase may be explained as follows. Since the ionic-covalent interchain bonds are weaker than the covalent intrachain ones, the in-plane compressibility should be larger than that along the *c* axis. This is supported by the x-ray data by Demishev [127], who obtained gradual reduction of the *a/c* ratio with increased pressure, and by the measurements of elastic constants of thallium selenide [129] showing larger compressibility in the *a,b* plane in comparison to that along the *c* axis. Therefore the pressure-induced reduction in the Tl - Tl distance under pressure in the *a,b* plane should be larger that that along the *c* axis. In the TlSe crystal the *a* parameter is the doubled $Tl^+$ - $Tl^{3+}$ bond length in the plane lattice, and *c* is the doubled $Tl^{3+}$ - $Tl^{3+}$ distance in the chains along the *c* axis; at that, *c/a* < 1. Thus one is led to a conclusion that the change of $E_g$ under pressure, $dE_g / dP$, is mainly dominated by the deformation of the interchain Tl – Se – Tl bond, and band's closing under pressure is caused by the increase in the interchain overlap of the electron wave functions. Though, as noticed above, the two-dimensional lattice with alternated tri- and univalent ions is not metallic (in contrast to a plane lattice with equivalent doubly-charged ions), one can expect an increase in electron hopping and gradual transformation to the metallic state due to the reduction in the Tl – Tl distance. Therefore we suggest an electronic nature of the above transition under pressure. We note that metallization in InTe is accompanied by structural phase transition under pressure. One could expect similar behavior in TlSe as well. Therefore it is strange that metallization of TlSe occurs in the TlSe-I phase around P = 2.7 GPa, too far from the structural phase transition at P = 21 GPa observed by Demishev et al. [127].

We would like to notice that the aforementioned increase in the electron hopping and gradual transformation to the metallic state maybe obtained also by heating of the sample. To this end, it is worth to mention the study of the thallium-selenium phase diagram by Morgant et al. [130], who reported on a new cubic centered polymorphic form of TlSe at high temperature, above ~ 473 K. Temperature-induced increase in electron hopping between $Tl^{1+}$ and $Tl^{3+}$ ions, facilitated by the aforementioned wave function overlap, should equalize the electronic configurations of the Tl atoms, and such a valence averaging might finally cause the phase transition to the cubic phase. This high temperature phase seems to be an analog of the pressure-induced TlSe II (CsCl-type) phase with equivalent Tl ions observed by Demishev et al. [127], which is expected to exhibit metallic-like conductivity due to removal of the structural constraints upon electron transfer between $Tl^{1+}$ to $Tl^{3+}$ ions. Thus we are led to a conclusion on the electronic nature of the aforementioned phase transitions in the mixed valence compounds TlSe and TlS.

Rabinal et al [56] also reported on the measurements of the electrical resistivity components, $\rho_\parallel$ and $\rho_\perp$, in the $TlInX_2$ (X = Se, Te) chain-type single crystals under high quasi-hydrostatic pressure up to 7 GPa at room temperature (Figures 24-26). The measurements of $\rho_\parallel$ were carried out at high pressure and down to liquid nitrogen temperature. Both crystals reveal nearly the same pressure coefficient of the electrical activation energy parallel to the *c* axis, $d(\Delta E_\parallel)/dP = -2.9\ 10^{-10}$ eV/Pa, which results from the narrowing of the band gap under pressure. The electrical resistivities of $TlInSe_2$ at ambient conditions are $\rho_\parallel = 99.6\ \Omega$ cm, and $\rho_\perp = 24.9$ k$\Omega$ cm, respectively. Under high pressure, both these values are reduced continuously (Figure 24a), reaching values as low as $5.5 \times 10^{-4}$ and $1.0 \times 10^{-2}$ $\Omega$ cm, respectively, around 5.0 GPa. $TlInTe_2$ has $\rho_\parallel = 199.6\ \Omega$ cm and $\rho_\perp = 33.2\ \Omega$ cm at ambient conditions, and exhibits a similar high-pressure resistivity behavior (Figure 24b), i.e. continuous metallization under pressure. In this crystal, $\rho_\parallel$ and $\rho_\perp$ drop down to $2.0 \times 10^{-2}$ and $5.0 \times 10^{-3}$ $\Omega$ cm, respectively, around 5.0 GPa.

The chain-type $TlInSe_2$ and $TlInTe_2$ crystals exhibit anisotropy $\rho_\parallel/\rho_\perp = 0.004$ and 6.0 at ambient conditions [56]. While the former can be attributed to 1D behavior, the latter definitely can not be. Under high pressure, both crystals show anisotropic resistivity (Figure 24); however, while $TlInSe_2$ shows $\rho_\parallel/\rho_\perp < 1$, $TlInTe_2$, in opposite, exhibits $\rho_\parallel/\rho_\perp > 1$. In other words, the former crystal shows higher conductivity along the chain axis (i.e., one-dimensional-like behavior), while the latter one – in the *a,b* plane. Resistivity measurements of $TlInSe_2$ and $TlInTe_2$ crystals down to liquid nitrogen temperature [56] showed that $\rho_\parallel$ as a function of temperature obeys Arrhenius law (Figure 25)

$$\rho_\parallel = \rho_0 \exp(\Delta E/kT) , \qquad (6)$$

where $\rho_0$ is the pre-exponential factor and *ΔE* is the activation energy for electrical conduction. At that, the slope of the ln $\rho_\parallel$ (1/T) curves depends on the applied pressure. Rabinal et al. [56] showed such dependence up to 1.2 GPa only. Finally, the reduction of the band gap with increased pressure (Figure 26) should lead to metallization of the compounds. The authors discuss the measured properties [56] along the band structure calculations by Gashimzade and Orudzhev [120]. Since, according to ref. [120], in TlSe-type structures the top of the valence band at the T point originates from the *s*-states of monovalent $Tl^+$ ions and from the bonding orbitals of Se ions, and the bottom of the conduction band arises mainly from

*s*-lone pairs of Se ions, they conclude that the monovalent Tl⁺ ions play an important role in the establishment of the semiconductor nature of the compounds, and that the interactions between the *s* lone pairs, $p_x$ and $p_y$ orbitals of chalcogen ions, on the one hand, and the *s* states of the monovalent Tl⁺ ions on the other hand, are mainly responsible for the metallization of these compounds under pressure [56].

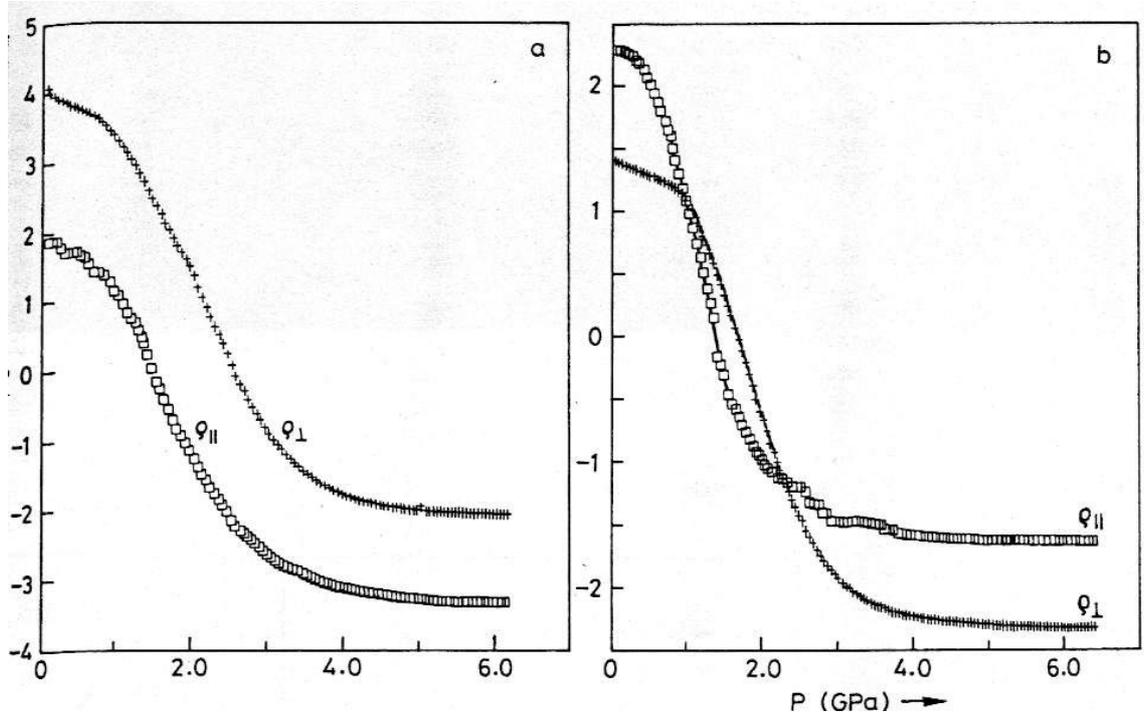

Figure 24. Pressure dependence of the parallel and perpendicular components of the electrical resistivities of (a) TlInSe$_2$ and (b) TlInTe$_2$ at ambient temperature (From [56]).

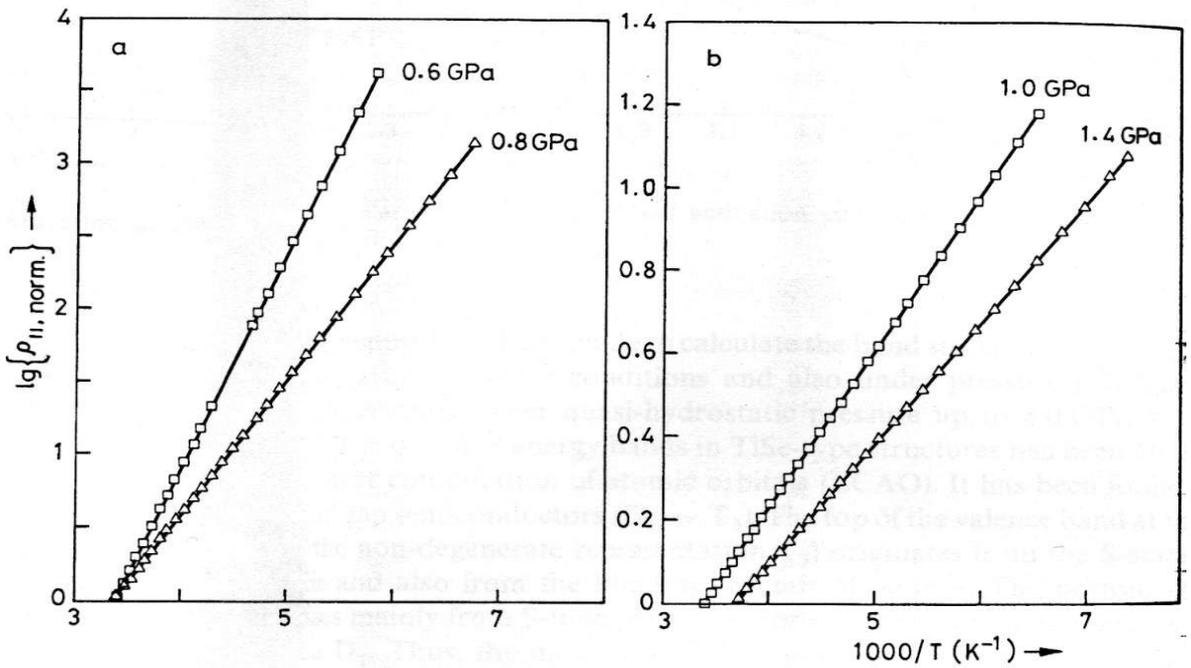

Figure 25. Normalized electrical resistivity $\rho_\parallel$ of TlInSe$_2$ as a function of temperature at different pressures. (From [56]).

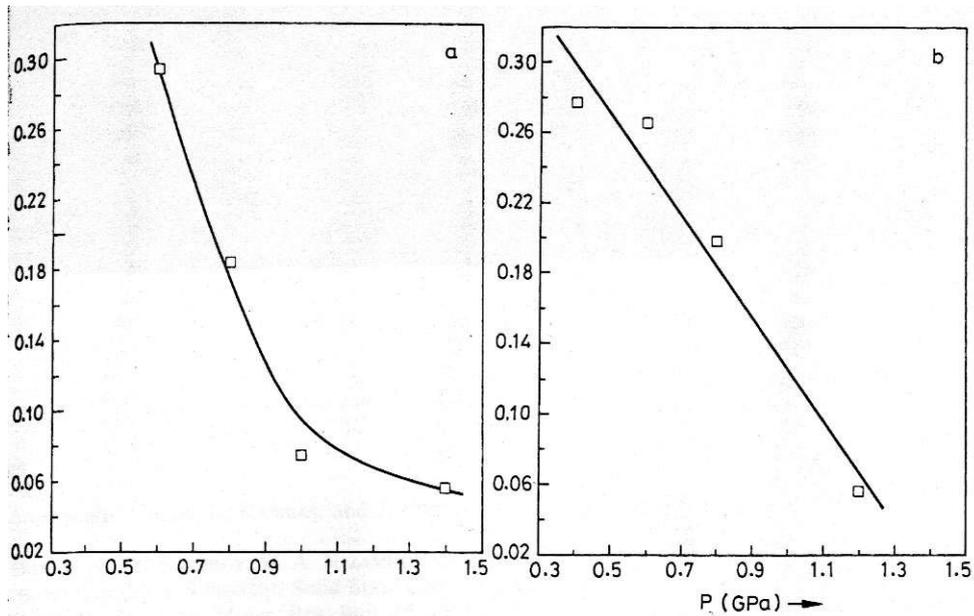

Figure 26. Pressure dependence of the electrical activation energy, $\Delta E_\parallel$, for (a) $TlInSe_2$ and (b) $TlInTe_2$. (From [56]).

Completing this section, I would also like to notice the interesting paper by Parthasarathy et al. [131] who reported on pressure-induced transitions in amorphous thallium-selenium alloys. In this paper, the electrical resistivity of bulk semiconducting amorphous $Tl_xSe_{100-x}$ alloys with $0 \leq x \leq 25$ has been investigated up to a pressure of 14 GPa and down to liquid nitrogen temperature using a Bridgman anvil device. All compounds undergo a discontinuous pressure-induced semiconductor-metal transition (Figure 27). The value of the electrical conductivity of the high-pressure phase lies in the range from 100 to 1000 $\mu\Omega$ cm. The temperature coefficient of the samples in the metallic phase ranges from 0.80 $\mu\Omega$ cm K$^{-1}$ for the $Tl_5Se_{95}$ sample to 10.52 $\mu\Omega$ cm K$^{-1}$ for the $Tl_{25}Se_{75}$ sample. The transition pressure decreases with Tl content. The variation of the *d.c.* conductivity as a function of temperature obeys the Arrhenius relation. X-ray diffraction studies show that the high-pressure phase is the crystalline phase. The pressure-induced crystalline products were identified to be a mixture of Se having a hexagonal structure with *a* = 4.37 Å and *c* = 4.95 Å, and TlSe having a tetragonal structure with *a* = 8.0 Å and *c* = 7.0 Å. Thus the metallic-like conductivity is seemed to be observed in the tetragonal TlSe-type phase. The X-ray diffraction studies of the samples subjected to pressures less than the transition pressure do not show any crystallinity. So the pressure-induced glass-to-crystal transition occurs simultaneously with the semiconductor-metal transition.

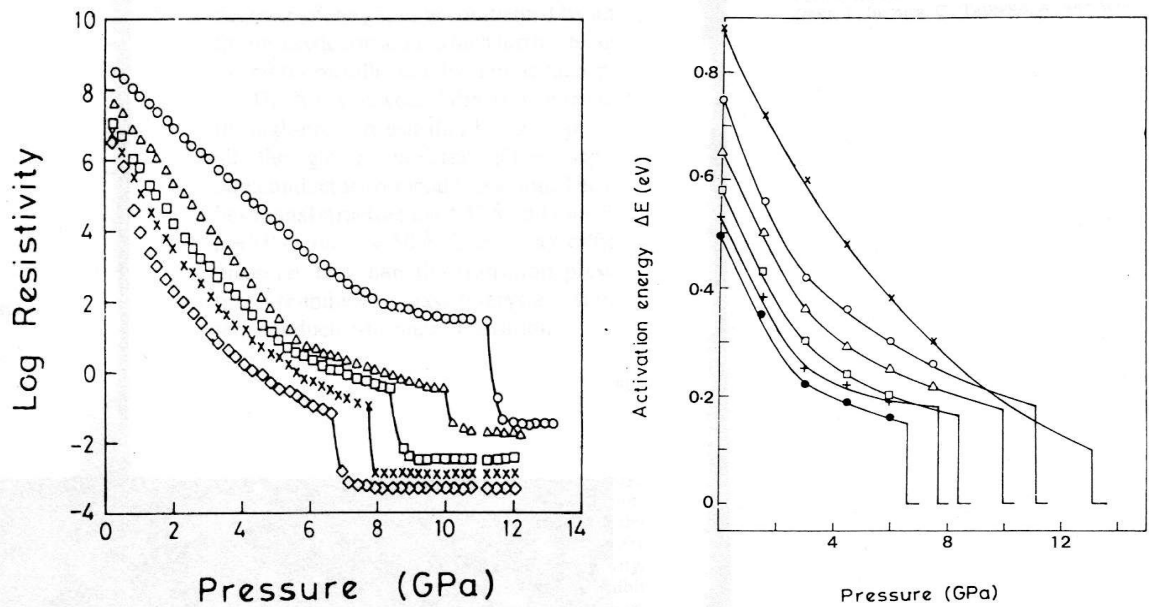

Figure 27. Pressure dependence of the electrical conductivity (left panel) and activation energy (right panel) of $Tl_xSe_{100-x}$ alloys. Data presented for x values of: 5 (circles); 10 (triangles); 15 (squares); 20 (crosses); and 25 (diamonds) on the left panel and for x values of: 0 (crosses); 5 (circles); 10 (triangles); 15 (squares); 20 (pluses); and 25 (filled circles) on the right panel. (From [131]).

Table 16. Phase transition pressures for binary and ternary chain-type thallium chalcogenide crystals.

| Compound | Method of investigation | Phase transition pressure, GPa | Reference |
|---|---|---|---|
| TlSe | calculation | 2.2 – 2.7 (uniaxial pressure) <br> 5 (hydrostatic pressure) | 120 |
| TlSe | x-ray | 21 | 127 |
| TlS | x-ray | 5, 10 and 25 | 127 |
| TlSe | electrical resistivity | ~ 2.7 | 54 |
| $TlInSe_{0.2}S_{1.8}$ | optical absorption | 0.6 | 123 |
| $TlInSe_{1.42}S_{0.6}$ | optical absorption | 0.72 | 125 |
| $TlInTe_2$ | Raman spectra | 7 | 126 |
| $TlInSe_2$ | electrical resistivity | ~ 5 | 56 |
| $TlInTe_2$ | electrical resistivity | ~ 5 | 56 |
| $Tl_xSe_{100-x}$ | electrical resistivity | From ~ 7 (x = 25) to ~ 11.5 (x = 5) | 131 |
| InTe | x-ray | 5 and 15 | 16 |

And finally, let us mention the pressure-induced properties of InTe that is isostructural to TlSe. At ambient conditions, the chemically distinct In$^+$ and In$^{3+}$ ions occupy two different crystallographic positions preventing free transfer of electrons from the In$^+$ to the In$^{3+}$. Upon application of hydrostatic pressure of 3 GPa and temperature in the 670 – 770 K range, InTe transforms to the NaCl (B1) structure [16,132-137]. All cations are equivalent in this modification and are coordinated by six Te$^{2-}$ ions. The structural constraint upon electron transfer from In$^{1+}$ to In$^{3+}$ ions is thus removed and the compound exhibits metallic conductivity. Moreover, Geller et al [132,133] reported that cubic (B1) InTe is superconductor below 2.18 K, while Banus et al. [137] observed the superconducting transition at 3.5 K. Chattopadhyay et al. [16] investigated InTe using high pressure X-ray diffraction technique up to 34 GPa. They discovered that InTe undergoes a pressure induced first-order phase transition from the tetragonal TlSe-type (B37) to the NaCl-type (B1) phase at about 5 GPa. The latter is just the aforementioned pressure-induced NaCl-type phase that exhibits metallic conductivity caused by removal of the structural constraints upon electron transfer from In$^{1+}$ to In$^{3+}$ ions. A further pressure induced first-order phase transition from the NaCl-type to the CsCl-type (B2) phase was discovered at about 15 GPa. The linear thermal expansion coefficients along the *a* and *c* axes in the temperature range 297 – 505 K were found to be $1.99 \times 10^{-5}$ and $1.99 \times 10^{-5}$ $K^{-1}$, respectively. Although *a* and *c* change considerably with temparature, the ratio c/a was found to be constant (0.85) in the temperature range investigated.

For the convenience of the readers, all data on phase transition pressures for binary and ternary chain-type thallium chalcogenide crystals are collected in Table 16.

### *6.2. Layered crystals*

Applied pressure enhances the interlayer coupling and could therefore lead to the pressure-induced phase transitions, which may be either reversible or irreversible. The data on pressure-induced phase transitions in the layered $TlMX_2$ crystals are somewhat contradictory. The first report on a possible phase transition in such compounds under pressure was published by Vinogradov et al. [138], who measured the room temperature Raman spectra of single crystals $TlGaS_2$ and $TlGaSe_2$ under various hydrostatic pressures up to 0.725 and 0.927 GPa for the first and second crystal, respectively. In $TlGaSe_2$, the new band appears in a narrow pressure range from 0.5 to 0.55 GPa. This effect is reversible. In this pressure range discontinuities of several Raman bands were observed. (We notice, however, that most of these discontinuities seem to be within the accuracy of the measurements). $TlGsSe_2$ shows some change in Raman spectra and decrease in the intensity of all Raman bands in the vicinity of 0.5 GPa. The authors claim that the observed features can serve as a proof of a structural phase transformation. However, Henkel et al. [6] ascertained that an appearance and disappearance of Raman modes with increased pressure are due to a crossing of bands with $A_g$ and $B_g$ symmetry and are therefore not sufficient to indicate the occurrence of a structural phase transition.

Allakhvrdiev et al. [139] observed some features in the fundamental absorption edge of $TlGaS_2$, $TlGaSe_2$ and $TlInS_2$ crystals under pressure and tried to explain them suggesting phase transitions in the pressure range of 0.3 to 0.9 GPa.

Prins et al. [140] measured the optical transmission spectra of $TlGaS_2$, $TlGaSe_2$ and $TlInS_2$ crystals up to 13 GPa and plotted the values of the pressure coefficient of the transmission spectra at 30% transmission against pressure (Figure 28). They observed several pressure-induced changes in the sign of the pressure coefficient that were attributed to the phase transitions. For $TlGaS_2$, they postulated existence of three phases: phase I in the pressure range from 0 to ~ 0.8 GPa, phase II in the pressure range from ~ 1.2 to ~ 3.1 GPa, and phase III above ~ 5 GPa. For $TlGaSe_2$, four phases were proposed: phase I in the pressure range from 0 to ~ 1 GPa, phase II from ~ 1.1 to ~ 4.4 GPa, phase III between ~ 5.6 and ~ 8.5 GPa, and phase IV from ~ 9.7 to ~ 14.8 GPa. For $TlInS_2$, the authors proposed five phases: phase I in the pressure range from 0 to ~ 1 GPa, phase II from ~ 1.0 to ~ 2.9 GPa, phase III between ~ 3.4 and ~ 6.3 GPa, phase IV between ~ 7.6 and ~ 10.6 GPa, and phase V above ~ 12.1 GPa. To our knowledge, such phase multiformity was not confirmed by the other investigators. Anyhow, for making more reliable conclusions, confirmations coming from the accurate X-ray diffraction (XRD) measurements are needed.

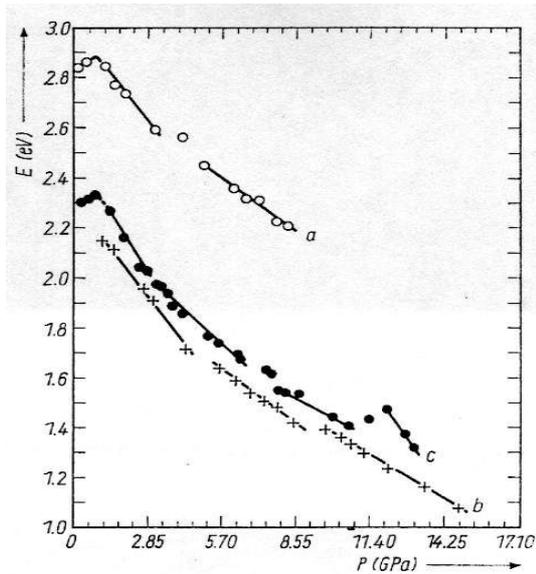

Figure 28. Pressure dependence of the 30% transition point $E_{30}$ of (a) TlGaS$_2$, (b) TlGaSe$_2$, (c) TlInS$_2$. (From [140]).

Ves [141] investigated the effect of hydrostatic pressure on the lowest direct gaps $E_a$ ($\Gamma 2v \rightarrow \Gamma 3,2c$) and $E_b$ ($\Gamma 2v \rightarrow \Gamma 3,1c$, $\Gamma 1c$) of thallium gallium selenide (TlGaSe$_2$) by optical-absorption measurements in a diamond-anvil cell for thin (3.5-20 μm) samples at room temperature and at pressures up to 12.0 GPa. Sublinear and linear red shifts were observed for the $E_a$ and $E_b$ gaps, respectively. The results are indicative of a possible phase transition at about 1.85 GPa.

Recently Perez et al. [142] reported on Raman scattering measurements under pressure performed in a single crystal of TlGaS$_2$. By means of the spectra deconvolution, the mode-Gruneisen parameters of the external and internal vibration modes were calculated. A first order phase transition at 7.4 GPa and at 10 GPa for external and internal modes, respectively, was observed. However, measurements of the refractive index $n$ of TlGaS$_2$ as a function of pressure $P \leq 20$ GPa in a diamond anvil cell, carried out by Contreras et al. [143], indicated a possible structural phase transition at 13 GPa. Such conclusion was made based on the observed change in the slope of the pressure dependence of the refractive index from positive to negative at pressure $\leq 13$ GPa. At that, the authors interpreted the negative slope at high pressure as a sign of "the typical behavior of 3D solid".

The only pressure-dependent XRD study of layered compound was made by Range et al [144] who reported on a quenchable high-pressure phase of TlGaSe$_2$ isotypical to the chain TlSe type and belonging to the tetragonal space group I4/mcm with $a = 8.053$ Å and $c = 6.417$ Å at T = 600 C and P = 2 GPa. Treatment of monoclinic TlInS$_2$-I at 300 C and 3 GPa [145] gave the hexagonal-rhombohedral modification TlInS$_2$-II in the space group $R\bar{3}m$ -$D^5_{3d}$ with $a = 3.83$, $c = 22.23$ Å, and $Z = 3$. It has α-NaFeO$_2$-type structure and is metastable. At 400 C and 3 GPa, both TlInS$_2$-I and -II are transformed into hexagonal TlInS$_2$-III in the space group P6$_3$/mmc-$D^4_{6h}$ with $a = 3.83$, $c = 14.88$ Å, and $Z = 2$ [145].

Table 17. Phase transition pressures for ternary layer-type thallium chalcogenide crystals.

| Compound | Method of investigation | Phase transition pressure, GPa | Reference |
|---|---|---|---|
| $TlGaS_2$ | Raman spectra | 0.5 | 138 |
| $TlGaS_2$ | optical absorption | 0.3 – 0.9 | 139 |
| $TlGaS_2$ | transmission spectra | from 0.8 to 1.2 and from 3.1 to 5 | 139 |
| $TlGaSe_2$ | Raman spectra | 0.5 | 138 |
| $TlGaSe_2$ | optical absorption | 0.3 – 0.9 | 139 |
| $TlGaSe_2$ | transmission spectra | from 1 to 1.1; from 4.4 to 5.6; and from 8.5 to 9.7 | 140 |
| $TlInS_2$ | optical absorption | 0.3 – 0.9 | 139 |
| $TlInS_2$ | transmission spectra | 1; from 2.9 to 3.4; from 6.3 to 7.6; 12.1 | 140 |
|  | XRD | 3 GPa and 300 C, 3 GPa and 400 C | 144,145 |

For the convenience of the readers, all the data on phase transition pressures for the ternary layer-type thallium chalcogenide crystals are collected in Table 17. One can find out some scattering of the measured transition pressures. Allakhvrdiev [146] noticed that the existence of polytypes (due to a variety of layer-packing in the *c* direction) maybe a reason for some controversies concerning the optical and other properties of these crystals. However, the main problem is that the effects discussed above have been obtained by the optical measurements, and that the occurrence of the phase transitions in the layered crystals has never been confirmed by the XRD measurements (except for ref. [144,145]). In our opinion, assignment of the observed features of the absorption and transmission spectra to phase transitions is of insufficient reliability. To produce a strong evidence of the pressure-induced phase transitions in the layered crystals, further investigations, particularly detailed and accurate XRD measurements, are needed. Also, one must be sure of no destruction of the single crystals under the high-pressure experiment.

We note that in the layer-type compounds $Tl^+$ and $Ga^{3+}/In^{3+}$ ions are located in the different sublattices and exhibit rather weak bonding between anion layers and cation chains. This circumstance prevents free transfer of electrons between $Tl^+$ and $Ga^{3+}/In^{3+}$. Therefore no metal-semiconductor transitions, characteristic of the chain $TlMX_2$ structures, were observed in the layered compounds under review.

## 7. Temperature-induced phase transitions and incommensurability at ambient pressure

In this section, we will consider phase transitions in TlX and $TlMX_2$ crystals that are observed due to temperature variation at the ambient pressure. Let us start our discussion with the chain-like crystals.

### *7.1. Tetragonal (chain-type) TlSe and TlS crystals*

It was already noticed above tha Morgant et al. [130,147], who studied phase diagram of thallium selenides using thermoanalytical, X-ray diffraction and metallographic examinations, reported on the observation of a new polymorphic cubic centered form of the solid TlSe at high temperature, around 470 K. The two forms, tetragonal α–form with $a$ = 8.02 Å, $c$ = 7.00 Å and cubic β–form with $a$ = 6.187 Å, are separated by a two-phase (α + β) region near 470 K. Occurrence of the aforementioned β-form was confirmed by Romermann et al [148]. Eventual properties of this phase were briefly discussed in section 6. In the mixed valence compound TlSe, or $Tl^{1+}Tl^{3+}Se_2^{-2}$, the reported structural transformation could involve electron hopping between $Tl^{1+}$ and $Tl^{3+}$ ions, since such hopping should equalize the electronic configurations of the Tl atoms. Therefore, though the low temperature phase with alternating univalent and trivalent ions is semiconducting, one can expect that the high temperature phase with equivalent ions will be metallic due to electron hopping between the $Tl^{1+}$ and $Tl^{3+}$ sites, by analogy with the pressure-induced phases in TlSe that exhibits metallic conductivity caused by removal of the structural constraints upon electron transfer from $Tl^{1+}$ to $Tl^{3+}$ ions. We note, however, that Brekow et al [149], Kurbanov et al [150], Mamedov et al [151-153], Aliev et al [154] and Rzaev et al [129] who measured specific heat, thermal expansion and elastic constants of TlSe, did not report on the temperature-induced phase transitions.

Information about such phase transitions in the chain-like TlS is absent.

### 7.2. Chain-type TlInSe$_2$ and TlInTe$_2$ crystals

Heat capacity measurements of a chain-like TlInSe$_2$ crystal (Figure 29, curve 2), carried out by Mamedov et al [155], showed no anomalies in the temperature range 4.2 to 300 K. However, Alekperov et al [156], who measured temperature dependence of the heat capacity, lattice parameters, and photoconductivity in this crystal, reported on a phase transition in the range of 135-184 K. Aliev et al [157] and Allakhverdiev et al [158] also reported that TlInSe$_2$ exhibits some anomalies and features, which were tentatively attributed to structural transitions yielding incommensurate phases. From the contradictory data mentioned above, one can conclude that the information about phase transitions in this compound is inconsistent.

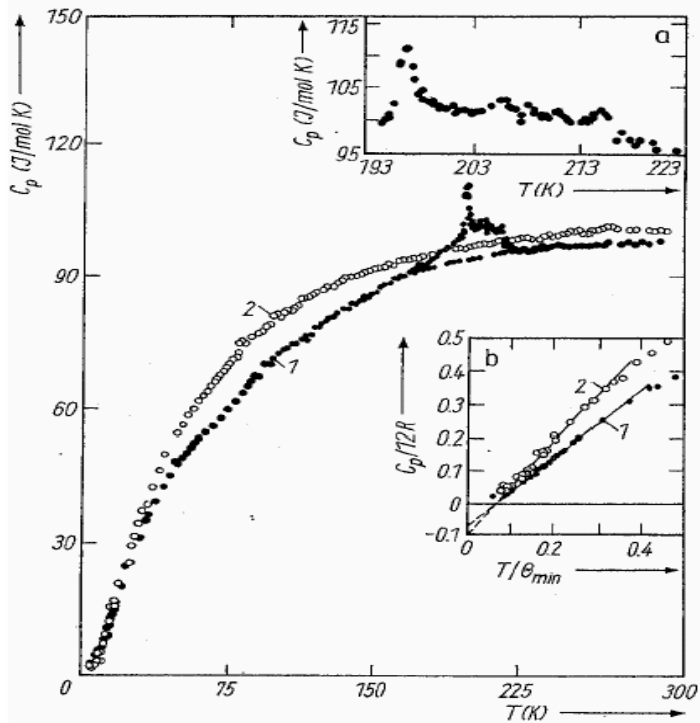

Figure 29. Temperature dependence of the heat capacities of (1) TlInS$_2$ and (2) TlInSe$_2$ (From [155]).

Heat capacity measurements [159] at temperatures from 5 to 300 K and powder x-ray diffraction [160] and transport [57] measurements from 80 to 300 K showed that TlInTe$_2$ does not undergo phase transitions in the above temperature regions.

### *7.3. Chain-type TlGaTe$_2$: phase transitions and incommensurate state*

Aliev et al [161] reported on x-ray diffraction, calorimetric and transport measurements of the semiconductor crystal TlGaTe$_2$ exhibiting chain-type structure. X-ray diffraction reveals unusual temperature behavior of the lattice parameters (Figure 30). The temperature dependence of the lattice parameter *a* (curve 1 at the left panel of Figure 30) shows several invariant regions (110-160 K, 180-210 K and 240-270 K), between which *a* increases monotonically. The *a*(T) dependence exhibits a strong anomaly in the temperature range 90 – 110 K. The temperature dependence of the intensities of the (200) and (211) Bragg reflections (curve 2, left panel of Figure 30) correlates with the *a*(T) dependence. Such a behavior is characteristic of the complete devil's staircase. Heat capacity ($C_p$) measurements (middle panel of Figure 30) show an anomaly at T = 98.5 K that was attributed to the phase transition. Electric conductivity (measured on cooling) exhibits an anomaly near 93 K with a change of 38%. The authors [161] concluded that (i) TlGaTe$_2$ undergoes a second-order phase transition at T = 98.5 K and (ii) thermal behavior of the lattice parameters and of the intensities of the Bragg reflections in the temperature region from 110 to 290 K reflects the periodic modulation characteristic of an incommensurate phase.

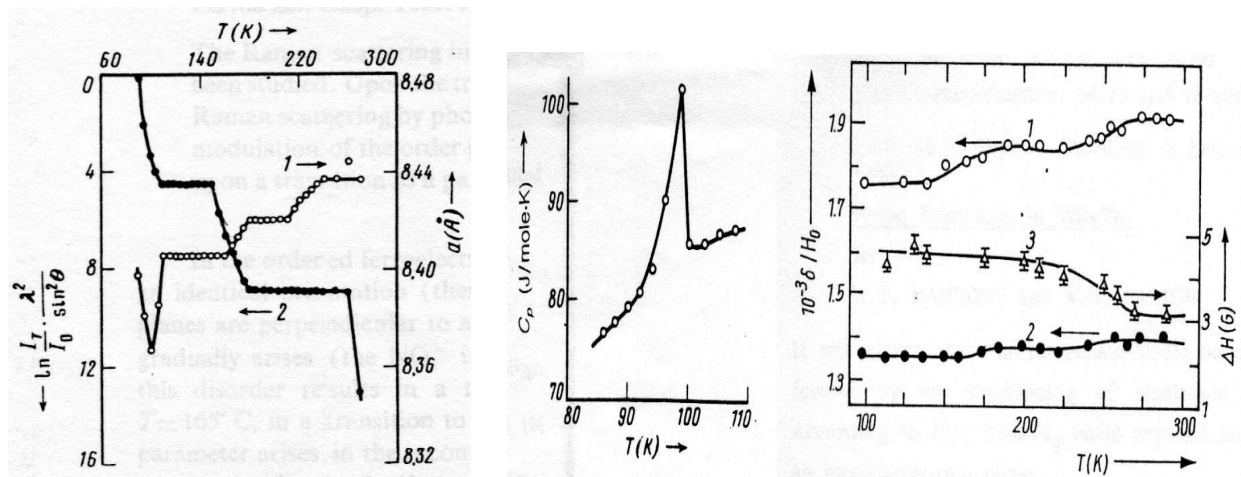

Figure 30. Left panel: Temperature dependence of the lattice parameter *a* (1) and ln ($I_T$ / $I_{90\,K}$) ($\lambda^2$/sin$^2\theta$) (2) in TlGaTe$_2$ (From [161]). Middle panel: Temperature dependence of the specific heat $C_p$ of TlGaTe$_2$ in the region of the anomaly. (From [161]). Right panel: Temperature dependence of magnetic shielding tensor components $\sigma_\parallel$ (1) and $\sigma_\perp$ (2) and $^{205}$Tl NMR linewidth (3) (From [96,163]).

Occurrence of the aforementioned phase transitions in chain semiconductor TlGaTe$_2$ has been confirmed by means of XRD [160,162] and NMR measurements [96,163]; at that, no anomalies were observed in dielectric measurements [162]. Therefore one can suggest that the transition is not of a ferroelectric type. Temperature dependent NMR measurements of the $^{205}$Tl chemical shielding of this crystal, carried out by Panich et al [96,163] (right panel of Figure 30), revealed several plateaus. The observed dependences of the shielding tensor components $\sigma_\parallel(T)$ and $\sigma_\perp(T)$ correlate with the $a(T)$ and $I(T)$ curves obtained by Aliev et al [161]. NMR measurements in the frequency range from 19 to 45 MHz does not reveal magnetic field dependence of the line width, indicating that the occurrence of the structurally inequivalent thallium atoms with different shielding constants in the low temperature phase is questionable. Assuming that the spatially modulated structure suggested by Aliev et al [161] nevertheless occurs, the XRD and NMR data can be reconciled on condition that the difference in the shielding parameters of Tl atoms in rather low magnetic fields used in the NMR experiments is less than the line width.

Summarizing the aforementioned data on the crystals with the chain-type structure, we are led to conclusion that the phase transitions were reliably established for the chain-type TlGaTe$_2$ crystal only, while their existence in the other crystals is still questionable.

### *7.4. Phase transition in tetragonal TlTe.*

To our knowledge, phase transition in the semi-metallic TlTe was discovered by Jensen et al [69] who observed a jump in the Hall coefficient and a hump in the resistivity by a factor of 2.5 at T = 170 K. Then this transition has been studied in detail by Stöwe [18] by means of the temperature dependent XRD measurements. As it was mentioned in section 2, at ambient temperature TlTe crystallizes in the space group I4/mcm. The crystal structure reveals univalent Tl$^+$ cations and a polytelluric counterpart with linear equidistant Te chains in the [001] direction. One-half of the chains is unbranched; the other one consists of linear [Te$_3$] units stacked cross-shaped one upon the other. Stöwe showed [18] that at T = 172 K, one-half of the branched chains transforms by a Peierls distortion into a linear chain with alternating distances of 2.855 and 3.302 Å. By this transformation the unit cell volume is doubled and the space group changed into P4$_2$/nmc with $a$ = 18.229 and $c$ = 6.157 Å at 157 K. Since the other half of the branched chains and the unbranched chains remain equidistant, it was expected that the compound keeps its semimetallic behavior, in accordance with experimental resistivity data in the literature.

### *7.5. Ferroelectric phase transition and incommensurate state in the layered crystal TlS*

Dielectric and XRD measurements of the monoclinic modification of semiconductor thallium sulfide, carried out by Kashida et al [11,164,165], revealed that this crystal undergoes successive structural phase transitions at 318.6 K and 341.1 K (Figure 31, left panel) with an intermediate incommensurate phase. The transition at 341.1 K is of the second order, while that at 318.6 K is of the first order, showing a temperature hysteresis of 1 K [165]. The room temperature phase was found to have a commensurate structure; the satellite reflections observed at $q_c = (0,0,1/4)$ suggest fourfold lattice modulation along the $c$ axis. The modulation is caused by cooperative distortion of trigonal prisms around the univalent $Tl^{1+}$ ions. A study of the polarization loop showed that this phase is ferroelectric [11]. The intermediate phase between $T_c = 318.6$ K and $T_i = 341.1$ K was found to be incommensurate [11,164,165], with the satellite reflections shifted to $q_i = (0.04, 0, 1/4)$. The high temperature phase above $T_i = 341.1$ K is paraelectric, where the satellite reflections disappear, and the compound shows commensurate structure that belongs to the space group C2/c, having the $TlGaSe_2$ - type structure.

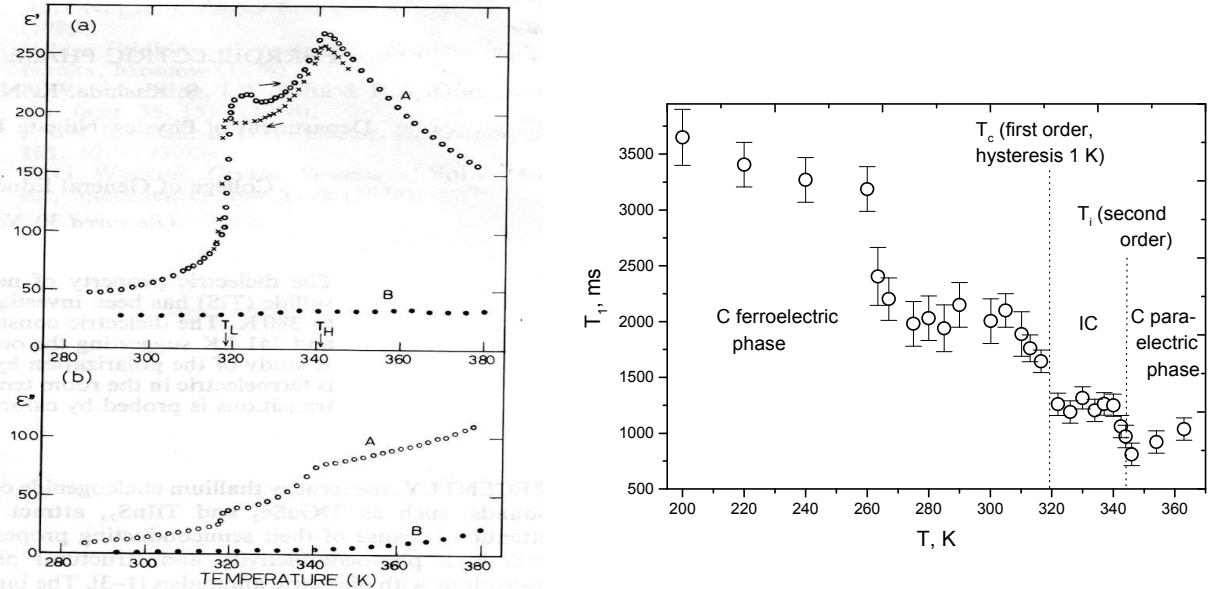

Figure 31. Left panel: Temperature dependence of the real (a) and imaginary (b) parts of the dielectric constant of the monoclinic TlS, measured at 1 MHz. Open circles – sample A on heating; crosses – sample A on cooling; filled circles – sample B on heating. (From [11]).
Right panel: Temperature dependence of the $^{205}$Tl spin-lattice relaxation time $T_1$ measured at the resonance frequency 197.2 MHz. (From [166]).

The above conclusions on phase transitions are supported by the $^{205}$Tl NMR measurements [166]. Temperature dependence of the $^{205}$Tl spin-lattice relaxation time $T_1$ (Figure 31, right panel) reveals phase transitions at 319 K and 341 K. The plateau in $T_1(T)$ between these temperatures is a characteristic property of the classical incommensurate phase, corresponding to the temperature-independent phason-driven spin-lattice relaxation. An increase in $T_1$ at ~ 265 K is indicative of the third phase transition.

Besides the pioneer works of Kashida et al [11,164,165], there were several other attempts to prepare and to study monoclinic thallium monosulfide. Sardarly et al [167] used powder x-ray diffraction, differential thermal (DTA) and micro-structural analyses to study phase diagram of the Tl-S system in the range of TlS + 0, 2, 4 and 6 % S, corresponding to the monoclinic phase of TlS. The TlS + 4% S specimen have been measured in more detail and showed phase transformations at 290 and 352 K. The higher phase transition temperature $T_i$ (352 K instead of 341 K) and assumed occurrence of the tetragonal phase at 353 K contradicts the aforementioned reliable and well-repeated data of Kashida [11,164,165] and is probably due to the non-stehiometry of the samples of Sardarly et al [167]. Aliev et al [168] reported on preparation, structure, and electrical properties of TlS single crystals with an excess of sulfur (4 at. %). They assert observation of monoclinic TlGaSe$_2$-type phase after annealing the sample at 258 K, of tetragonal phase with the space group $P4_12_12$ after annealing the sample at T = 323 and 373 K K, and TlSe-type tetragonal phase of the space group $I4/mcm$ after annealing at T = 423 K. The transport and dielectric measurements of thallium monosulfide show an increase in the electrical conductivity on heating and a hump in the dielectric permittivity in the temperature range from 401 to 411 K. The authors suggest that these findings indicate a phase transition into a superionic conduction state.

### *7.6. Ferroelectric phase transition and incommensurate state in the layered crystal TlInS$_2$. Relaxor behavior of doped and irradiated TlInS$_2$ compounds*

Historically, layered TlInS$_2$ and TlGaSe$_2$ crystals crystals were the first low-dimensional semiconductors in which a series of phase transitions with modulated structures have been discovered. In this and following sections, we present a detailed review of the above properties of these crystals, starting with TlInS$_2$. To set forth below, we will use the standard designations, i.e. temperature $T_i$ will correspond to a high temperature normal-incommensurate transition, and $T_c$ - to a low temperature "lock-in" transition to a lower temperature commensurate phase.

Phase transitions in TlInS$_2$ have been discovered in 1980's by Volkov et al. [169], Vakhrushev et al. [170] (Figure 32) and Aliev et al. [171] who observed anomalous behavior of this crystal by means of dielectric, optical, dilatometric and neutron diffraction measurements. Then these transitions were studied in detail by the other authors [3,172-208] using heat capacity, x-ray diffraction and other techniques. As early as the first measurements [169,171] showed that a spontaneous polarization occurs in TlInS$_2$ below $T_c$, these structural phase transitions were attributed to transitions of paraelectric-ferroelectric type, with presumably an intermediate incommensurate phase. Neutron diffraction pattern obtained by Vakhrushev et al. [170] in the temperature range of 200 - 216 K corresponds to an incommensurate phase with $q_{IC}$ = ($\delta$, $\delta$, 0.25) and $\delta$ = 0.012. Below T = 200 K, the authors obtained a change in the structure modulation, accompanied by the appearance of a commensurate structure with $q_1$ = (0, 0, 0.25), though with some small deviation from commensurability. Further cooling results in the final lock-in phase transition to the commensurate phase with $q_1$ = (0, 0, 0.25) that is the evidence of a quadrupling of the unit cell along the $c$ axis. Vakhrushev et al. [170] reported that the low-temperature transition at $T_i \sim$ 170 K shows a hysteresis and thus is the first order transition, while two others are nearly second-order transitions. Detailed single-crystal x-ray diffraction and dielectric measurements of TlInS$_2$ have been made by Kashida and Kobayashi [3]. Upon cooling, the dielectric constant exhibits two maxima at 209 and 200 K and a shoulder at 197 K (Figure 33, left panel), corresponding to successive phase transitions. In the paraelectric phase, the temperature dependence of the dielectric constant is well fit by the Curie-Weiss curve. At higher temperature, however, owing to the semiconductor nature of the crystal, the *dc* conductivity increases, and the dielectric constant deviates from the Curie-Weiss law. X-ray diffraction showed successive structural phase transitions at 194 and 214 K with the intermediate incommensurate phase.

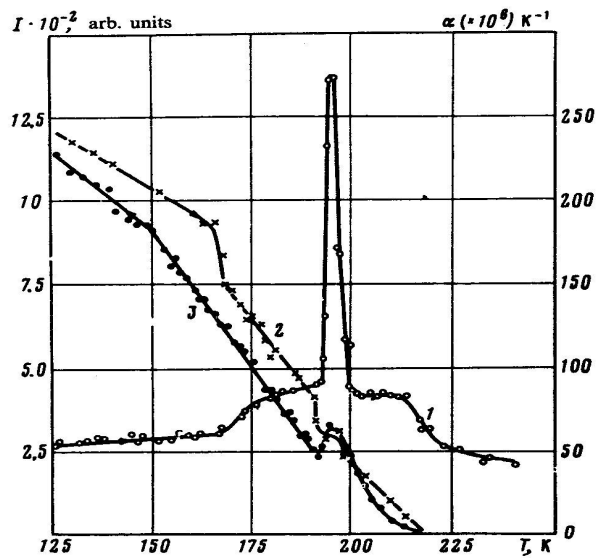

Figure 32. Temperature dependence of the thermal-expansion coefficient α (curve 1) and neutron diffaction scattering intensity at the point [1; 1; 1.25] (curves 2, 3). (From [170]).

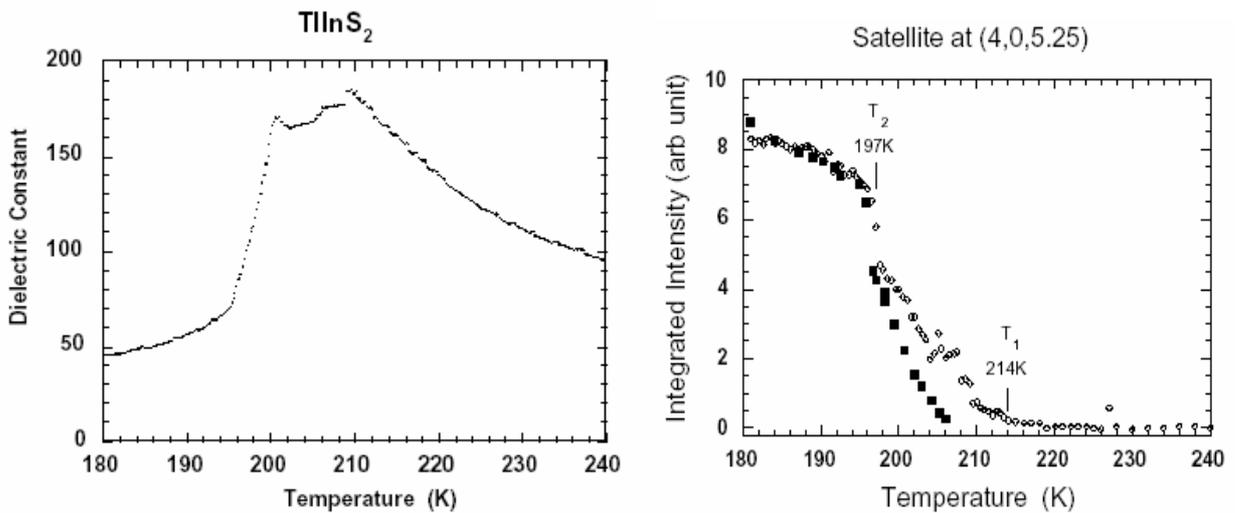

Figure 33. Left panel: Temperature dependence of the real part of the dielectric constant in TlInS$_2$ measured along the [1,1,0] axis. Peak at 210 K and the shoulder at 197 K correspond to transitions to the intermediate- and the lowest-temperature phases, respectively. Right panel: Temperature dependence of the integrated intensity of the satellite peak in TlInS$_2$, measured around the (4, 0, 5.25) Bragg point. The full squares show the data for the spontaneous polarization normalized at 85 K. (From [3]).

Right panel of Figure 33 shows temperature dependence of the integrated intensity of the diffraction peaks. When the crystal is cooled down to T = 214 K, satellite reflections appear. The satellite maps revealed that the modulation mode, which characterizes the intermediate phase, has $q_i$ = ($\delta$, 0, 0.25) with $\delta \approx$ 0.044. It was shown that the previously reported satellite reflections at $q$ = ($\delta$, $\delta$, 0.25) should be ascribed to twins. In the temperature range between 214 and 197 K, the satellite intensity increases almost linearly on cooling. A structural model for the incommensurate phase was presented. Below 197 K, the satellite intensity increases steeply, indicating a second phase transition to the low-temperature phase. In this phase, the satellite reflections are shifted to a commensurate position with $q_c$ = (0, 0, 0.25), reflecting quadrupling of the lattice period along the *c* axis. Kashida and Kobayashi [3] discussed the origin of the structural phase transitions in TlInS$_2$ that can be ascribed to small displacements of atoms from the positions that they occupy in the high-temperature phase. Since the dielectric constant measured along the *c*-axis shows little anomaly at the phase transition points, the authors [3] expected the relevant displacements to be in the *a–b* plane. More detailed analysis showed that the displacements are parallel to the *a*-axis, and that the modulation wave that characterizes the intermediate phase is not of the fundamental q$_{1/4}$ - type but is of the optical third harmonic q$_{3/4}$ - type. Analogous model was suggested by the same authors in order to explain phase transitions in the layered TlS compound. The origin of the aforementioned phase transitions will also be discussed below in the end of this section.

Banys [173] reported on x-ray study of the structure of paraelectric and incommensurate phases in TlInS$_2$. Room temperature paraelectric phase is believed to be C2/c, but some of the (*h,h,l*) Bragg reflections, those with odd *l*, were broadened along the c* direction. This was explained to occur due to the stacking sheets perpendicular to the c* direction, suggesting that on average a stacking fault occurs in the single crystal once for every twenty layers. Between 214 and 195 K, the crystal has an incommensurate structure with a wave vector (δ,δ, 1/4), where δ = 0.02 reciprocal lattice units. Plyushch and Sheleg [174] reported on a x-ray diffraction investigation of TlInS$_2$ that showed an incommensurate phase with modulation wave vector $q_i$ = (1/4 ± δ)c* in the temperature range 196 – 214 K. Superstructure reflections of long-period polytypic modifications were not detected.

Specific heat measurements of the single crystal and powder TlInS$_2$ made by Krupnikov et al. [175] revealed a number of anomalies assigned to the phase transitions, though some of these peaks are inexpressive. The authors proposed a coexistence of two monoclinic modifications of TlInS$_2$ and suggested that the temperature variation results in step-like changes in the concentration of these phases, manifesting in jumps in the $C_p(T)$ curves. They speculated that the transformation between the micro-phases gives rise to an incommensurate structure. Abdullaev et al. [176] interpreted the obtained temperature dependence of the heat capacity of TlInS$_2$ in the range 196.9-214.9 K as an occurrence of a devil's staircase.

Ozdemir et al [177-179] reported on variations of the electrical conductivity and thermally stimulated current in TlInS$_2$ under successive thermocyclings between the commensurate and incommensurate phases and concluded that this process results in a shift in the commensurate-incommensurate phase transition temperature and in remarkable changes in the density of discommensurations. The results were explained in the model of a disordered-like system of coexisting incommensurate and commensurate states with discommensurations that influence the carrier transport. Aliev et al [29] and Youssef [180] detected very small variations of the electric conductivity under the static conditions in the regions of the phase transitions, 190 – 199 K, 200 – 212 K and 217 – 223 K. However, under dynamic conditions, Youssef [180] observed the pronounced anomalous behavior of the electrical conductivity at 192, 198, 201 and 217 K. Also the anomalies at temperatures 175, 205, and 209 K, corresponding, in the author's opinion, to ferroelectric and structural phase transition, have been detected.

Suleimanov et al. [181] suggested that the transitions in TlInS$_2$, observed at 201 and 204 K, are due to a splitting of commensurate-incommensurate phase transition into two closely set transitions. Dielectric properties of TlInS$_2$ below the ferroelectric phase transition temperature $T_c$ were regarded to the appearance of a chaotic state. Salaev et al [182] reported on a splitting of the "lock-in" transition at $T_c$ and suggested a coexistence of polar regions (domains) of the ferroelectric and the $T_{c1}$-$T_{c2}$ phases over a wide range of temperature. A metastable behavior of dielectric permittivity observed at thermo-cycling was explained by existence of a chaotic state below the ferroelectric phase transition temperature $T_{c2}$ = 201 K.

An opportunity of the incommensurate-incommensurate phase transition in the sequence of structural phase transitions in the layered crystal TlInS$_2$ has been demonstrated theoretically by Gadzhiev et al [183]. Allakhverdiev et al [184] reported that the dielectric susceptibility exhibits three peaks at 201, 206 and 216 K and a shoulder at 195 K (Figure 34), attributed to the successive phase transitions in the TlInS$_2$ crystals. They considered the transitions at $T_{c1}$ = 204 K and $T_{c2}$ = 201 K as incomplete lock-in transitions, the transitions at $T_i'$ = 206 K – as an additional incommensurate transition, while phase transition at 195 K as the final lock-in transition. Peak at 216 K was attributed to the normal-to-incommensurate phase transition. They also explained the observed peculiarities in the temperature behavior of the dielectric properties in the model of two inequivalent sublattices.

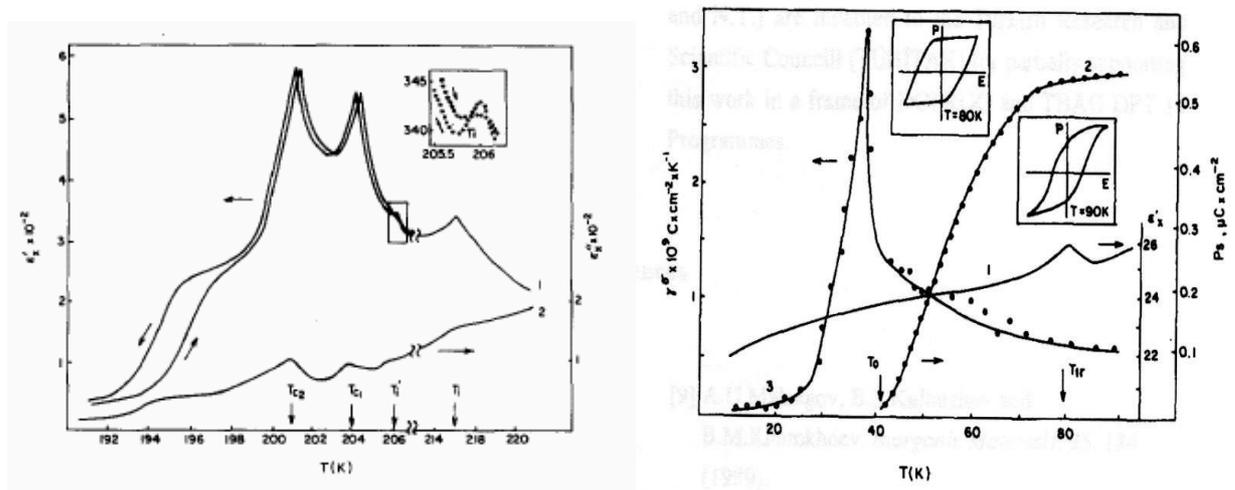

Figure 34. Left panel: Temperature dependences of real ε' (1) and imaginary ε'' (2) parts of dielectric susceptibility of TlInS$_2$. Vertical arrows show phase transition temperatures: $T_i$ - transition to incommensurate phase; $T_i'$ - transition at which a change of incommensuration takes place; $T_{c1}$ - transition to antipolar state; $T_{c2}$ - transition to the ferroelectric state. Arrows on ε'(T) curves indicate measurements on heating and cooling. The inset shows magnified part of ε' near $T_{i'}$ = 206 K. Right panel: Temperature dependences of ε' (1), of spontaneous polarization $P_s$ (2) and of pyroelectric coefficient $\gamma^\sigma$ (3) of TlInS$_2$. The inserts show the dielectric hysteresis loops taken at 80 and 90 K at E = 8 kV/cm. (From [184]).

Low temperature phase transition in TlInS$_2$ at T = 79 K (Figure 34) was observed in the dielectric susceptibility measurements of Allakhverdiev et al [158,184]; at that, temperature dependence of the spontaneous polarization and pyroelectric coefficient did not reveal any anomaly at 79 K. Investigation of the origin of this transition has not yet been performed. The spontaneous polarization $P_s$ was shown to vanish at T = 42 K [184]; all changes in $P_s$ take place in the plane of the layers only. A pronounced maximum of the pyroelectric coefficient was observed at T ~ 38 K.

Banys et al [185] reported on a pinning effect on the microwave dielectric properties and the soft mode in TlInS$_2$ and TlGaSe$_2$ ferroelectrics. They call these crystals to be new proper ferroelectrics. They obtained that on cooling, TlInS$_2$ undergoes a second-order phase transition from a paraelectric monoclinic phase to an incommensurate phase at $T_i$ = 214 K and a first-order lock-in transition to a ferroelectric phase at $T_c$ = 202 K. The incommensurate structure was considered to be of type II. In these crystals there is a strongly over-damped soft ferroelectric mode, whose frequency is extremely low (drops to the millimeter wave region) in the vicinity of the phase transitions, T$_i$ and T$_c$, and causes dielectric microwave dispersion and high contribution to the static dielectric permittivity. At T$_i$ the soft mode splits into an acoustic-like phason and an optic-like amplitudon modes. The phason in the real crystals is pinned, strongly over-damped and is active in the dielectric spectra. It reveals itself as a relaxor. It was shown that in the less defective crystals the frequency of the phason is about $10^7$ Hz. Crystal imperfections result in the pinning of the phason and increase the gap in the phason spectrum. Even small concentrations (0.5 to 1%) of impurities increase the frequency of the phason to $10^9$ Hz. The low-frequency soft mode is responsible for the high value of the static dielectric permittivity at $T > T_i$ of these semiconductive ferroelectrics. The existence of a phason branch in the incommensurate phase and the shift of its wave vector to zero on heating increases the contribution of the phason to the static permittivity at $T_c$ so that ε($T_c$) > ε($T_i$). Because of the pinning effect, different impurities spread or narrow the incommensurate phase and express or suppress the anomalies of permittivity at $T_c$ and $T_i$.

Mikailov et al [186] found that maxima in the temperature dependence of the dielectric constants, indicating (to their opinion) two incommensurate and two commensurate phase transitions of TlInS$_2$ crystal, are shifted to lower and higher temperatures, respectively, under external bias electric field. The authors suggested a theoretical model of improper and proper ferroelectric phase transitions with incommensurate structure that includes existence of two order parameters and two polar sublattices in TlInS$_2$. Then they measured [187] time dependences of the dielectric constant in TlInS$_2$ (Figure 35) and obtained the presence of two characteristic relaxation parameters with different temperature behavior. This finding was interpreted as occurrence of a chaotic state accompanied by a coexistence of different commensurate ferroelectric structures in the temperature range from 194 to 200 K; at that, dielectric anomaly at 195 K was considered as a phase transition accompanied by the destruction of the improper ferroelectric polarization. Furthermore, along the experimental temperature dependent dielectric susceptibility measurements in the incommensurate phase of TlInS$_2$, Mikailov et al. [188] have outlined a theoretical approach based on the above two-sublattice model and on hypothesis of coexistence of improper and proper ferroelectricity, as well as the coexistence of type I and type II incommensurate substructures in the same compound. The calculated temperature dependence of the dielectric constants in the incommensurate phase of TlInS$_2$ showed an agreement between the theory and experiment. Gadjiev [189] developed a theory of a sequence of phase transitions of high-symmetry-incommensurate - commensurate phase controlled by competing the order parameters, and calculated the temperature dependence of dielectric constants, which were compared with the experimentally obtained data for the TlGaSe$_2$ and TlInS$_2$ crystals.

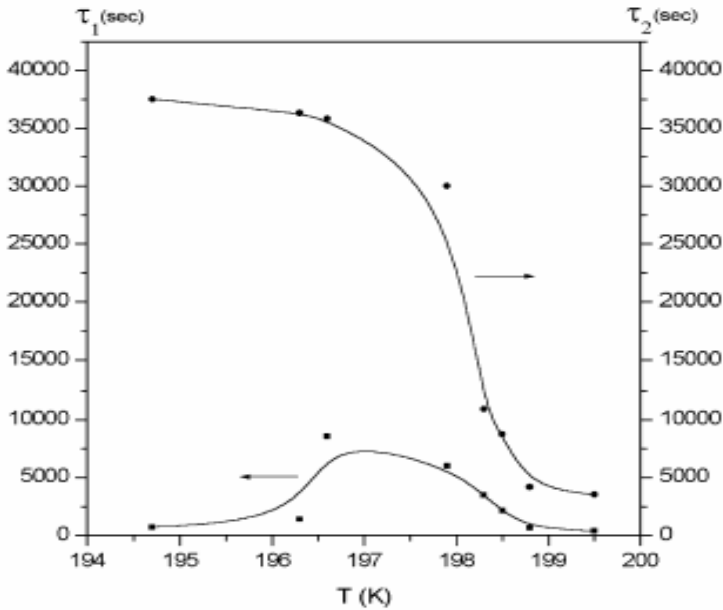

Figure 35. Temperature dependence of the relaxation times $\tau_1$ and $\tau_2$ of the coexisting phases in the proposed metastable chaotic state of TlInS$_2$. (From [187]).

Mikailov et al [190,191] reported on EPR studies of the TlInS$_2$ crystals doped with paramagnetic Cr$^{3+}$ and Fe$^{3+}$ ions. Remarkable temperature dependencies of the linewidth and resonance field of the main absorption peak in EPR spectra of Cr$^{3+}$ ion were attributed to the successive phase transitions between 200 and 220 K. Considerable changes in the EPR spectra (strong line broadening, splitting, and appearance of additional peaks) were also observed below 80 K, i.e., at the low-temperature phase transition. These transformations seem to be caused by tetragonal distortions of the InS$_4$ tetrahedra. An iron-doped TlInS$_2$ single crystal showed the fine structure of the EPR spectra of the paramagnetic Fe$^{3+}$ ions. The spectra were interpreted as the transitions among the spin multiplets (S=5/2, L=0) of the Fe$^{3+}$ ion, which is split in the local crystal field of orthorhombic symmetry. Experimental results indicate that iron ions substitute indium ions at the center of the InS$_4$ tetrahedrons, and the rhombic distortion of the crystal field is caused by Tl ions located in the trigonal cavities between the tetrahedral complexes.

Generation of the optical second harmonic in the vicinity of phase transitions in TlInS$_2$ has been reported by Gorelik et al [192], Ibragimov et al [193-195] and Allakhverdiev et al [196]. Gadjiev [197] has made calculations of the linear and nonlinear optical properties of soliton regime in the incommensurate phase of TlInS$_2$. The transmission coefficient in the modulated incommensurate phase was shown to be an oscillation function of temperature. It was shown that the intensity of the second harmonic in the incommensurate phase increases with decreasing temperature. Temperature dependent polarized transmission intensity study in strongly converging and parallel light beams has been made for off-zone-center incommensurate semiconductors-ferroelectrics TlInS$_2$, TlGaSe$_2$ and TlGaS$_2$ by Mamedov et al [198]. Temperature dispersion of the dielectric axes of TlInS$_2$ was shown to include forbidden rotation that might be associated with a lower than monoclinic symmetry or spatially dispersed microscopic domains. Unusual effect with temperature behavior encountered in soliton-like incommensurate phases was observed for TlInS$_2$. Wavelength dispersion of the largest partial birefringence and ferroelectric structural phase transition through an incommensurate phase were visualized in acute bisectrix light figures of biaxial TlInS$_2$, TlGaSe$_2$ and TlGaS$_2$ ternary compounds [199].

Table 18. Phase transition temperatures of TlInS$_2$ determined by different authors.

| Phase transition temperatures, K | Method of investigation | References |
|---|---|---|
| 189 and 213 | dielectric constant/losses | 169 |
| 170, 195-202, 216-220 | Neutron diffraction, thermal expansion | 170 |
| 204, 216 | dielectric constants, spontaneous polarization, birefringence | 171 |
| 200 | thermal expansion | 172 |
| 173.4, 196.9, and 214.9 | Heat capacity | 155, 176 |
| 189, 195 and 213 | Brillouin scattering | 202 |
| 189 and 213 | Brillouin scattering | 203 |
| 192-198, 200-202, 203.6-206.5, 206-209, 214 | Dielectric constant, spontaneous polarization, a.c. electrical conductivity | 204 |
| 170, 194 and 214 | x-ray diffraction | 205 |
| ~ 80 K | dielectric permittivity | 158 |
| 201 and 204 | dielectric permittivity | 181 |
| 202, 221 | electrical conductivity | 29 |
| 79, 195, 201, 204, 206 | Dielectric susceptibility | 184 |
| 189, 220 | electrical conductivity | 207 |
| 194, 214<br>197, 200, 209 | x-ray diffraction,<br>dielectric constant | 3<br>3 |
| 195, 214 | x-ray scattering | 173 |
| 202, 214 | microwave dielectric measurements | 185 |
| 79, 201, 204, 206 | Second harmonic generation | 196 |
| 195, 201 | Dielectric constants | 187 |
| 79, 204, 216 | ESR | 190 |
| 156, 166, 173, 192, 202, 207, 216, 222, 227.5, 244, 253, 258.5 | Specific heat (single crystal) | 175 |
| 193, 202, 208, 220, 226, 232, 253.5, 263.5 | Specific heat (powder sample) | 175 |

Summarizing, we see that despite some discrepancies in the number of the observed anomalies and in their positions in the temperature scale, the majority of the investigators agree that there are four anomalies in the temperature dependence of the various physical properties of $TlInS_2$ near the temperatures of 79 K, 170, 194-197 K and 214 K. The intermediate phase between two latter transitions is incommensurate. Some data indicate splitting of one of the transitions to two closely set transitions around 200 K. Two-sublattice model of the incommensurate phase is discussed. The phase transition temperatures reported by different authors are collected in Table 18.

In addition, we would like to mention very interesting recent works of Sardarly et al [209-216] who doped $TlInS_2$ crystals with ~ 0.1 at. % Cr, Mn, Yb, Sm, Bi, Fe, Ge or La. The authors have shown that even slight deviations from stoichiometry have a significant effect on the dielectric properties of the ferroelectric $TlInS_2$ crystals. The most interesting effect [209-214] is that the layered semiconductor ferroelectric $TlInS_2$, doped with Fe, Mn, Cr and Ge, exhibits a pronounced relaxor behavior that is accompanied by the occurrence of polar nano-regions (PNRs), or nanodomains. For example, doping of the $TlInS_2$ crystals with Ge yields change in the temperature dependence of dielectric susceptibility: instead of the peak corresponding to the phase transition at 196 K, two broad peaks occur; at that, the low temperature peak at 142 K corresponds to the transition from the relaxor (nano-domain) state to the ferroelectric state. Similar relaxor behavior has also been observed in the γ-irradiated $TlInS_2$ crystals [215,216].

In contrast to the "classic" relaxors, the layered relaxors are characterized by dispersion-free optical mode in the (001) direction and occurrence of the incommensurate state that seems to coincide with the relaxor state. In the incommensurate phase, the Brillouin zone is determined by the superlattice period. This state is characterized by a multifold splitting of the energy bands, which results in a line-like spectrum of the density of states. Besides, the doping and irradiation result in the appearance of the defects of the size of 10 nm [215]. It was shown that these nano-regions exhibit a behavior similar to that of quantum dots, characterized by the energy spectrum quantization. Furthermore, the doping atoms form some capture levels in the forbidden gap of the semiconductor thus the conductivity is driven by carrier tunneling through the potential barriers [215,216].

### *7.7. Ferroelectric phase transition and incommensurability in the layered crystal $TlGaSe_2$*

In 1983, Volkov et al [217,218] reported on sub-millimeter dielectric spectroscopy measurements of $TlGaSe_2$, which showed that this compound exhibits successive phase transitions at ~107 and ~120 K with an intermediate phase that was assumed to be incommensurate. Dielectric measurements [219] revealed ferroelectric character of the low-temperature phase and paraelectric character of the high-temperature phase. Volkov et al [217] showed that $TlGaSe_2$ exhibits soft mode behavior (Figure 36, left panel) that is typical of ferroelectric crystals with incommensurate phases.

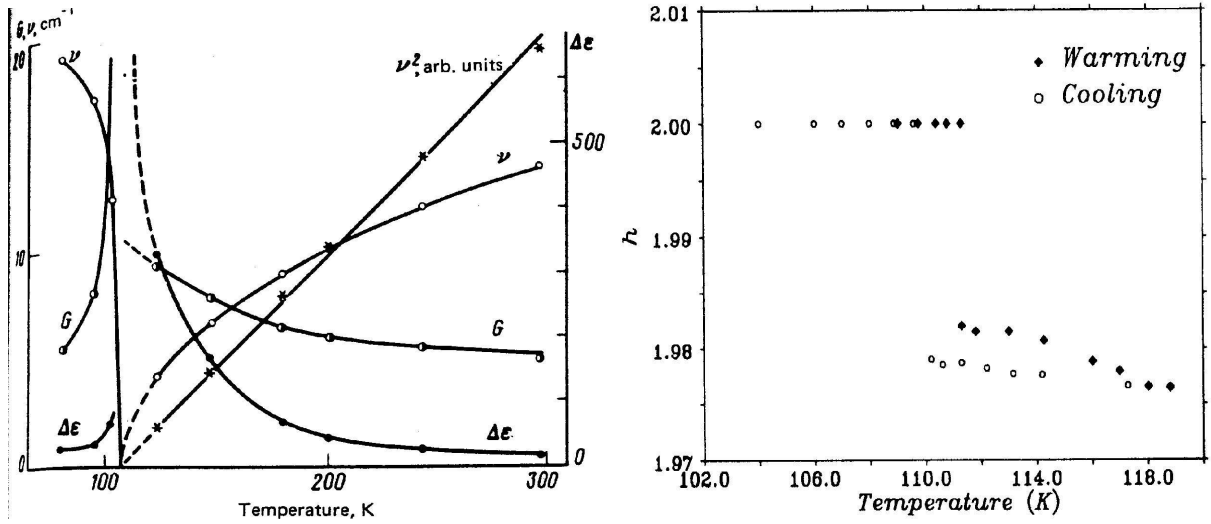

Figure 36. Left panel: Frequency ν, damping G and dielectric contribution Δε of the soft mode (From [217]). Right panel: Temperature dependence of the position of the incommensurate satellite near the position (1.978, 1.978, 12.25) along the [$h, h$, 12.25] direction (From [221]).

After Volkov et al [217,218], the transitions and incommensurate states in TlGaSe$_2$ have extensively been studied [219-247] using a number of different techniques. Transition temperatures $T_c$ and $T_i$, reported by most of the authors, were found between 107-110 K ($T_c$) and 117-120 K ($T_i$), respectively. The phase transition at $T_c$ has been reported to be of the first order, while that at $T_i$ is of the second order [219,221]. Detailed study of the low-temperature commensurate phase has not yet been performed.

X-ray diffraction measurements in a single crystal of TlGaSe$_2$ [221] showed that the phase between 117 and 110 K is incommensurate and characterized by a modulation wave vector (δ, δ, 1/4), where δ ≈ 0.02 in reciprocal lattice units. Reduction of the temperature from 117 to 110 K yields some decrease in δ, until it jumps discontinuously to zero at $T_c$ =110 K to produce a commensurate phase (Figure 36, right panel). On heating from below 110 K, the structure remains commensurate up to 111.3 K, showing a hysteresis in the transition temperature. As the scattering from the incommensurate modulation was very weak, McMorrow et al [221] were not able to comment on the structural distortion giving rise to the incommensurate peaks. The scattering from the low temperature commensurate ferroelectric phase indicates a quadrupling of the unit cell along the $c$ axis compared to that of the high temperature phase. This phase was assigned to the space group $Cc$ [219]. Recent single crystal neutron scattering investigation of TlGaSe$_2$ by Kashida et al. [222] showed the existence of an incommensurate state between 107 and 118 K with a modulation wave vector (δ, 0, 1/4), where δ = 0.04. In the low temperature phase, the satellite reflections appear at the commensurate position with $q_c$ = (0, 0, ±0.25). The authors concluded [222] that the incommensurate structure in TlGaSe$_2$ maybe ascribed to the type of displacements whose wave vector is parallel to the monoclinic $c$ axis. The direction of the modulation throughout the incommensurate phase was found to be almost temperature independent. The lack of the center of symmetry of the diffraction spots indicates that the modulation is not represented by a simple standing wave type displacement 2 cos(kz) = exp(ikz) + exp(-ikz). The authors suggested that the modulation includes a local deformation of the unit cells. The structure of discommensurations remained unknown.

In addition to the aforementioned phase transitions, some authors have also reported on anomalies attributed to the phase transitions around 101-103 K [172,223-226], 200-215 K [227] and 240-250 K [223-225, 228-235]. Dielectric measurements of Allakhverdiev et al. [158] and Mikhailov et al [235] allowed to observe the low-temperature phase transition at T = 65 K. The phase transition temperatures obtained by different authors are collected in Table 19.

Table 19. Phase transition temperatures of TlGaSe$_2$ determined by different authors

| Phase transition temperature, K | Method of investigation | Reference |
|---|---|---|
| 107 and 120 | Sub-millimeter dielectric spectroscopy | 217,218 |
| 108.9 and 118.4 | heat capacity | 220 |
| 110 and 117 | x-ray scattering | 221 |
| 110 and 120 | Dielectric, heat capacity, IR reflectivity, and XRD measurements | 219 |
| 107 and 118 | Neutron scattering | 222 |
| 65 | dielectric permittivity | 158 |
| 107-120.5, 252.5 | heat capacity, NMR | 228 |
| 108 and 118 | NMR | 236, 237 |
| 102, 111, and 120 | thermal expansion | 172 |
| 101 and 246 | optical absorption, heat capacity | 223 |
| 101, 106, 109, 113, 117, 253, 340 | heat capacity | 224 |
| 194 and 247.5 | transmission oscillations | 229 |
| 105, 117, and 200-215 | Photoconductivity, optical absorption | 227 |
| 112, 124, 132, and 248<br>110, 120, 130, and 245 | Thermal expansion, $c$(T)<br>Thermal expansion, $a$(T) | 230 |
| 247.5 | NMR | 231 |
| 250 | luminescence | 232 |
| 95-107, 122, 240-250 | Dielectric measurements | 233 |
| 250 | luminescence | 234 |
| 103, 110, 119, and 246 | Acoustic emission | 225 |
| 107, 114 | thermal expansion | 208 |
| 65, 108, 115, 242 | Dielectric measurements | 235 |
| 103, 110 | Dielectric measurements | 226 |
| 108, 118, ~220 (?) | NMR | 238 |

Recent $^{69,71}$Ga and $^{205}$Tl NMR measurements of powder and single crystalline samples of TlGaSe$_2$ [236-238] detected phase transitions at 108 and 118 K and the incommensurate state [238] between these temperatures. Thallium spectra were found to be more complicated than expected from the structural data. When magnetic field $B_0$ is applied along the $c$ axis, $^{205}$Tl NMR spectra [238] show two lines coming from the two inequivalent sites in the structure of TlGaSe$_2$. However, at $B_0 \perp c$ three lines are observed above ~220 K (Figure 37). They are evidently caused by different perpendicular components $\sigma_\perp$ of the chemical shielding tensor of thallium nuclei and reflect different electronic surrounding of the nuclei. Usually such distinct lines come from the inequivalent atomic sites [239]; however, in order to reconcile the diffraction and NMR data, one can suggest that some structurally equivalent Tl nuclei in the unit cell exhibit different orientations of the principal axes of the chemical shielding tensor in the $a,b$ plane. The spectra transformation in the temperature range from 198 to 290 K (Figure 37, left panel) does not seem to be characteristic of a phase transition, thus a suggestion about phase transition around 220K is questionable. At that, we note that Abutalybov et al [227] reported on eventual phase transition around 200-215 K in TlGaSe$_2$.

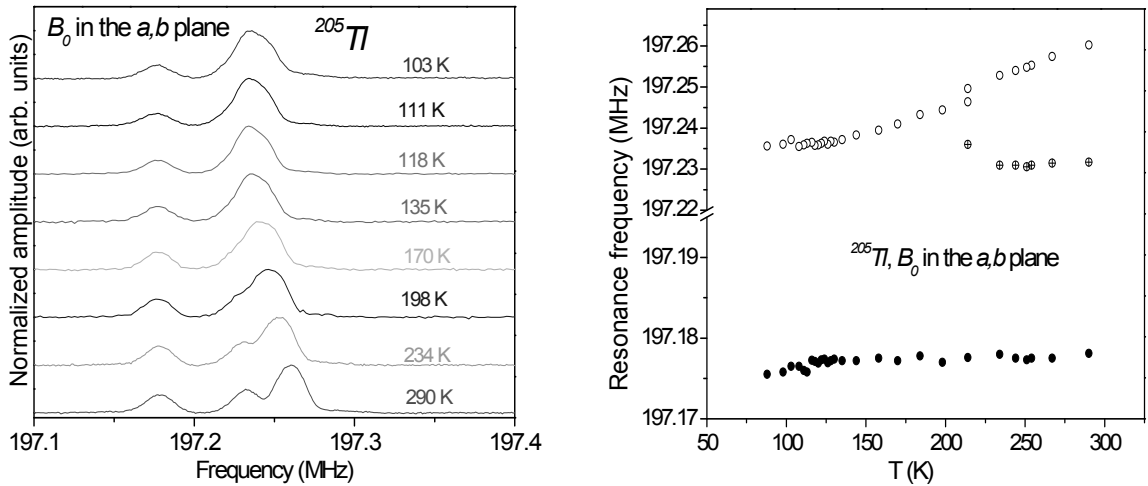

Figure 37. Left panel: $^{205}$Tl spectra of TlGaSe$_2$ single crystal at different temperatures for $B_0 \perp c$. Right panel: dependence of the $^{205}$Tl resonance frequencies in TlGaSe$_2$ single crystal for $B_0 \perp c$ (From [238]).

Exciton spectroscopy and dielectric measurements of TlGaSe$_2$ made by Alekperov [240] showed that the incommensurate phase region may be considered as a coexistence of two spatially dispersed media with different dielectric constants and different behavior. Mikailov et al. [188] reported a coexistence of type I and type II incommensurate structures in TlInS$_2$, which is isostructural to TlGaSe$_2$. Then, Mikailov et al [226] interpreted their dielectric data on TlGaSe$_2$ suggesting a coexistence of two polar sublattices. Senturk et al [241] have measured the frequency and time dependence of the $ac$ conductivity within incommensurate phase of TlGaSe$_2$ and found out that its time dependence exhibits two different conductivity relaxation times above 120 K. The authors assigned these two relaxation times to two different incommensurate orderings in the temperature range 120 - 242 K. Dielectric measurements [235,241,242] reveal analogous behavior, showing two relaxation mechanisms associated with two relaxation sites, also assigned to the occurrence of two different types of the incommensurate phases in the range 115 – 242 K. At that, the authors [235,241, 242] ignore the well-known fact that incommensurate phases in the temperature ranges of 120-242 K and 160-242 K have never been observed. In our opinion, the authors [235,241, 242] did not obtain solid proves of the occurrence of the incommensurate state in these regions, and their assignment seems to be ambiguous.

Memory effects in thallium gallium selenide have been obtained by Gololobov et al [225], Senturk et al [242,243], Mikailov et al [244,245], Babaev et al [246] and Aliyev et al [247]. The latter author investigated non-equilibrium properties of the incommensurate phase and observed an influence of the pre-history of the heat treating of the crystal, i.e. annealing at a fixed temperature in the region of incommensurate phase, on the dielectric properties in the vicinity of the incommensurate-commensurate phase transition. Memory effects were shown to cause a change in the temperature range of the existence of the incommensurate phase in $TlGaSe_2$.

### 7.8. Possible phase transitions in $TlGaS_2$

Information about phase transitions in $TlGaS_2$ [171,205,220,248-254] (Table 20) is inconsistent. Specific heat measurements of Krupnikov et al. [248] showed six anomalies in the range of 73 – 187 K. At that, doping of the $TlGaS_2$ single crystal with $Nd_2S_3$ ($Nd^{3+}$ ions are suggested to occupy the $Tl^{1+}$ vacancies) suppresses all phase transitions except for the first around 75 K. It was suggested that the above phase transition sequence in pure $TlGaS_2$ is due to the crystal defects. Aydinli et al. [249] observed the anomalies in the temperature dependences of low and high frequency phonon modes at ~ 180 and 250 K that were attributed to phase transitions. However, these transitions are not well pronounced in a narrow temperature range as it is seen for TlGaSe$_2$ and TlInS$_2$. It was supposed that the phase transitions are caused by the deformation of GaS$_4$ tetrahedra rather than by slippage of Tl atom channels in [110] and [1$\bar{1}$0] directions. These transitions do not seem to be ferroelectric; or, if ferroelectricity in the low temperature phases of TlGaS$_2$ exists, it is much weaker in comparison with TlGaSe$_2$ and TlInS$_2$. No evidence of the soft mode behavior in TlGaS$_2$ has been found. Mal'sagov et al. [205], Ates et al. [250] and Dzhafarova et al. [251] also observed some anomalies in their XRD, optical absorption and Raman scattering measurements, which were attributed to phase transitions in TlGaS$_2$. However, specific heat study by Abdullaeva [220] et al. showed that the temperature dependence of $C_p(T)$ in TlGaS$_2$ crystals reveals a monotonically increased function on heating that has no abrupt anomalies in the range 4.2 K - 300 K. Aliev et al. [171], who made dielectric and optical measurements, and Durnev et al. [252] who measured Raman and IR spectra of TlGaS$_2$, also did not report on phase transitions in this crystal. Raman study of TlGaS$_2$ by Gorban et al. [253] revealed absence of phase transitions in the temperature range 1.8 - 300 K. The same conclusion results from the transport measurements of Kashida [254]. Therefore we conclude that the data on phase transitions in TlGaS$_2$ are very conflicting. In order to ascertain whether phase transitions occur in this crystal, further investigations are required.

Table 20. Phase transition temperatures (K) of $TlGaS_2$ determined by different authors.

| Crystal | Phase transition temperatures | Method of investigation | References |
|---|---|---|---|
| $TlGaS_2$ | 73.5, 91, 101, 114, 133.5, 187 | Specific heat | [248] |
| $TlGaS_2$ | 75 | Specific heat ($Nd_2S_3$ doped) | [248] |
| $TlGaS_2$ | 180 and 250 | Polarized Raman scattering | [249] |
| $TlGaS_2$ | 120, 180, 220 and 280 | Optical absorption | [250] |
| $TlGaS_2$ | 90 | Raman scattering | [251] |
| $TlGaS_2$ | 121 | x-ray diffraction | [205] |
| $TlGaS_2$ | no phase transitions | Specific heat | [220] |
| $TlGaS_2$ | no phase transitions | Dielectric and optical measurements | [171] |
| $TlGaS_2$ | no phase transitions | Raman and IR spectroscopy | [252] |
| $TlGaS_2$ | no phase transitions | Raman spectra | [253] |

| TlGaS$_2$ | no phase transitions | Electrical mearuments | [254] |

### *7.9. Phase transitions and soft modes in mixed ternary crystals*

Some mixed crystals (solid solutions) also show soft mode behavior and undergo structural phase transitions [255-259]. For example, dielectric spectroscopy measurements at 5-30 cm$^{-1}$ by Volkov et al [255] showed that the vibrational spectra of the TlGaSe$_2$-TlGaS$_2$ mixed crystals exhibit low-frequency lattice excitations with indications of soft modes. The behavior was studied as a function of temperature and percent concentration of S and Se. Aldzhanov et al [256] measured the heat capacities of mixed crystals TlGaS$_2$-TlGaSe$_2$ and TlInS$_2$-TlInSe$_2$ in the temperature range 4 - 350 K and reported that for TlGaSe$_{1.8}$S$_{0.2}$ the anomaly occurs at unexpected temperature ~ 154 K. Duman et al [257] investigated optical absorption in layered semiconductor single crystals of TlGaSe$_{2(1-x)}$S$_{2x}$ and suggested that abrupt changes in the energy spectrum, observed in the temperature intervals 90 - 100, 100, 100 - 120, 160 - 180, 220 - 240, and 240 - 250 K, may be attributed to the phase transitions. Tekhanovich et al [258] measured specific heat of TlInS$_{1.8}$Se$_{0.2}$ in the temperature range 60-300 K (Figure 38) and observed a clear maximum at $T_c$ = 190 K, corresponding to the phase transition between incommensurate and low-temperature commensurate phases. In the vicinity of the $T_i$ point, specific heat displayed a scattering of the $C_p$ values obtained in different series of the measurements, thus $T_i$ was roughly estimated as ~ 213 K. The authors came to a conclusion that the replacement of the light sulfur atoms by heavier selenium atoms yields a shift of the phase transition temperatures toward lower temperatures by about 5 K. Surprisingly, that Golubev et al [259], who studied vibrational spectra of the TlInSe$_{2(1-x)}$S$_{2x}$ mixed single crystals, found no indication of a phase transition at 4.2 - 600 K.

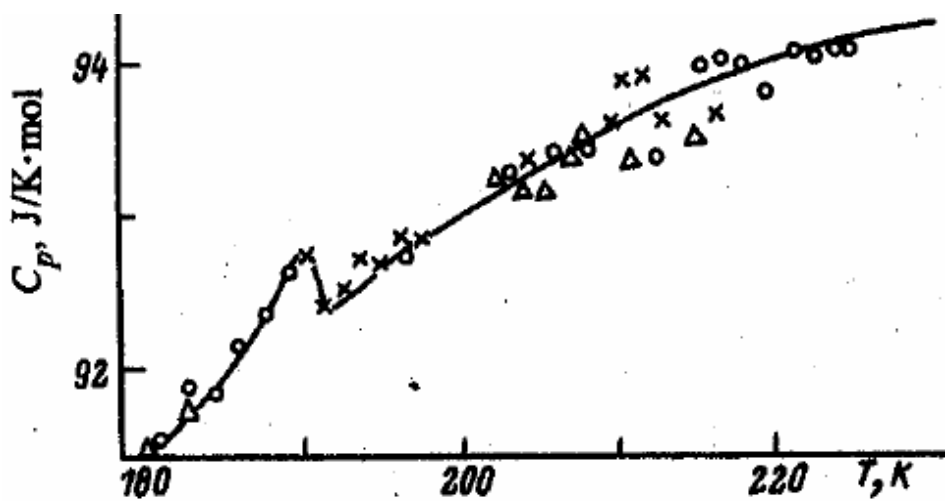

Figure 38. Temperature dependence of the specific heat of TlInS$_{1.8}$Se$_{0.2}$. (From [258]).

*7.10. Electronic origin of phase transitions in the layered crystals TlGaSe$_2$, TlInS$_2$ and TlS*

As it was noticed above, the experimental data reveal that phase transition into the feroelectric state in the entitled crystals is associated with the displacement of thallium atoms in the *a,b* plane. In this paragraph, we show that this effect is caused by the specific electronic structure of the above-mentioned layered crystals and discuss the electronic factors of the phase transitions in these compounds. Hochheimer et al [219] and Yee and Albright [117] were the first who have drawn attention on this problem investigating the bonding and structure of thallium gallium diselenide. Yee and Albright [117] found that bonding between Tl and Se is reasonably covalent and the region around the Fermi level consists primarily of Tl 6*s* states antibonding to Se lone pairs. Two structural deformations have been discussed for the ferroelectricity associated with TlGaSe$_2$. One hypothesis involves a D$_{2d}$ squashing motion of the Ga$_2$Se$_4$ unit. (We note that such deformation mechanism was also proposed by Burlakov et al [260], who discussed a substantial deformation of the Ga$_4$Se$_{10}$ polyhedron in the low-temperature phase, which could cause a spontaneous polarization). However, Yee and Albright [117] have found no evidence for a double well potential in such a model, and the energy required for deformation was found to be quite steep. On the other hand, a soft, double well potential exists for the Tl atoms to slide away from a trigonal prismatic to (3 + 3) environment [117]. This in turn reinforces Tl$^+$ - Tl$^+$ interactions in TlGaSe$_2$, which are turned into a net bonding situation by Tl *p* mixing into filled Tl *s* orbitals. Many years ago Orgel [101] predicted that the distortion around a Tl$^+$ environment would cause Tl 6*p* orbitals to mix into the filled Tl 6*s* levels. The formation of the directed lone pairs at Tl$^+$ centers then creates a ferroelectric material. Drawing from this theory, Hochheimer et al [219] have proposed that the ferroelectricity in TlGaSe$_2$ is caused by the stereo-chemically active electron lone pair configuration of the Tl$^+$ ion, which is therefore the heart of the displacive transition. They suggested that phase transition is accompanied by shifts of the Tl atoms parallel to the [110] and [1$\bar{1}$0] directions. With these shifts, the Tl atoms move in the trigonal prisms and change their coordination from CN = 6 into CN = 3 + 3, which allows the formation of stereo-chemically active electron lone pair configuration at the Tl$^+$ ions. The powder x-ray investigations [219] indicate that that the inversion centers are lost at the transition, which is absolutely necessary for the occurrence of ferroelectricity. The authors [219] suggest that the electron lone pair of the Tl$^+$ cation becomes stereo-chemically active at the transition temperature, which leads to asymmetric shifts of the Tl$^+$ cations in the (001) plane, thus destroying the inversion center and making the ferroelectric transition in TlGaSe$_2$ to be of the displacive type. To our knowledge, TlGaSe$_2$ provides the first example for ferroelectricity caused by the aforementioned electronic mechanism predicted by Orgel [101].

Yee and Albright [117] estimated a value of ~0.8 Å for Tl atom slippage along the channels. They pointed out that the mixing of virtual Tl $p$ states into orbitals around Fermi level is absolutely critical for this process. The intermixing provides the driving force to create the double-well potential. Hybridized lone pairs are created that point in the opposite direction to Tl slippage. A unique consequence is that Tl-Tl bonding in their model is increased although the Tl-Tl distance remains constant. According to the above calculations, slippage of Tl atoms in the channels parallel to the $a$ direction while the Tl atoms remained fixed in the channels parallel to the $b$ direction, and *vice versa*, results in an energy lowering of half (within 0.04 kcal/mol per formula unit) of the total energy. Thus, slippage in either direction is predicted to be not strongly correlated. Incommensurate phases may be formed via non-correlated Tl slippages. This may explain the existence of oscillations of the heat capacity versus temperature around the phase transition region [228], a devil's ladder, which can be interpreted to arise from the intervention of incommensurate phases.

No doubts that the ferroelectric phase transitions in the isostructural TlInS$_2$ and TlGaSe$_2$ are driven by the same mechanism. Kashida and Kobayashi [3] ascribed the origin of the structural phase transitions in TlInS$_2$ to small displacements of atoms from the positions that they occupy in the high-temperature high-symmetry phase. Detailed analysis showed that the displacements are parallel to the $a$-axis. Analogous model was suggested by Kashida et al [11,113,165] to explain the phase transitions in the layered monoclinic TlS compound. In that case, the data analysis showed that at the phase transition temperature the Tl$^+$ ions and the apical S ions move reversibly normal to the channel direction (i.e., differently from TlInS$_2$ and TlInS$_2$ where the Tl$^+$ ions move along the channel). For monoclinic TlS, the microscopic mechanism of the structural phase transition was also argued in terms of the calculated electronic band structure [106]. As the microscopic origin of ferroelectricity, stereochemical instability of the Tl 6$s^2$ lone pair electrons [219,117] was also suggested, that is, a distortion around a Tl$^{1+}$ ion would cause Tl 6$p$ orbitals to mix into the filled Tl 6$s$ orbital, and thus stabilizes the polar structure by extending the band gap. Thus the ferroelectric phase transition in monoclinic TlS is of electronic origin. XRD study of monoclinic TlS [13] showed that, in the low temperature phase, actually the Tl$^{1+}$ ion and the apical S ions move in the opposite directions normal to the Tl$^{1+}$ channels.

## 8. Conclusion

The reader can conclude that intensive ongoing investigations of the TlX and TlMX$_2$ crystals resulted in a satisfactory understanding of a number of properties of the above compounds, such as electronic structure, transport properties, and mechanism of the phase transitions, though the results of different studies sometimes disagree and require further accurate measurements, such as parallel XRD and transport measurements in studying the pressure-induced phase transitions. The structural details of the incommensurate and low temperature commensurate phases are still not well understood. Next, little is known about the relationship between the electronic properties and the structural features, and some macroscopic properties. In this regard, it is of interest to determine whether the observation of the non-linear effects in the *I-V* characteristics, as well as the occurrence of a metastable chaotic state, are accompanied by the coexistence of different phases and can be understood in terms of the structural and electronic features of these materials.

Recent discovery of the relaxor behavior of the doped and irradiated layered TlMX$_2$ crystals shows considerable promise for both science and applications. Next promising direction is preparation and investigation of nanoparticles of binary (A$^{III}$B$^{III}$) and ternary A$^{III}$B$^{III}$C$_2^{VI}$ chalcogenides [104,261]. We also note that thallium compounds are very useful in studying of some interesting physical effects, such as nuclear spin diffusion, at the atomic level [262].

## 9. Acknowledgements

I thank Prof. S. Kashida for helpful discussions on the structure and phase transitions in the TlX and TlMX$_2$ compounds and for providing me with the band structure calculations of TlSe prior publication. I am grateful to the American Physical Society, the Physical Society of Japan, Elsevier Publishing, Wiley-VCH Verlag GmbH, IOP Publishing and Taylor&Francis Publishers who granted me permission to reproduce the figures in this paper. Figure 14 is cited by courtesy of S. Kashida.

**10. References**
1. Müller D, Poltmann F E and Hahn H 1974 *Z. Naturforsch*. **29b** 117-118
2. Müller D and Hahn H 1978 *Z. Anorg. Allg. Chem*. **438** 258-272
3. Kashida S and Kobayashi Y 1999 *J. Phys. Condens. Matter* **11** 1027
4. Delgado GE, Mora AJ, Perez FV and Gonzalez J 2007 *Physica B* **391** 385
5. Gasanly NM, Marvin BN, Sterin KE, Tagirov VI and Khalafov ZD 1978 *Phys. Status Solidi B* **86** K49
6. Henkel W, Hochheimer HD, Carlone C, Werner A, Yes S and v. Schnering HG 1982 *Phys. Rev.* B **26** 3211
7. Ketelaar JAA, Hart WH, Moerel M and Polder D 1939 *Z. Kristallogr*. **101** 396
8. Hahn H and Klingler W 1949 *Z. Anorg. Chem*. **260** (1949) 110
9. Müller D, Eulenberger G and Hahn H 1973 *Z. Anorg. Allg. Chem*. **398** 207
10. Bradtmöller S, Kremer RK and Böttcher P 1994 *Z. anorg. allg.Chem*. **620** 1073
11. Kashida S, Nakamura K and Katayama S 1992 *Solid State Commun*. **82** 127
12. Kashida S and Nakamura K 1994 *J. Solid State Chem*. **110** 264
13. Nakamura K and Kashida S 1993 *J. Phys. Soc. Japan* **62** 3135
14. Shannon RD 1976 *Acta Cryst*. **A 32** 751
15. Schubert K, Dore E and Kluge M 1955 *Z. Metallkunde* **46** 216
16. Chattopadhyay T, Santandrea RP and von Schnering HG 1985 *J. Solid State Chem.* **46** 351
17. Toure AA, Kra G, Eholie R, Olivier-Fourcade J and Jumas JC 1990 *J. Solid State Chem*. **87** 229
18. Stöwe K 2000 *J. Solid State Chem*. **149** 123
19. Gasanly NM, Ozkan H, and Culfaz A 1993 *Phys. Stat. Sol.* (a) **140** K1.
20. Gasanly NM 2006 *J. Korean Phys. Soc.* **48** 914.
21. Guseinov GD, Abdinbekov SS, Godzhaev MM and Agamaliev DG. 1988 *Izv. Akad. Nauk SSSR, Neorg. Mater*. **24** 144
22. Aliyev RA, Guseinov GD, Najafov AI and Aliyeva MK 1985 *Bull. Soc. Chim. France* **2** 142
23. Bakhyshov AE and Akhmedov AM 1979 *Izv. Akad. Nauk SSSR, Neorg. Mater.* **15** 417
24. Babanly MB and Kuliev AA 1977 *Azerbaidzhanskii Khimicheskii Zhurnal* **4** 110
25. Bidizinova SM, Guseinov GD, Guseinov GG and Zargarova MI 1973 *Azerbaidzhanskii Khimicheskii Zh.* **2** 133.
26. Vinogradov EA, Gasanly NM, Goncharov AF, Dzhavadov BM and Tagirov VI 1980 *Sov. Phys. Solid State* **22** 526 (1980 *Fiz. Tverd. Tela (Leningrad)* **22** 899)
27. Hatzisymeon KG, Kokkou SC, Anagnostopoulos AN and Rentzeperis PI 1998 *Acta Cryst*. **B 54** 358
28. Mustafaeva SN, Aliev VA and Asadov MM 1998 *Phys. Solid State* **40** 41 (1998 *Fiz. Tverd. Tela (St. Petersburg)*, **40**, 48).
29. Aliev VA, Bagirzade EF, Gasanov NZ and Guseinov GD 1987 *Phys. Status Solidi A* **102** K109
30. Mustafaeva SN, Aliev VA and Asadov MM 1998 *Phys. Solid State* **40** 561 (1998 *Fiz. Tverd. Tela (St. Petersburg)*, **40**, 612).
31. Hanias M, Anagnostopoulos AN, Kambas K and Spyridelis J 1992 *Mater. Res. Bull*. **27** 25
32. Kashida S, Saito T, Mori M, Tezuka Y and Shin S 1997 *J. Phys: Condens. Matter* **9** 10271
33. Katayama S, Kashida S and Hori T 1993 *Jpn. J. Appl. Phys*. 32, Suppl. 32-3, 639
34. Bakhyshov AD, Musaeva LG, Lebedev AA and Jakobson MA 1975 *Sov. Phys. Semicond*. **9** 1021 (1975 *Fiz. Tekh. Poluprovodn*. **9** 1548).
35. Allakhverdiev KR, Sardarly RM, Wondre F and Ryan JF 1978 *Phys. Status Solidi B* **88** K5
36. Allakhverdiev KR, Mammadov TG, Suleymanov RA and.Gasanov NZ 2003 *J. Phys.: Condens. Matter* **15** 1291
37. Abay B, Guder HS, Efeoglu H and Yogurtcu YK 2001 *Phys. stat. sol. (b)* **227** 469
38. Abutalybov GI, Aliev AA, Larionkina LS, Nelman-zade IK and Salaev EY 1984 *Soviet Phys. Solid State*, **26** 846
39. Hanias M, Anagnostopoulos AN, Kambas K and Spyridelis J 1989 *Physica B* **160** 154
40. Nagat AT, Gamal GA, Gameel YH and Mohamed NM 1990 *Phys. Status Solidi A* **119** K47
41. Karpovich IA, Chervova AA and Demidova LI 1974 *Izvestiya Akademii Nauk SSSR, Neorg. Mater.* **10** 2216
42. Bakhyshov AE, Lebedev AA, Khalafov ZD and Yakobson MA 1978 *Soviet Phys. Semocond*. **12** 320 (1978 *Fizika Tekhnika Poluprovodnikov* **12**, 580).
43. Godzhaev MM, Guseinov GD, Abdinbekov SS, Alieva MK and Godzhaev VM 1986 *Mater. Chem. Phys.* **4** 443
44. Godzhaev MM, Guseinov GD and Kerimova EM *Izv. Akad. Nauk SSSR, Neorg. Mater.* **23** 2087
45. Mooser E and Pearson WB 1956 *Phys.Rev*. **101** 492
46. Itoga RS and Kannewurf CR 1971 *J. Phys. Chem. Solids* **32** 1099
47. Allakhverdiev KR, Gasymov SG, Mamedov TG, Salaev EY, Efendieva IK 1982 *Phys. Status Solidi B* **113** K127
48. Pickar PB and Tiller HD 1968 *Phys. Status Solidi* **29** 153
49. Nayar PS, Verma JKD and Nag BD 1967 *J. Phys. Soc. Jpn.* **23** 144
50. Guseinov GD and Akhundov GA 1965 *Doklady Akademii Nauk Azerbaidzhanskoi SSR* **21** 8
51. Hussein SA and Nagat AT 1989 *Crystal Research and Technology* **24** 283
52. Abdullaev NA, Nizametdinova MA, Sardarly AD and Suleymanov RA 1992 *J. Phys. Condens. Matter* **4** 10361

53. Allakhverdiev KR, Gasymov SG, Mamedov T, Nizametdinova MA and Salaev EY 1983 *Sov. Phys. Semicond.* **17** 131 (1983 *Fizika Tekhnika Poluprovodnikov* **17**, 203).
54. Rabinal MK, Asokan S, Godazaev MO, Mamedov NT and Gopal ESR 1991 *Phys. Status Solidi B* **167** K97
55. Nagat AT 1989 *J. Phys.: Condensed Matter* **1** 7921
56. Rabinal MK, Titus SSK, Asokan S, Gopal ESR, Godzaev MO and Mamedov NT 1993 *Phys. Status Solidi B* **178** 403
57. Godzhaev EM, Zarbaliev MM, Aliev SA 1983 *Izv. Akad. Nauk SSSR, Neorg. Mater*. **19** 374
58. Guseinov GD, Mooser E, Kerimova EM, Gamidov RS, Alekseev IV and Ismailov MZ 1969 *Phys. Status Solidi* **34** 33
59. Nagat AT, Gamal GA and Hussein SA 1991 *Crystal Research & Technology* **26** 19.
60. Allakhverdiev KR, Bakhyshov NA, Guseinov SS, Mamedov TG, Nizametdinova MA and Efendieva IK 1988 *Phys. Status Solidi B* **147** K99
61. Hanias MP and Anagnostopoulos AN 1993 *Phys. Rev. B* **47** 4261
62. Hanias MP, Anagnostopoulos AN, Kambas K and Spyridelis J 1991 *Phys. Rev. B* **43** 4135
63. Hanias MP, Kalomiros JA, Karakotsou C, Anagnostopoulos AN and Spyridelis J 1994 *Phys. Rev. B* **49** 16994
64. Watzke O, Schneider T and Martienssen W 2000 *Chaos, Solitons and Fractals* **11** 1163
65. Abdullaev AG and Aliev VK 1980 *Mater. Res. Bull*. **15** 1361
66. Hussein SA *Crystal Research and Technology* 1989 **24** 635
67. Pal S and Bose DN 1996 *Solid State Commun*. **97** 725
68. Parlak H, Ercelebi C, Gunal I, Ozkan H and Gasanly NM 1996 *Crystal Research & Technology* **31** 673
69. Jensen JD, Burke JR, Ernst DW and Allgaier RS 1972 *Phys. Rev. B* **6** 319
70. Ikari T and Hashimoto K 1978 *Phys. Status Solidi (b)* **86** 239
71. Kalkan N, Hanias MP and Anagnostopoulos A 1992 *Mater. Res. Bull.* **27** 1329
72. Karpovich IA, Chervova AA, Demidova LI, Leonov EI and Orlov VM 1972 *Izv. Akad. Nauk SSSR, Neorg. Mater.* **8** 70
73. Bakhyshev AE, Aliev RA, Samedov SR, Efendiev ShM and Tagirov VI 1980 *Fiz. Tekh. Poluprovodn.* **14** 1661
74. Bakhyshev AE, Natig BA, Safuat B, Samedov SR and Abbasov ShM 1990 *Sov.Phys.Semicond.* **24** 828 (1990 *Fiz. Tekh. Poluprovodn.* **24** 1318).
75. Kalomiros JA, Kalkan N, Hanias M, Anagnostopoulos AN and Kambas K 1995 *Solid State Commun.* **96** 601
76. Kalkan N, Kalomiros JA, Hanias M and Anagnostopoulos AN 1996 *Solid State Commun*. **99** 375
77. Kerimova EM, Mustafaeva SN, Kerimov RN and Gadzhieva GA 1999 *Inorg. Mater*. **35** 1123
78. Qasrawi AF and Gasanly NM 2003 *Phys. Status Solidi A* **199** 277
79. Samedov SR and Baykan O 2003 *Intern. J. Infrared and Millimeter Waves* **24** 231
80. Ashraf IM, Abdel-Rahman MM and Badr AM 2003 *J. Phys. D: Applied Physics* **36** 109
81. Ashraf IM 2004 *J. Phys. Chem. B* **108** 10765
82. Porte L and Tranquard A 1980 *J. Solid State Chem.* **35** 59.
83. Kilday DG, Niles DW, Margaritondo G and Levy F 1987 *Phys. Rev. B* **35** 660
84. Okazaki K, Tanaka K, Matsuno J, Fujimori A, Mattheiss LF, Iida S, Kerimova E and Mamedov N 2001 *Phys. Rev. B* **64** 045210
85. Mimura K, Wakita K, Arita M, Mamedov N, Orudzhev G, Taguchi Y, Ichikawa K, Namatame H and Taniguchi M 2007 *J. Electron Spectrosc.&Related Phenomena* **156-158** 379
86. Kholopov EV, Panich AM and Kriger YuG 1983 *Sov. Phys. JETP* **57**, 632 (1983 *Zh. Eksp. Teor. Fiz*. **84** 1091)
87. Kramers HA 1934 *Physica* **1** 184
88. Van Vleck JH 1948 *Phys. Rev.* **74** 1168
89. Bloembergen N and Rowland TJ 1955 *Phys. Rev.* **97** 1679
90. Karimov YS and Schegolev IF 1962 *Soviet Phys. JETF* **14** 772 (1961 *Zh. Eksperim. Teor. Fiz*. **41** 1082).
91. Saito Y 1966 *J. Phys. Soc. Jpn*. **21** 1072
92. Panich AM, Belitskii IA, Gabuda SP, Drebushchak VA and Seretkin YV 1990 *J. Struct. Chem.* **31** 56 (1990 *Zh. Strukturnoi Khimii* **31** 69).
93. Panich AM and Doert Th 2000 *Solid State Commun*. **114** 371
94. Panich AM and Gasanly NM 2001 *Phys. Rev. B* **63** 195201
95. Panich AM and Kashida S 2002 *Physica B* **318** 217
96. Panich AM 1989 *Sov. Phys. Solid State* **31** 1814 (1989 *Fizika Tverdogo tela* **31**, 279).
97. Panich AM, Gabuda SP, Mamedov NT and Aliev SN 1987 *Sov. Phys. Solid State* **29** 2114 (1987 *Fiz. Tverd. Tela (Leningrad)* **29** 3694).
98. Panich AM and Kashida S 2004 *J. Phys. Condens. Matter* **16** 3071
99. Panich AM 2004 *Appl. Magn. Reson.* **27** 29
100. Carrington A and McLachlan AD *Introduction to Magnetic Resonance.* NY: Harper&Row 1967.
101 Orgel LE 1959 *J. Chem. Soc*. **4** 3815
102. Chesnut DB 2003 *Chem. Phys.* **291** 141.
103. Gasanly NM, Akinoglu BG, Ellialtioglu S, Laiho R and Bakhyshov AE 1993 *Physica B* **192** 371.

104. Panich AM, Shao M, Teske CL, Bensch W 2006 *Phys. Rev. B* **74** 233305
105. Panich AM, Teske CL, Bensch W, Perlov A, Ebert H 2004 *Solid State Commun.* **131** 201
106. Gashimzade FM and. Orudzhev GS 1980 *Dokl. Akad. Nauk Azerb. SSR* **36** 18
107. Gashimzade FM and. Orudzhev GS 1981 *Sov. Phys. Semicond.* **15** 757 (1981*Fiz. Tekh. Poluprovodn.* **15**, 1311).
108. Gashimzade FM and. Guliev DG 1985 *Phys. Status Solidi B* **131** 201
109. Orudzhev GS, Efendiev SM and Dzhakhangirov ZA 1995 *Sov. Phys. Solid State* **37** 152 (1995 *Fizika Tverdogo Tela* **37** 284).
110. Orudzhev G, Mamedov N, Uchiki H, Yamamoto N, Iida S, Toyota H, Gojaev E and Hashimzade F 2003 *J. Phys. Chem. Solids* **64** 1703
111. Ellialtoglu S, Mete E, Shaltaf R, Allakhverdiev K, Gashimzade F, Nizametdinova M and Orudzhev G 2004 *Phys. Rev.* B **70** 195118
112. Kashida S, Electronic band structure of TlSe. *Unpublished results*.
113. Shimosaka W and Kashida S 2004 *J. Phys. Soc. Japan* **73** 1532
114. Abdullaeva SG, Mamedov NT and Orudzhev GS 1983 *Phys. Status Solidi* B **119** 41
115. Abdullaeva SG and Mamedov NT 1986 *Phys. Status Solidi B* **133** 171
116. Kashida S, Yanadori Y, Otaki Y, Seki Y and Panich AM 2006 *Phys. Status Solidi (a)* **203** 2666
117. Yee KA and Albright TA 1991 *J. Am. Chem. Soc.* **113** 6474
118. Janiak C and Hoffmann R 1990 *J. Amer. Chem. Soc.* **112** 5924
119. Wagner FR and Stöwe K 2001 *J. Solid State Chem.* **157** 193
120. Gashimzade FM and Orudzhev GS 1981 *Phys. Status Solidi B.* **106** K67
121. Valyukonis GR, Medeishis AS and Shileika AY 1982 *Sov. Phys. Semicond.* **16** 730 (1982 *Fiz. Tekh. Poluprovodn.* **16** 1137).
122. Allakhverdiev KR, Babaev SS, Bakhyshov NA, Mamedov TG, Salaev EY and Efendieva IK 1984 *Phys. Status Solidi B* **126** K139
123. Allakhverdiev KR, Babaev SS, Bakhyshov NA, Mamedov TG, Salaev EY and Efendieva IK 1984 *Sov. Phys. Semicond* **18** 1068
124. Allakhverdiev KR and Ellialtıoglu S 2001 In: *Frontiers of High Pressure Research II: Application of High Pressure to Low-Dimensional Novel Electronic Materials, NATO Science Series II* (Dordrecht: Kluwer) v. 48, p 119
125. Kerimova E, Mustafaeva S, Guseinova D, Efendieva I,. Babaev S. Mamedov TG, Mamedov TS, Salaeva Z and Allkhverdiev K 2000 *Phys. Status Solidi (a)* **179** 199
126. Ves S 1990 *Phys. Status Solidi B* **159** 699
127. Demishev GB, Kabalkina SS and Kolobyanina TN 1988 *Phys. Status Solidi A* **108** 89
128**.** Allakhverdiev KR, Gasymov SG, Mamedov TG, Nizametdinova MA and Salaev EY 1982 Proc. Vses. Conf. Fiz. Poluprovodn., Baku, USSR **1** 82
129. Rzaev KI and Orudzheva SO 1970 Izv. Akad. Nauk Azerb. SSR, ser. Fiz.-Tekh. & Matemat. Nauk **3** 76
130. Morgant G, Legendre B, Maneglier-Lacordaire S and Souleau C 1981 *Annales de Chimie (France)* **6** 315
131. Parthasarathy G, Asokan S, Naik GM and Krishna R 1987 *Philos. Mag. Lett.* **56** 191
132. Geller S, Jayaraman A and Hull GW Jr 1965 *J. Phys. Chem. Solids* **26** 353
133. Geller S, Jayaraman A and Hull GW Jr 1964 *Appl. Phys. Lett.* **4** 35
134. Darnell AJ and Libby WF 1964 *Phys. Rev.* **135** 1453
135. Sclar CB, Carrison LC and Schwartz CM 1964 Science **143** 352
136. Bommel MD, Darnell AJ, Libby WF, Tit BR and Yencha AJ 1963 Science **141** 714
137. Banus MD, Hanneman RE, Strongin M and Gooen K 1963 Science **142** 662
138. Vinogradov EA, Zhizhin GN, Mel'nik NN, Subbotin SI, Panfilov VV, Allakhverdiev KR, Salaev EY and Nani RK 1979 *Phys. Status Solidi B* **95** 383
139. Allakhverdiev KR, Mamedov TG, Panfilov VV, Shukyurov MM and Subbotin SI 1985 *Phys. Status Solidi B* **131** K23
140. Prins AD, Allakhverdiev KR, Babaev SS, Guseinov SS, Mekhtiev EI, Shirinov MM, Tagiev MM and Dunstan DJ 1989 *Phys. Status Solidi B* **151** 759
141. Ves S 1989 *Phys. Rev. B* **40** 7892
142. Perez FV, Cadenas R Power C, Gonzalez J and Chervin CJ 2007 *J. Appl. Phys.* **101** 063534
143. Contreras O, Power C, Gonzalez J and Chervin JC 2003 *Revista Mexicana de Fisica* **49** (Supl.3) 186
144. Range KJ, Maheberd G and Obenland S 1977 Z. Naturforsch. **32** B 1354
145. Range KJ, Engert G, Mueller W and Weiss A 1974 Z. Naturforsch. **29** B 181
146. Allakhverdiev KR In: *Frontiers of High Pressure Research II: Application of High Pressure to Low-Dimensional Novel Electronic Materials, NATO Science Series II* (Dordrecht: Kluwer) Vol. 48, p. 99
147. Morgant G, Legendre B and Souleau C 1982 *Fr. Ann. Chim.* (Paris) **7** 301
148. Romermann F, Febltelai Y, Fries SG and Blachnik R 2000 *Intermetallics* **8** 53
149. Brekow G, Meissner M, Scheiba M, Tausend A and Wobig D 1973 *J. Physics C* **6** L462

150. Kurbanov MM, Godzhaev EM, Guliev LA and Nagiev AB 1982 *Soviet Phys. Solid State* **24** 154 (1982 *Fizika-Tverdogo-Tela* **24**, 274).
151. Mamedov KK, Aldzhanov MA, Mekhtiev MI and Kerimov IG 1980 *J. Engineering Phys.* **39** 1310 (1980 *Inzhenerno-Fizicheskii Zhurnal* **39**, 1005).
152. Mamedov KK, Aldzhanov MA, Kerimov IG and Mekhtiev MI 1978 *Izv. Akad. Nauk Az. SSR, Ser. Fiz.-Tekh. Mat. Nauk* **1** 71
153. Mamedov KK, Kerimov IG, Kostryukov VN and Mekhtiev MI 1967 *Sov. Phys. Semicond.* **67** 363 (1967 *Fiz. Tekh. Poluprovodn.* 1 441).
154. Aliev AM, Nizametdinova MA and Shteinshraiber VY 1981 *Phys.Status Solidi B* **107** K151
155. Mamedov KK, Abdullaev AM and Kerimova EM 1986 *Phys. Status Solidi A* **94** 115
156. Alekperov OZ, Aljanov MA and Kerimova EM 1998 *Turk. J. Phys.* **22** 1053
157. Aliev VA and Aldzhanov MA 1998 *Inorg. Mater* **34** 207
158. Allakhverdiev KR, Salaev FM, Mikhailov FA and Mamedov TS 1992 *Sov. Phys. Solid State* **34** 1938 (1992 *Fizika Tverdogo Tela (Sankt-Peterburg)* **34** 3615).
159. Aldzhanov MA and Mamedov KK 1985 *Fiz. Tverd. Tela* (Leningrad) **27** 3114
160. Banys J, Wondre FR and Guseinov G 1990 *Mater. Lett.* **9** 269
161. Aliev VA, Aldganov MA and Aliev SN 1987 *JETP Lett.* **45** 534 (1987 *Pis'ma Zh. Eksp. Teor. Fiz.* **45** 418).
162. Godzhaev EM and Kafarova DM 2004 *Inorg. Mater.* **40** 924
163. Mamedov NT and Panich AM 1990 *Phys. Stat. Sol (a)* **117** K15
164. Kashida S, Nakamura K and Katayama S 1993 *J. Phys: Condens. Matter* **5** 4243
165. Kashida S 1994 *Ferroelecrtrics* **151** 165
166. Panich AM and Kashida S (*unpublished results*).
167. Sardarly RM, Abdullaev AP, Guseinov GG, Nadzhafov AI and Eyubova NA 2000 *Crystallogr. Rep.* **45** 551 (2000 *Kristallografiya* **45,** 606).
168. Aliev VP, Gasimov SG, Mammadov Tg, Nadjafov AI and Seyidov My *2006 Phys. Solid State* **48** 2322 (2006 *Fizika Tverdogo Tela* **48** 2194).
169. Volkov AA, Goncharov YG, Kozlov GV, Allakhverdiev KR and Sardarly RM 1983 *Fiz. Tverd. Tela (Leningrad)* **25** 3583
170. Vakhrushev SB, Zhdanova VV, Kvyatkovskii BE, Okuneva NM, Allakhverdiev KR, Aliev RA and Sardarly RM 1984 *JETF Lett.* **39** 291 (1984 *Pis'ma Zh. Eksp. Teor. Fiz.* **39** 245).
171. Aliev RA, Allakhverdiev KR, Baranov AI, Ivanov NR and Sardarly RM 1984 *Sov. Phys. Solid State* **26** 775 (1984 *Fiz. Tverd. Tela (Leningrad)* **26** 1271).
172. Abdullaev NA, Allakhverdiev KR, Belenkii GL, Mamedov TG, Suleimanov RA and Sharifov YN 1985 *Solid State Commun.* **53** 601
173. Banys J 1999 *Liet. Fiz. Z.* **39** 33
174. Sheleg AU, Iodkovskaya KV, Rodin SV and Aliev VA 1997 *Phys. Solid State* **39** 975 (1997 *Fiz. Tverd. Tela (S.-Peterburg)* **39** 1088).
175. Krupnikov ES, Aliev FY and Orudzhev RG 1992 *Sov. Phys. Solid State* **34** 1574 (1992 *Fiz. Tverd. Tela (S.-Peterburg)* **34**, 2935).
176. Abdullaev AM, Kerimova EM and Famanova AK 1994 *Inorg. Mater.***30** 887
177. Ozdemir S, Suleymanov RA and Civan E 1995 *Solid State Commun.* **96** 757
178. Ozdemir S, Suleymanov RA, Civan E and Firat, T 1996 *Solid State Commun.* **98** 385
179. Ozdemir S and Suleymanov RA 1997 *Solid State Commun.* **101** 309
180. Youssef SB 1995 *Physica A* **215** 176
181. Suleimanov RA, Seidov MY, Salaev FM and Mikailov FA 1993 *Soviet Phys. Solid State* **35** 177 (1993 *Fiz. Tverd. Tela (S.-Peterburg)* **35**, 348).
182. Salaev FM, Allakhverdiev KR and Mikailov FA 1992 *Ferroelectrics* **131** 163
183. Gadzhiev BR, Seidov MGY and Abdurakhmanov VR 1996 *Phys. Solid State* **38** 1 (1996 *Fiz. Tverd. Tela (S.-Peterburg)* **38** 3).
184. Allakhverdiev KR, Turetken N, Salaev FM and Mikailov FA 1995 *Solid State Commun.* **96** 827.
185. Banys J, Brilingas A and Grigas J 1990 *Phase Transitions* **20** 211
186. Mikailov FA, Basaran E and Senturk E 2001 *J. Phys.: Condens. Matter* **13** 727
187. Mikailov FA, Basaran E and Senturk E 2002 *Solid State Commun.* **122** 161
188. Mikailov FA, Basaran E, Mammadov TG, Seyidov MY and Senturk E 2003 *Physica B* **334** 13
189. Gadjiev BR 2004 Los Alamos Nat. Lab. arXiv:cond-mat/0403667 http://xxx.lanl.gov/pdf/cond-mat/0403667
190. Mikailov FA, Rameev BZ, Kulibekov AM, Senturk E and Aktas B 2003 J. *Magn. Magn. Mater.* **258** 419
191. Mikailov FA, Rameev BZ, Kazan S, Yildiz F, Mammadov TG and Aktas B 2004 *Phys. Status Solidi C* **1** 3567
192. Gorelik VS, Agal'tsov AM and Ibragimov TD 1988 *J. Appl. Spectroscopy* **49** 661
193. Ibragimov TD and Aslanov II 2002 *Solid State Commun.* **123** 339
194. Ibragimov TD, Sardarly RM and Aslanov II 2001 *J. Appl. Spectrosc.* **68** 711
195. Ibragimov TD 2003 *J. Appl. Spectrosc* **70** 99

196. Allakhverdiev K, Salaev E, Salaeva Z, Onari S, Kulibekov A and Mamedov T 2001 *Phase Transitions* **73** 579
197. Gadjiev BR 2003 *Ferroelectrics* **291** 111
198. Mamedov N, Shim Y and Yamamoto N 2002 *Japanese J. Appl. Phys.1* **41** 7254
199. Yamamoto N, Mamedov N, Shinohara T and Kunie A 2002 *J. Cryst. Growth* **237-239** 2023
200. Burlakov VM, Vinogradov EA, Nurov S, Gasanly NM and Ismailov YG 1985 *Soviet Physics Solid State* **27** 2025 (*Fiz. Tverd. Tela (Leningrad)* **27** 3365)
201. Allakhverdiev KR, Babaev SS, Bakhyshov NA and Mamedov TG 1985 *Fiz. Tverd. Tela (Leningrad)* **27** 3699
202. Laiho R, Levola T, Sardarly RM, Allakhverdiev KR, Sadikov IS and Tagiev MM 1987 *Solid State Commun*. **63** 1189
203. Laiho R and Sardarly RM 1987 *Ferroelectrics* **80** 185
204. Allakhverdiev KR, Baranov AI, Mamedov TG, Sandler VA and Sharifov YN 1988 *Ferroelectr. Lett*. **8** 125
205. Mal'sagov AU, Kul'buzhev BS and Khamkhoev BM 1989 *Izv. Akad. Nauk SSSR, Neorg. Mater*. **25** 216
206. Suleimanov RA, Seidov MY, Salaev FM and Mamedov TS 1992 *Sov. Phys. Solid State* **34** 976 (1992 *Fiz. Tverd. Tela (S.-Peterburg)* **34** 1829).
207. Gamal GA 1997 *Cryst. Res. Technol*. **32** 561
208. Plyushch OB and Sheleg AU 1999 *Crystallography Reports* **44** 813 (*Kristallografiya* **44** 873).
209. Sardarly RM, Samedov OA, Nadzhafov AI and Sadykhov IS 2003 *Phys.Solid State* **45** 1137 (2003 *Fizika Tverdogo Tela* **45** 1085).
210. Sardarly RM, Samedov OA, Sadykhov IS and Aliev VA 2003 *Phys. Solid State* **45** 1118 (2003 *Fiz. Tverd. Tela (S.-Peterburg)* **45**, 1067).
211. Sardarli RM, Samedov OA and Sadigov IS 2004 *Ferroelectrics* **298** 275
212. Sardarly RM, Samedov OA, Sadykhov IS 2004 *Inorganic Materials* **40** 1018 (2004 *Neorg. Mater*. **40** 1).
213. Sardarly RM, Samedov OA, Sadykhov IS 2004 *Phys. Solid State* **46** 1917 (2004 *Fiz. Tverd. Tela (S.-Peterburg)* **46** 1852).
214. Sardarli A, Filanovsky IM, Sardarli RM, Samedov OA, Sadigov IS and Aslanov AI 2003 *Can. J. Optoelectronics&Advanced Mater*. **5** 741
215. Sardarly RM, Samedov OA, Sadykhov IS, Nadzhafov AI and Salmanov FT 2005 *Phys. Solid State* **47** 1729 (2005 *Fizika Tverdogo Tela* **47** 1665).
216. Sardarly RM, Mamedov NT, Wakita K, Shim Y, Nadjafov AI, Samedov OA and Zeynalova EA 2006 *Phys. Status Solidi A* **203** 2845
217. Volkov AA, Goncharov YG, Kozlov GV, Lebedev SP, Prokhorov AM, Aliev RA and Allahverdiev KP 1983 Sov. *JETP Lett*., **37** 615 (1983 *Pis'ma Zh. Eksp. Teor. Fiz*. **37** 517).
218. Volkov AA, Goncharov YG, Kozlov GV and Sardarly RM 1984 *JETP Lett*. **39** 351 (1984 *Pis'ma Zh. Eksp. Teor. Fiz*. **39** 293).
219. Hochheimer HD, Gmelin E, Bauhofer W, von Schnering-Schwarz C, von Schnering HG, Ihringer J and Appel W 1988 *Z. Phys. B – Condensed Matter* **73** 257
220. Abdullaeva SG, Abdullaev AM, Mamedov KK and Mamedov NT 1984 *Sov. Phys. Solid State* **26** 375 (1984 *Fiz. Tverd. Tela (Leningrad)* **26** 618).
221. McMorrow DF, Cowley RA, Hatton PD and Banys J 1990 *J. Phys. Condens. Matter* **2**, 3699
222. Kashida S and Kobayashi Y 1998 *J. Korean Phys. Soc*. **32** S40
223. Allakhverdiev KR, Aldzanov MA, Mamedov TG and Salaev EY 1986 *Solid State Commun*. **58** 295
224. Aldzhanov MA, Guseinov NG and Mamedov ZN 1987 *Phys. Status Solidi A* **100** K145
225. Gololobov YP and Perga VM 1992 *Fiz. Tverd. Tela (S.-Peterburg)* **34** 115
226. Mikailov FA, Basaran E, Senturk E, Tumbek L, Mammadov TG and Aliev VP 2004 *Solid State Commun.* **129** 761
227. Abutalybov GI, Larionkina LS and Ragimova NA 1989 *Fiz. Tverd. Tela (Leningrad)* **31** 312
228. Mamedov NT, Krupnikov ES and Panich AM 1989 *Sov. Phys. Solid State* **31** 159 (1989 *Fiz. Tverd. Tela (Leningrad)* **31** 290).
229. Abdullaeva SG, Mamedov NT, Mamedov SS and Mustafaev FA 1987 *Fiz. Tverd. Tela (Leningrad)* **29** 3147
230. Aliev VA 1990 *Kristallografiya* **35** 506
231. Gabuda SP, Kozlova SG and Mamedov NT 1990 *Sov. Phys. Solid State* **32** 995 (1990 *Fiz. Tverd. Tela (Leningrad)* **32** 1708)
232. Gololobov YP, Shilo SA and Yurchenko IA 1990 *Ukr. Fiz. Zh*. **35** 1721
233. Belyaev AD, Gololobov YP and Dubrova KS 1991 *Ukr. Fiz. Zh*. **36** 1258
234. Gololobov YP, Shilo SA and Yurchenko IA 1991 *Fiz. Tverd. Tela (Leningrad)* **33** 2781
235. Mikailov FA, Basaran E, Sentuerk E, Tuembek L, Mammadov TG and Aliev VP 2003 *Phase Transit.* **76** 1057
236. Panich AM, Ailion D, Kashida S and Gasanly N 2004 XIII Int. Conf.Hyperfine Interactions - XVII Int. Symp. Nuclear Quadrupole Interaction, Bonn, Germany, 23 - 27 August 2004, *Program and Abstracts, p. O-G-5*.
237. Panich AM, Ailion D, Kashida S and Gasanly N 2004 The 1st EENC-AMPERE Joint Meeting, Lille, France, 6-11 September 2004. *Program, p.16, PO-193.*
238. Panich AM and Kashida S *submitted*

239. Abragam A, 1961, *The Principles of Nuclear Magnetism*: Oxford, Clarendon Press.
240. Alekperov OZ 2003 *J. Phys. Chem. Solids* **64** 1707
241. Senturk E, Tumbek L, Salehli F and Mikailov FA 2005 *Cryst. Res. Technol.* **40** 248
242. Senturk E 2006 *Physics Letters A* **349** 340
243. Senturk E and Mikailov FA 2006 *Cryst. Res.Technol.* **41** 1131
244. Mikailov FA, Basaran E, Tuembek L, Sentuerk E and Mammadov TG 2005 *J.Non-Crystalline Solids* **351** 2809
245. Mikailov FA, Sentuerk E, Tuembek L, Mammadov TG and Mammadov TS 2005 *Phase Transitions* **78** 413
246. Babaev SS, Basaran E, Mammadov TG, Mikailov FA, Salehli FM, Seyidov MHY and Suleymanov RA 2005 *J. Phys.: Condens. Matter* **17** 1985
247. Aliyev VP, Babayev SS, Mammadov TG, Seyidov MHY Suleymanov RA 2003 *Solid State Commun.* **128** 25
248. Krupnikov ES and Abutalybov GI 1992 *Sov. Phys. Solid State* **34** 1591 (1992 *Fizika Tverdogo Tela (St-Peterburg)* **34** 2964).
249. Aydinli A, Elliatioglu R, Allakhverdiev KR Ellialtioglu S, Gasanly NM 1993 *Solid State Commun.* **88** 387
250. Ates A, Gurbulak B, Yildirim M, Dogan S, Duman S, Yildirim T and Tuzemen S 2002 *Turkish J. Phys.* **26** 127
251. Dzhafarova SZ, Ragimova NA and Abutalybov GI 1991 *Phys. Status Solidi A* **126** 501
252. Durnev YI, Kul'buzhev BS, Torgashev VI and Yuzyuk YI 1989 *Izv. Akad. Nauk SSSR Ser. Fiz.* **53** 1300
253. Gorban I, Okhrimenko OB and Guseinov GD 1991 *Ukr. Fiz. Zh.* **36** 357
254. Kashida S, *private communication*
255. Volkov AA, Goncharov YG, Kozlov GV, Allakhverdiev KR and Sardarly RM 1984 *Fiz. Tverd. Tela (Leningrad)* **26** 2754
256. Aldzhanov MA, Guseinov NG, Mamedov ZN and Abdurragimov AA 1987 *Dokl.Akad.Nauk Azerb. SSR* **43** 23
257. Duman S and Guerbulak B 2005 *Physica Scripta* **72** 79
258. Tekhanovich NP, Sheleg AU, Aliev VA 1992 *Sov. Phys. Solid State* **34** 1038 (1992 *Fiz. Tverd. Tela (St. Petersburg*) **34** 1946).
259. Golubev LV, Vodop'yanov LK, Allakhverdiev KR and Sardarly RM 1980 *Fiz.Tverd. Tela (Leningrad)* **22** 2529
260. Burlakov VM, Nurov S and Ryabov AP 1988 *Fiz. Tverd. Tela (Leningrad)* **30** 3616
261. Ni Y, Shao M, Wu Z, Gao F and Wei X 2004 *Solid State Commun*. **130** 297
262 Panich AM, Teske CL and Bensch W 2006 *Phys.Rev. B* **73** 115209